\documentclass{article}
\usepackage{epsf}
\usepackage{longtable}
\usepackage{epubtk}

\def\al{\alpha}
\def\be{\beta}
\def\ga{\gamma}
\def\de{\delta}
\def\ep{\epsilon}

\def\et{\eta}
\def\th{\theta}

\def\ka{\kappa}
\def\la{\lambda}

\def\si{\sigma}

\def\ps{\psi}
\def\om{\omega}

\def\De{\Delta}

\def\lrD{\stackrel{\leftrightarrow}{D}}
\def\lrDnu{\stackrel{\leftrightarrow}{D^\nu}}

\begin{document}

\title{Modern Tests of Lorentz Invariance}

\author{
\epubtkAuthorData{David Mattingly}
                 {University of California at Davis}
                 {mattingly@physics.ucdavis.edu}
                 {}
}

\date{}
\maketitle

\begin{abstract}
Motivated by ideas about quantum gravity,  a tremendous amount of
effort over the past decade has gone into testing Lorentz
invariance in various regimes.  This review summarizes both the
theoretical frameworks for tests of Lorentz invariance and
experimental advances that have made new high precision tests
possible. The current constraints on Lorentz violating effects
from both terrestrial experiments and astrophysical observations
are presented. \\
\\
\end{abstract}

\epubtkKeywords{Lorentz invariance, quantum gravity phenomenology}

\newpage

\section{Introduction} \label{sec:Introduction}
Relativity has been one of the most successful theories of the
last century and is a cornerstone of modern physics. This review
focuses on the modern experimental tests of one of the fundamental
symmetries of relativity, Lorentz invariance.  Over the last
decade there has been tremendous interest and progress in testing
Lorentz invariance. This is largely motivated by two factors.
First, there have been theoretical suggestions that Lorentz
invariance may not be an exact symmetry at all energies. The
possibility of four dimensional Lorentz violation has been
investigated in different quantum gravity models (including string
theory~\cite{Kostelecky:1988zi, Ellis:1999yd}, warped brane
worlds~\cite{Burgess:2002tb}, and loop quantum
gravity~\cite{Gambini:1998it}), although no quantum gravity model
predicts Lorentz violation conclusively. Other high energy models
of spacetime structure, such as non-commutative field theory, do
however explicitly contain Lorentz
violation~\cite{Douglas:2001ba}.  High energy Lorentz violation
can regularize field theories, another reason it may seem
plausible. Even if broken at high energies Lorentz symmetry can
still be an attractive infrared fixed point, thereby yielding an
approximately Lorentz invariant low energy
world~\cite{Chadha:1982qq}. Other ideas such as emergent gauge
bosons~\cite{Bjorken:2001pe,Kraus:2002sa,Jenkins:2003hw,
Chkareuli:2004ci}, varying moduli~\cite{Damour:1994zq},
axion-Wess-Zumino models~\cite{Andrianov:1998ay}, analogues of
emergent gravity in condensed
matter~\cite{Barcelo:2005fc,Novello:2002qg}, ghost
condensate~\cite{Arkani-Hamed:2003uy}, space-time varying
couplings~\cite{Kostelecky:2002ca, Bertolami:2003qs}, or varying
speed of light cosmologies~\cite{Moffat:1992ud, Magueijo:2003gj}
also incorporate Lorentz violation.   The ultimate fate of Lorentz
invariance is therefore an important theoretical question.

We shall primarily focus on quantum gravity induced Lorentz
violation as the theoretical target for experimental tests. If
Lorentz invariance is violated by quantum gravity, the natural
scale one would expect it to be strongly violated at is the Planck
energy of $\approx 10^{19}$ GeV. While perhaps theoretically
interesting, the large energy gap between the Planck scale and the
highest known energy particles, the trans-GZK cosmic rays of
$10^{11}$ GeV (not to mention accelerator energies of $\sim 1$
TeV), precludes any direct observation of Planck scale Lorentz
violation.

Fortunately, it is very likely that strong Planck scale Lorentz
violation yields a small amount of violation at much lower
energies.  If Lorentz invariance it is violated at the Planck
scale, there must be an interpolation to the low energy, (at least
nearly) Lorentz invariant world we live in.  Hence a small amount
of Lorentz violation should be present at all energies. Advances
in technology and observational techniques have dramatically
increased the precision of experimental tests, to the level where
they can be sensitive to small low energy residual effects of
Planck scale Lorentz violation.  These experimental advances are
the second factor stimulating recent interest in testing Lorentz
invariance. One should keep in mind that low energy experiments
cannot directly tell us whether or not quantum gravity is Lorentz
invariant. Rather, they can only determine if the ``state'' that
we live in is Lorentz violating. For example, it is possible that
quantum gravity might be Lorentz invariant but contain tensor
fields that acquire a vacuum expectation value at low
energies~\cite{Kostelecky:1988zi}, thereby spontaneously breaking
the symmetry. Experiments carried out at low energies would
therefore see Lorentz violation, even though it is a good symmetry
of the theory at the Planck scale. That said, any discovery of
Lorentz violation would be an important signal of beyond standard
model physics.

There are currently a number of different theoretical frameworks
in which Lorentz symmetry might be modified, with a parameter
space of possible modifications for each framework. Since many of
the underlying ideas come from quantum gravity, which we know
little about, the fate of Lorentz violation varies widely between
frameworks. Most frameworks explicitly break Lorentz invariance,
in that there is a preferred set of observers or background field
other than the metric~\cite{Colladay:1998fq, Arkani-Hamed:2003uy}.
However others try to deform the Poincar\'{e} algebra, which would
lead to modified transformations between frames but no preferred
frame (for a review see ~\cite{Kowalski-Glikman:2004qa}). These
latter frameworks lead to only ``apparent'' low energy Lorentz
violation. Even further complications arise as some frameworks
violate other symmetries, such as CPT or translation invariance,
in conjunction with Lorentz symmetry.  The fundamental status of
Lorentz symmetry, broken or deformed,  as well as the additional
symmetries makes a dramatic difference as to which experiments and
observations are sensitive. Hence the primary purpose of this
review is to delineate various frameworks for Lorentz violation
and catalog which types of experiments are relevant for which
framework. Theoretical issues relating to each framework are
touched on rather briefly, but references to the relevant
theoretical work are included.

Tests of Lorentz invariance span atomic physics, nuclear physics,
high-energy physics, relativity, and astrophysics.  Since
researchers in so many disparate fields are involved, this review
is geared towards the non-expert/advanced graduate level, with
descriptions of both theoretical frameworks and
experimental/observational approaches. Some other useful starting
points on Lorentz violation are~\cite{Amelino-Camelia:2005qa,
Will:1993ns, Kosteleckybook,Jacobson:2005bg}. The structure of
this review is as follows. An general overview of various issues
relating to the interplay of theory with experiment is given in
section \ref{sec:Issues}. The current theoretical frameworks for
testing Lorentz invariance are given in sections
\ref{sec:KinematicFrameworks} and \ref{sec:DynamicalFrameworks}. A
discussion of the various relevant results from earth based
laboratory experiments, particle physics, and astrophysics is
given in sections \ref{sec:ExperimentsEarth} and
\ref{sec:ExperimentsAstro}. Limits from gravitational observations
are in section \ref{sec:GravitationalObservations}.  Finally, the
conclusions and prospects for future progress are in section
\ref{sec:Conclusions}. Throughout this review $\eta^{\alpha
\beta}$ denotes the Minkowski $(+\,-\,-\,-)$ metric. Greek indices
will be used exclusively for spacetime indices whereas Roman
indices will be used in various ways. Theorists' units $\hbar=c=1$
are used throughout. $E_{Pl}$ denotes the (approximate) Planck
energy of $10^{19}$ GeV.

\section{General Considerations} \label{sec:Issues}
\subsection{Defining Lorentz violation} \label{sec:WhatIsLV}
\subsubsection{Lorentz violation in field theory}
Before we discuss Lorentz violation in general, it will be useful
to detail a pedagogical example that will give an intuitive feel
as to what ``Lorentz violation'' actually means. Let us work in a
field theory framework and consider a ``bimetric'' action for two
massless scalar fields $\Phi, \Psi$
\begin{equation} \label{eq:bimetric}
S=\frac {1} {2} \int \sqrt{-g} d^4x \left( g^{\alpha \beta}
\partial_\alpha \phi
\partial_\beta \phi + (g^{\alpha \beta} + \tau^{\alpha \beta}) \partial_\alpha \psi
\partial_\beta \psi\right)
\end{equation}
where $\tau^{\alpha \beta}$ is some arbitrary symmetric tensor,
not equal to $g^{\alpha \beta}$.  Both $g^{\al \be}$ and
$\tau^{\al \be}$ are fixed background fields.  At a point, one can
always choose coordinates such that $g^{\al \be}=\eta^{\al \be}$.
Now, consider the action of local Lorentz transformations at this
point, which we define as those transformations for which
$\eta_{\al \be}$ is invariant, on $S$.\footnote{Note that since
there are two metrics, $\eta^{\alpha \beta}$ and $\eta^{\alpha
\beta} + \tau^{\alpha \beta}$ there can be two different sets of
transformations that leave one of the metrics invariant.  In this
sense there are two Lorentz groups.} $S$ is a spacetime scalar, as
it must be to be well-defined and physically meaningful. Scalars
are by definition invariant under all passive diffeomorphisms
(where one makes a coordinate transformation of \textit{every}
tensor in the action, background fields included). A local Lorentz
transformation is a subgroup of the group of general coordinate
transformations so the action is by construction invariant under
``passive'' local Lorentz transformations.   This implies that as
long as our field equations are kept in tensorial form we can
freely choose what frame we wish to calculate in.  Coordinate
invariance is sometimes called ``observer Lorentz invariance'' in
the literature~\cite{Kostelecky:2001xz} although it really has
nothing to do with the operational meaning of Lorentz symmetry as
a physical symmetry of nature.

Lorentz invariance of a physical system is based upon the idea of
``active'' Lorentz transformations, where we only transform the
\textit{dynamical fields} $\phi$ and $\psi$. Consider a Lorentz
transformation of $\phi$ and $\psi$,
\begin{eqnarray}
\phi'(x)=\phi((\Lambda^{-1})^\mu_\nu x^\nu) \\ \nonumber
\psi'(x)=\psi((\Lambda^{-1})^\mu_\nu x^\nu)
\end{eqnarray}
where $\Lambda^\mu_\nu$ is the Lorentz transformation matrix,
$x'^\mu=\Lambda^\mu_\nu x^\nu$. The derivatives transform as
\begin{eqnarray}
\partial_\nu \phi'(x) =(\Lambda^{-1})^\mu_\nu \partial_\mu
\phi((\Lambda^{-1})^\al_\be x^\be)\\ \nonumber
\partial_\nu \psi'(x) =(\Lambda^{-1})^\mu_\nu \partial_\mu
\psi((\Lambda^{-1})^\al_\be x^\be)
\end{eqnarray}
from which one can easily see that $\eta^{\al \be} \partial_\al
\phi'(x) \partial_\be \phi'(x)=\eta^{\al \be} \partial_\al \phi(x)
\partial_\be \phi(x)$ since by definition $\eta^{\al \be}
(\Lambda^{-1})^\mu_\be (\Lambda^{-1})^\nu_\al = \eta^{\mu \nu}$.
The $\eta^{\al \be}$ terms are therefore Lorentz invariant.
$\tau^{\al \be}$ is not, however, invariant under the action of
$\Lambda^{-1}$ and hence the action violates Lorentz invariance.
Equations of motion, particle thresholds, etc. will all be
different when expressed in the coordinates of relatively boosted
or rotated observers.

Since in order for a physical theory to be well defined the action
must be a spacetime scalar, breaking of active Lorentz invariance
is the \textit{only} physically acceptable type of Lorentz
violation.  Sometimes active Lorentz invariance is referred to as
``particle'' Lorentz invariance~\cite{Kostelecky:2001xz}. We will
only consider active Lorentz violation and so shall drop any
future labelling of ``observer'', ``particle'',``active'', or
``passive'' Lorentz invariance.  For the rest of this review,
Lorentz violation always means active Lorentz violation. For
another discussion of active Lorentz symmetry in field theory
see~\cite{Peskin:1995ev}. Since we live in a world where Lorentz
invariance is at the very least an excellent approximate symmetry,
$\tau^{\alpha \beta}$ must be small in our frame.  In field
theoretical approaches to Lorentz violation, a frame in which all
Lorentz violating coefficients are small is called a
\textit{concordant frame}~\cite{Kostelecky:2000mm}.

\subsubsection{Modified Lorentz groups}
Almost all models for Lorentz violation fall into the framework
above, where there is a preferred set of concordant frames
(although not necessarily a field theory description). In these
theories Lorentz invariance is broken; there is a preferred set of
frames where one can experimentally determine that Lorentz
violation is small.  A significant alternative that has attracted
attention is simply modifying the way the Lorentz group acts on
physical fields. In the discussion above, it was assumed that
everything transformed linearly under the appropriate
representation of the Lorentz group.  On top of this structure,
Lorentz non-invariant tensors were introduced that manifestly
broke the symmetry but the group action remained the same.  One
could instead modify the group action itself in some manner.  A
partial realization of this idea is provided by so-called ``doubly
special
relativity''~\cite{Amelino-Camelia:2000mn,Kowalski-Glikman:2004qa}
(DSR), which will be discussed more thoroughly in Sec.
\ref{subsec:DSR}. In this scenario there is still Lorentz
invariance, but the Lorentz group acts non-linearly on physical
quantities.  The new choice of group action leads to a new
invariant energy scale as well as the invariant velocity $c$
(hence the name doubly special). The invariant energy scale
$\lambda_{DSR}$ is usually taken to be the Planck energy. There is
no preferred class of frames in these theories, but it still leads
to Lorentz ``violating'' effects. For example, there is a
wavelength dependent speed of light in DSR models. This type of
violation is really only ``apparent'' Lorentz violation. The
reader should understand that it is a violation only of the usual
\textit{linear} Lorentz group action on physical quantities.

\subsection{Kinematics vs. dynamics} \label{subsec:kinvsdyn}
A complete physical theory must obviously include dynamics.
However, over the years a number of kinematic frameworks have been
developed for testing Lorentz violation that do not postulate a
complete dynamics~\cite{Robertson, MS, THeu,
Amelino-Camelia:1997gz}. Furthermore, some proposals coming from
quantum gravity are at a stage where the low energy kinematics are
partially understood/conjectured, but the corresponding dynamics
are not understood (a good example of this is
DSR~\cite{Kowalski-Glikman:2004qa}). Hence until these models
become more mature, only kinematic tests of Lorentz invariance are
truly applicable. Strictly enforced, this rule would preclude any
use of an experiment that relies on particle interactions, as
these interactions are determined by the dynamics of the theory.
Only a select few observations, such as interferometry,
birefringence, Doppler shifts, or time of flight are by
construction insensitive to dynamics. However, the observational
situation is often such that tests that use particle interactions
can be applied to theories where only the kinematics is
understood.  This can be done in astrophysical threshold
interactions as long as the dynamics are assumed to be not
drastically different from Lorentz invariant physics (see section
\ref{subsec:ThreshIntro}). In terrestrial experiments, one must
either recognize that different experiments can give different
values with kinematic frameworks (for an example, see the
discussion of the Robertson-Mansouri-Sexl framework in section
\ref{subsec:RMS}) or embed the kinematics in a fully dynamical
model like the standard model extension (section
\ref{subsubsec:mSME}).

\subsection{The role of other symmetries}
There are many other symmetries that affect how Lorentz violation
might manifest itself below the Planck scale.  The standard model
in Minkowski space is invariant under four main symmetries, three
continuous and one discrete. There are two continuous spacetime
symmetries, Lorentz symmetry and translation symmetry, as well as
gauge and CPT symmetry. Supersymmetry can also have profound
effects on how Lorentz violation can occur. Finally, including
gravity means that we must take into account diffeomorphism
invariance. The fate of these other symmetries in conjunction with
Lorentz violation can often have significant observational
ramifications.

\subsubsection{CPT Invariance} \label{subsubsec:CPT} Lorentz symmetry
is intimately tied up with CPT symmetry in that the assumption of
Lorentz invariance is required for the CPT theorem \cite{Jost}.
Lorentz violation therefore allows for (but does not require) CPT
violation, even if the other properties of standard quantum field
theory are assumed.  Conversely, however, CPT violation implies
Lorentz violation for local field theories~
\cite{Greenberg:2002uu}. Furthermore, many observational results
are sensitive to CPT violation but not directly to Lorentz
violation.  Examples of such experiments are kaon decay (section
\ref{subsec:Mesons}) and $\gamma$-ray birefringence (section
\ref{subsec:Birefringence}), both of which indirectly provide
stringent bounds on Lorentz violation that incorporates CPT
violation. Hence CPT tests are very important tools for
constraining Lorentz violation. In effective field theory CPT
invariance can explicitly be imposed to forbid a number of
strongly constrained operators. For more discussion on this point
see section \ref{subsubsec:InducedLowEnergyLV}.

\subsubsection{Supersymmetry} \label{subsubsec:Supersymmetry}
SUSY, while related to Lorentz symmetry, can still be an exact
symmetry even in the presence of Lorentz violation. Imposing exact
SUSY provides another custodial symmetry that can forbid certain
operators in Lorentz violating field theories.  If, for example,
exact SUSY is imposed in the MSSM (minimal supersymmetric standard
model), then the only Lorentz violating operators that can appear
have mass dimension five or above~\cite{GrootNibbelink:2004za}. Of
course, we do not have exact SUSY in nature. The size of low
dimension Lorentz violating operators in a theory with Planck
scale Lorentz violation and low energy broken SUSY has recently
been analyzed in~\cite{Bolokhov:2005cj}. For more discussion on
this point see Sec. \ref{subsubsec:InducedLowEnergyLV}.

\subsubsection{Poincar\'{e} Invariance} \label{subsubsec:Poincare}
In many astrophysics approaches to Lorentz violation, conservation
of energy-momentum is used along with Lorentz violating dispersion
relations to give rise to new particle reactions.  Absence of
these reactions then yields constraints. Energy/momentum
conservation between initial and final particle states requires
translation invariance of the underlying spacetime and the Lorentz
violating physics. Therefore we can apply the usual conservation
laws only if the translation subgroup of the Poincar\'{e} group is
left unmodified. If Lorentz violation happens in conjunction with
a modification of the rest of the Poincar\'{e} group, then it can
happen that modified conservation laws must be applied to
threshold reactions.  This is the situation in DSR: all reactions
that are forbidden by conservation in ordinary Lorentz invariant
physics are also forbidden in DSR~\cite{Heyman:2003hs}, even
though particle dispersion relations in DSR would naively allow
new reactions. The conservation equations change in such a way as
to compensate for the modified dispersion relations (see section
\ref{subsec:DSR}). Due to this unusual (and useful) feature, DSR
evades many of the constraints on effective field theory
formulations of Lorentz violation.

\subsection{Diffeomorphism invariance and prior
geometry}\label{subsubsec:Diffeomorphism} If Lorentz violating
effects are to be embedded in an effective field theory, then new
tensors must be introduced that break the Lorentz symmetry (c.f.
the bimetric theory (\ref{eq:bimetric}) of Sec.
\ref{sec:WhatIsLV}). If we are considering only special
relativity, then keeping these tensors as constant is viable.
However, any complete theory must include gravity, of course, and
one should preserve as many fundamental principles of general
relativity as possible while still introducing local Lorentz
violation. There are three general principles in general
relativity relevant to Lorentz violation: general covariance
(which implies both passive and active diffeomorphism
invariance~\cite{Rovelli:2004tv}), the equivalence principle, and
lack of prior geometry.  As we saw in section \ref{sec:Issues},
general covariance is automatically a property of an appropriately
formulated Lorentz violating theory, even in flat space.  The fate
of the equivalence principle we deal with below in section
\ref{subsec:equiv}. The last principle, lack of prior geometry, is
simply a statement that the metric is a dynamical object on the
same level as any other field. Coupled with diffeomorphism
invariance this leads to conservation of matter stress tensors
(for a discussion see ~\cite{Carroll:2004st}).  However, a fixed
Lorentz violating tensor constitutes prior geometry in the same
way that a fixed metric would.  If we keep our Lorentz violating
tensors as fixed objects, we immediately have non-conservation of
stress tensors and inconsistent Einstein equations. As a specific
example, consider again the bimetric theory (\ref{eq:bimetric}).
We will include gravity in the usual way by adding the
Einstein-Hilbert Lagrangian for the metric. The resultant action
is
\begin{equation} \label{eq:bimetricwgrav}
S=\int d^4x \sqrt{-g}\left(\frac {1} {16 \pi G} R+ \frac {1} {2}
g^{\alpha \beta} \nabla_\alpha \phi \nabla_\beta \phi + \frac {1}
{2} (g^{\alpha \beta} + \tau^{\alpha \beta}) \nabla_\alpha \psi
\nabla_\beta \psi\right)
\end{equation}
and the corresponding field equations are
\begin{eqnarray} \label{eq:bimetriceqs}
\nabla^\alpha \nabla_\alpha \phi =0 \\
\nabla^\alpha \nabla_\alpha \psi + \nabla_\alpha (\tau^{\alpha \beta} \nabla_\beta \psi) =0  \\
\label{eq:bimetricEE} G_{\alpha \beta} = \nabla_\alpha \phi
\nabla_\beta \phi - \frac {1} {2} \nabla^\sigma \phi \nabla_\sigma
\phi g_{\alpha \beta} + \nabla_\alpha \psi \nabla_\beta \psi -
\frac {1} {2} \nabla^\sigma \psi \nabla_\sigma \psi g_{\alpha
\beta}-\frac{1} {2} \tau^{\rho \sigma}\nabla_\rho \psi
\nabla_\sigma \psi g_{\alpha \beta}.
\end{eqnarray}
Taking the divergence of (\ref{eq:bimetricEE}) and using the
$\phi,\psi$ equations of motion yields
\begin{equation} \label{eq:bimetrictaurestrict}
0=-\nabla_\beta \psi \nabla_\rho(\tau^{\rho \sigma} \nabla_\sigma
\psi) - \frac {1} {2} \nabla_\beta(\tau^{\rho\sigma} \nabla_\rho
\psi \nabla_\sigma \psi)
\end{equation}
since $\nabla^\alpha G_{\alpha \beta}$ vanishes by virtue of the
Bianchi identities.

The right hand side of (\ref{eq:bimetrictaurestrict}) does not in
general vanish for solutions to the field equations and therefore
(\ref{eq:bimetrictaurestrict}) is not in general satisfied unless
one restricts to very specific solutions for $\psi$. This is not a
useful situation, as we would like to have the full space of
solutions for $\psi$ yet maintain energy conservation.  The
solution is to make all Lorentz violating tensors
\textit{dynamical}~\cite{Kostelecky:2003fs,Jacobson:2001yj},
thereby removing prior geometry. If the Lorentz violating tensors
are dynamical then conservation of the stress tensor is
automatically enforced by the diffeomorphism invariance of the
action.  While dynamical Lorentz violating tensors have a number
of effects that are testable in the gravitational sector, most
researchers have concentrated on flat space tests of Lorentz
invariance where gravitational effects can be ignored. Hence for
most of this review we will treat the Lorentz violating
coefficients as fixed and neglect dynamics.  The theoretical
consequences of dynamical Lorentz violation will be analyzed only
in section \ref{subsubsec:LVandgravity}, where we discuss a model
of a diffeomorphism invariant ``aether'' which has received some
attention. The observational constraints on this theory are
discussed in section \ref{sec:GravitationalObservations}.

\subsection{Lorentz violation and the equivalence principle} \label{subsec:equiv}
Lorentz violation implies a violation of the equivalence
principle.  Intuitively this is clear: in order for there to be
Lorentz violation particles must travel on world-lines that are
species dependent (and not fully determined by the mass). In
various papers dealing with Lorentz violating dispersion relations
one will sometimes see the equivalence principle being cited as a
motivation for keeping the Lorentz violating terms equal for all
particle species. We now give a pedagogical example to show that
the equivalence principle is violated even in this case.  Consider
a dispersion modification of the form
\begin{equation}
E^2=m^2+p^2+ \frac {f^{(4)}} {E_{Pl}}|p|^4
\end{equation}
for a free particle and assume $f^{(4)}$ is independent of
particle species. If we assume Hamiltonian dynamics at low energy
and use the energy as the Hamiltonian, then for a non-relativistic
particle in a weak gravitational field we have
\begin{equation}
H=m+\frac{p^2} {2m} + f^{(4)} \frac {p^4} {2m E_{Pl}^2} + V(x)
\end{equation}
where $V(x)$ is the Newtonian gravitational potential $m\Phi(x)$.
Applying Hamilton's equations to solve for the acceleration yields
\begin{equation}
\ddot{x}^{i}=- \frac {\partial \Phi} {\partial x^i} \bigg{(} 1+6
f^{(4)} \frac {m^2 \dot{x}^i \cdot \dot{x}^i} {E_{Pl}} \bigg{)}
\end{equation}
to lowest order in the Lorentz violating term.  From this
expression it is obvious that the acceleration is mass dependent
and the equivalence principle is violated (albeit slightly) for
particles of different masses with the same $f^{(4)}$. Of course,
if the $f^{(n)}$ terms are different, as is natural with some
Lorentz violating models~\cite{Ellis:2003if}, then it is also
obviously violated. As a consequence one cannot preserve the
equivalence principle with Lorentz violation unless one also
modifies Hamiltonian dynamics. Equivalence principle tests are
therefore able to also look for Lorentz violation and vice versa
(for an explicit example see~\cite{Alvarez:1995wi}). Other
examples of the relationship between equivalence principle
violation and Lorentz violation can be found
in~\cite{Haugan,Halprin:1999be, Shore:2004sh}.

\subsection{Systematic vs. Non-systematic violations}
\label{subsec:sysvsnonsys}%
Most tests of Lorentz violation deal with systematic Lorentz
violation, where the deviation is constant in time/space.  For
example, consider the modified dispersion relation $\om^2=k^2+
f^{(4)} k^4/E_{Pl}^2$ for a photon where $f^{(4)}$ is some fixed
coefficient. There is no position dependence, so the Lorentz
violating term is a constant as the particle propagates. However,
various models~\cite{Dowker:2003hb,Shiokawa:2000em,Ng:1993jb}
suggest that particle energy/momentum may not be constant but
instead vary randomly by a small amount.  Some authors have
combined these two ideas about quantum gravity, Lorentz violation
and stochastic fluctuations, and considered a stochastic violation
of Lorentz invariance characterized by a fluctuating
coefficient~\cite{Aloisio:2000cm, Ng:1999hm,
Amelino-Camelia:2002ws, Ellis:1999uh, Ford:1994cr}. We will
discuss non-systematic models in greater detail in section
\ref{subsec:nonsystematicdisp}.

\subsection{Causality and stability} \label{subsec:causality}
\subsubsection{Causality}
It is obvious that when we introduce Lorentz violation we have to
rethink causality - there is no universal light cone given by the
metric that all fields must propagate within.  Even with Lorentz
violation we must certainly maintain some notion of causality, at
least in concordant frames, since we know that our low energy
physics is causal.  Causality from a strict field theory
perspective is usually discussed in terms of
\textit{microcausality} which in turn comes from the cluster
decomposition principle: physical observables at different points
and equal times should be independently measurable.  This is
essentially a statement that physics is local.  We now briefly
review how microcausality arises from cluster decomposition. Let
$O_1(x), O_2(y)$ represent two observables for a field theory in
flat space. In a particular frame, let us choose the equal time
slice $t=0$, such that $x=(0,\vec{x}), y=(0,\vec{y})$ and further
assume that $\vec{x} \neq \vec{y}$. The cluster decomposition
principle then states that $O_1(x)$ and $O_2(y)$ must be
independently measurable. This in turn implies that their
commutator must vanish, $[O_1(x),O_2(y)]=0$. When Lorentz
invariance holds there is no preferred frame, so the commutator
must vanish for the $t=0$ surface of any reference frame.  This
immediately gives that $[O_1(x),O_2(y)]=0$ whenever $x,y$ are
spacelike separated, which is the statement of microcausality.
Microcausality is related to the existence of closed timelike
curves since closed timelike curves violate cluster decomposition
for surfaces that are pierced twice by the curves.  The existence
of such a curve would lead to a breakdown of microcausality.

Lorentz violation can induce a breakdown of microcausality, as
shown in~\cite{Kostelecky:2000mm}.  In this work, the authors find
that microcausality is violated if the group velocity of any field
mode is superluminal. Such a breakdown is to be expected, as the
light cone no longer determines the causal structure and notions
of causality based on ``spacelike'' separation would not be
expected to hold. However, the breakdown of microcausality does
not lead to a breakdown of cluster decomposition in a Lorentz
violating theory, in contrast to a Lorentz invariant theory.  Even
if fields propagate outside the light cone, we can have perfectly
local and causal physics in \textit{some} reference frames.  For
example, in a concordant frame Lorentz violation is small, which
implies that particles can be only slightly superluminal. In such
a frame all signals are always propagated into the future, so
there is no mechanism by which signals could be exchanged between
points on the same time slice. If we happened to be in such a
concordant frame then physics would be perfectly local and causal
even though microcausality does not hold.

The situation is somewhat different when we consider gravity and
promote the Lorentz violating tensors to dynamical objects.  For
example in an aether theory, where Lorentz violation is described
by a timelike four vector, the four vector can twist in such a way
that local superluminal propagation can lead to energy-momentum
flowing around closed paths~\cite{Lim:2004js}. However, even
classical general relativity admits solutions with closed time
like curves, so it is not clear that the situation is any worse
with Lorentz violation. Furthermore, note that in models where
Lorentz violation is given by coupling matter fields to a
non-zero, timelike gradient of a scalar field, the scalar field
also acts as a time function on the spacetime.  In such a case,
the spacetime must be stably causal (c.f.~\cite{Wald:1984rg}) and
there are no closed timelike curves.  This property also holds in
Lorentz violating models with vectors if the vector in a
particular solution can be written as a non-vanishing gradient of
a scalar.

Finally, we mention that in fact many approaches to quantum
gravity actually predict a failure of causality based on a
background metric~\cite{Garay:1994en} as in quantum gravity the
notion of a spacetime event is not necessarily
well-defined~\cite{Padmanabhan:1987au}. A concrete realization of
this possibility is provided in Bose-Einstein condensate analogs
of black holes~\cite{Barcelo:2005fc}. Here the low energy phonon
excitations obey Lorentz invariance and
microcausality~\cite{Unruh:2003ss}. However, as one approaches a
certain length scale (the healing length of the condensate) the
background metric description breaks down and the low energy
notion of microcausality no longer holds.

\subsubsection{Stability}
In any realistic field theory one would like a stable ground
state. With the introduction of Lorentz violation, one must still
have some ground state.  This requires that the Hamiltonian still
be bounded from below and that perturbations around the ground
state have real frequencies.  It will again be useful to discuss
stability from a field theory perspective, as this is the only
framework in which we can speak concretely about a Hamiltonian.
Consider a simple model for a massive scalar field in flat space
similar to (\ref{eq:bimetric}),
\begin{equation} \label{eq:stab1}
S=\frac {1} {2} \int  d^4x (\eta^{\alpha \beta} + \tau^{\alpha
\beta})
\partial_\alpha \psi
\partial_\beta \psi - m^2 \psi^2
\end{equation}
where we now assume that in some frame $S$ the only non-zero
component of $\tau^{\alpha \beta}$ is $\tau^{00}$.  This
immediately leads to the dispersion law $(1+ \tau^{00}) E^2=p^2 +
m^2$.  We can immediately deduce from this that if $\tau^{00}$ is
small the energy is always positive in this frame (taking the
appropriate root of the dispersion relation). Similar statements
about energy positivity and the allowable size of coefficients
hold in more general field theory
frameworks~\cite{Kostelecky:2000mm}.  If the energy for every mode
is positive, then the vacuum state $|0_S>$ is stable.

As an aside, note that while the energy is positive in $S$, it is
not necessarily positive in a boosted frame $S'$.  If
$\tau^{00}>0$, then for large momentum $E<p$, yielding a spacelike
energy momentum vector. This implies that the energy $E'$ can be
less than zero in a boosted frame.  Specifically, for a given mode
$p$ in $S$, the energy $E'$ of this mode in a boosted frame $S'$
is less than zero whenever the relative velocity $v$ between $S$
and $S'$ is greater than $E/p$. The main implication is that if
$v$ is large enough the expansion of a positive frequency mode in
$S$ in terms of the modes of $S'$ (one can do this since both sets
are a complete basis) may have support in the negative energy
modes. The two vacua $|0_S>$ and $|0_{S'}>$ are therefore
inequivalent.  This is in direct analogy to the Unruh effect,
where the Minkowski vacuum is not equivalent to the Rindler vacuum
of an accelerating observer.  With Lorentz violation even
\textit{inertial} observers do not necessarily agree on the
vacuum.  Due to the inequivalence of vacua an inertial detector at
high velocities should see a bath of radiation just as an
accelerated detector sees thermal Unruh radiation.  A clue to what
this radiation represents is contained in the requirement that
$E'<0$ only if $v>E/p$, which is exactly the criteria for
\v{C}erenkov radiation of a mode $p$.  In other words, the vacuum
\v{C}erenkov effect (discussed in more detail in section
\ref{subsec:ThresholdsEFT}) can be understood as an effect of
inequivalent vacua.

We now return to the question of stability.  For the models in
section \ref{subsec:systematicdisp} with higher order dispersion
relations ($E^2=p^2+ m^2 + f^{(n)} p^n/E_{Pl}^{n-2}$ with $n>2$)
there is a stability problem for particles with momentum near the
Planck energy if $f^{(n)}<0$ as modes do not have positive energy
at these high momenta. However, it is usually assumed that these
modified dispersion relations are only effective - at the Planck
scale there is a UV completion that renders the fundamental theory
stable.  Hence the instability to production of Planck energy
particles is usually ignored.

So far we have only been concerned with instability of a quantum
field with a background Lorentz violating tensor.  Dynamical
Lorentz violating tensors introduce further possible
instabilities. In such a dynamical theory, one needs a version of
the positive energy theorem~\cite{Schoen,Witten:1981mf} that
includes the Lorentz violating tensors.  For aether theories, the
total energy is proportional to the usual ADM energy of general
relativity~\cite{Eling:2004dk}. Unfortunately, the aether stress
tensor does not necessarily satisfy the dominant energy condition
(although it may for certain choices of coefficients), so there is
no proof yet that spacetimes with a dynamical aether have positive
energy.  For other models of Lorentz violation the positive energy
question is completely unexplored.  It is also possible to set
limits on the coefficients of the aether theory by demanding that
the theory be perturbatively stable, which requires that
excitations of the aether field around a Lorentz violating vacuum
expectation value have real frequencies~\cite{Jacobson:2004ts}.

\newpage

\section{Kinematic frameworks for Lorentz violation} \label{sec:KinematicFrameworks}
\subsection{Systematic modified dispersion}
\label{subsec:systematicdisp} Perhaps the simplest kinematic
framework for Lorentz violation in particle based experiments is
to propose modified dispersion relations for particles, while
keeping the usual energy-momentum conservation laws. This was the
approach taken in much of the work using astrophysical phenomena
in the late 1990's. In a given observer's frame in flat space,
this is done by postulating that the usual Lorentz invariant
dispersion law $E^2=p^2+m^2$ is replaced by some function
$E^2=F(p,m)$.  In general the preferred frame is taken to coincide
with the rest frame of the cosmic microwave background. Since we
live in an almost Lorentz invariant world (and are nearly at rest
with respect to the CMBR), in the preferred frame $F(p,m)$ must
reduce to the Lorentz invariant dispersion at small energies and
momenta. Hence it is natural to expand $F(p,m)$ about $p=0$, which
yields the expression
\begin{equation} \label{eq:generaldisp1}
E^2=m^2+p^2+F^{(1)}_i p^i + F^{(2)}_{ij}p^ip^j +
F^{(3)}_{ijk}p^ip^jp^k+...
\end{equation}
where the constant coefficients $F^{(n)}_{ij..n}$ are dimensionful
and arbitrary but presumably such that the modification is small.
The order $n$ of the first non-zero term in
(\ref{eq:generaldisp1}) depends on the underlying model of quantum
gravity taken.  Since the underlying motivation for Lorentz
violation is quantum gravity, it is useful to factor out the
Planck energy in the coefficients $F^{(n)}$ and rewrite
(\ref{eq:generaldisp1}) as
\begin{equation} \label{eq:generaldisp}
E^2=m^2+p^2+ E_{Pl} f^{(1)}_i p^i + f^{(2)}_{ij}p^ip^j + \frac
{f^{(3)}_{ijk}} {E_{Pl}} p^ip^jp^k+...
\end{equation}
such that the coefficients $f^{(n)}$ are dimensionless.

In most of the literature a simplifying assumption is made that
rotation invariance is preserved.  In nature, we cannot have the
rotation subgroup of the Lorentz group strongly broken while
preserving boost invariance. Such a scenario leads immediately to
broken rotation invariance at every energy which is
unobserved.\footnote{In a field theory, broken rotation invariance
automatically yields broken boost invariance.  For example, if
rotation invariance is broken by coupling matter to a non-zero
spacelike four vector, the four vector is also not boost
invariant.} Hence, if there is strong rotation breaking there must
also be a broken boost subgroup. However, it is possible to have a
broken boost symmetry and unbroken rotation symmetry. Either way,
the boost subgroup must be broken. Phenomenologically, it
therefore makes sense to look first at boost Lorentz violation and
neglect any violation of rotational symmetry. If we make this
assumption then we have
\begin{equation} \label{eq:rotinvdisp}
E^2=m^2+p^2+ E_{Pl} f^{(1)} |p| + f^{(2)} p^2 + \frac {f^{(3)}}
{E_{Pl}}|p|^3+...
\end{equation}

There is no \textit{a priori} reason (from a phenomenological
point of view) that the coefficients in (\ref{eq:rotinvdisp}) are
universal (and in fact one would expect the coefficients to be
renormalized differently even if the fundamental Lorentz violation
is universal~\cite{Alfaro:2004aa}) . We will therefore label each
$f^{(n)}$ as $f^{(n)}_A$ where $A$ represent particle species.

\paragraph{Modified dispersion and effective field theory}
Effective field theory (EFT) is not applicable if one wishes to
stick to straight kinematics, however the EFT implications for
modified dispersion are so significant that they must be
considered.  As will be shown in detail in Sec. \ref{subsec:EFT},
universal dispersion relations cannot be imposed for all $n$ from
an EFT standpoint.  For example, rotationally invariant $n=1,3$
type dispersion cannot be imposed universally on
photons~\cite{Colladay:1998fq,Myers:2003fd}. The operators that
give rise to $n=1,3$ dispersion are CPT violating and induce
birefringence (the dispersion modifications change sign based on
the photon helicity). Since EFT requires different coefficients
for particles with different properties and there is no underlying
reason why all coefficients should be the same, it is
phenomenologically safest when investigating modified dispersion
to assume that each particle has a different dispersion relation.
After this general analysis is complete the universal case can be
treated with ease.

\subsection{Robertson-Mansouri-Sexl framework} \label{subsec:RMS}
The Robertson-Mansouri-Sexl framework~\cite{Robertson,MS,
Will:2001mx} is a well known kinematic test theory for
parameterizing deviations from Lorentz invariance. In the RMS
framework, there is assumed to be a preferred frame $\Sigma$ where
the speed of light is isotropic. The ordinary Lorentz
transformations to other frames are generalized to
\begin{eqnarray} \label{eq:MSxfm}
t'=a^{-1}(t-\vec{\ep} \cdot \vec{x})\\
\vec{x}'=d^{-1} \vec {x} - (d^{-1} - b^{-1}) \frac {\vec{v}
(\vec{v} \cdot \vec{x})} {v^2} + a^{-1} \vec{v} t
\end{eqnarray}
where the coefficients $a,b,d$ are functions of the magnitude $v$
of the relative velocity between frames. This transformation is
the most general one-to-one transformation that preserves
rectilinear motion in the absence of forces.  In the case of
special relativity, with Einstein clock synchronization, these
coefficients reduce to $a=b^{-1}=\sqrt{1-v^2},d=1$.  The vector
$\ep$ depends on the particular synchronization used and is
arbitrary.  Many experiments, such as those that measure the
\textit{isotropy} of the one way speed of light~\cite{Will:1991wm}
or propagation of light around closed loops, have observables that
depend on $a,b,d$ but not on the synchronization procedure.  Hence
the synchronization is largely irrelevant and we assume Einstein
synchronization.

The RMS framework is incomplete, as it says nothing about dynamics
or how given clocks and rods relate to fundamental particles. In
particular, the coordinate transformation of (\ref{eq:MSxfm}) only
has meaning if we identify the coordinates with the measurements
made by a particular set of clocks and rods. If we chose a
different set of clocks and rods, the transformation laws may be
completely different.  Hence it is not possible to compare the RMS
parameters of two experiments that use physically different clocks
and rods (for example, an experiment that uses a cesium atomic
clock versus an experiment that uses a hydrogen one). However, for
experiments involving a single type of clock/rod and light, the
RMS formalism is applicable and can be used to search for
violations of Lorentz invariance in that experiment. The RMS
formalism can be made less ambiguous by placing it into a complete
dynamical framework, such as the standard model extension of Sec.
\ref{subsubsec:mSME}.  In fact, it was shown
in~\cite{Kostelecky:2002hh} that the RMS framework can be
incorporated into the standard model extension.

Most often, the RMS framework is used in situations where the
velocity $v$ is small compared to $c$.  We therefore expand
$a,b,d$ in a power series in $v$,
\begin{eqnarray}
a=1+(\al_{RMS} - \frac{1} {2})v^2+...\\
b=1+(\be_{RMS} + \frac{1} {2})v^2+...\\
a=1+\de_{RMS} v^2+...
\end{eqnarray}
and will give constraints on the parameters
$\al_{RMS},\be_{RMS},\de_{RMS}$ instead.

\subsection{$c^2$ framework}
The $c^2$ framework~\cite{Will:2001mx} is the flat space limit of
the $TH\epsilon \mu$~\cite{THeu} framework.  This framework
considers the motion of electromagnetically charged test particles
in a spherically symmetric, static gravitational field.
$T,H,\epsilon,\mu$ are all parameters that fold in to the motion
of the particles which vary depending on the underlying
gravitational model. In the flat space limit, which is the $c^2$
formalism, the units are chosen such that limiting speed of the
test particles is one while the speed of light is given in terms
of the $TH \epsilon \mu$ parameters by $c^2=H/(T \epsilon \mu)$.
The $TH \epsilon \mu$ and $c^2$ constructions can also be
expressed in terms on the standard model
extension~\cite{Kostelecky:2002hh}.

\subsection{``Doubly special'' relativity} \label{subsec:DSR}
Doubly special relativity (DSR), which has only been extensively
studied over the past few years, is a novel idea about the fate of
Lorentz invariance.  DSR is not a complete theory as it has no
dynamics and generates problems when applied to macroscopic
objects (for a discussion see~\cite{Kowalski-Glikman:2004qa}).
Furthermore, it is not fully settled yet if DSR is mathematically
consistent or physically meaningful. Therefore it is somewhat
premature to talk about robust constraints on DSR from particle
threshold interactions or other experiments. One might then ask,
why should we talk about it at all? The reason is twofold.  First,
DSR is the subject of a good amount of theoretical effort and so
it is useful to see if it can be observationally ruled out. The
second reason is purely phenomenological. As we shall see in the
sections below, the constraints on Lorentz violation are
astoundingly good in the effective field theory approach. With the
current constraints it is difficult to fit Lorentz violation into
an effective field theory in a manner that is theoretically
natural yet observationally viable.

DSR, if it can eventually be made mathematically consistent in its
current incarnation, has one phenomenological advantage
 - it does not have a preferred frame. Therefore it evades most of
the threshold constraints from astrophysics as well as any
terrestrial experiment that looks for sidereal variations, while
still modifying the usual action of the Lorentz group. Since these
experiments provide almost all of the tests of Lorentz violation
that we have, DSR becomes more phenomenologically attractive as a
Lorentz violating/deforming theory.

So what is DSR?  At the level we need for phenomenology, DSR is a
set of assumptions that the Lorentz group acts in such a way that
the usual speed of light $c$ \textit{and} a new momentum scale
$E_{DSR}$ are invariant.  Usually $E_{DSR}$ is taken to be the
Planck energy - we also make this assumption.  All we will need
for this review are the Lorentz boost expressions and the
conservation laws, which we will postulate as true in the DSR
framework. For brevity we only detail the Magueijo-Smolin version
of DSR~\cite{Magueijo:2002am}, otherwise known as DSR2 - the
underlying conclusions for DSR1~\cite{Amelino-Camelia:2000mn}
remain the same.  The DSR2 boost transformations are most easily
derived from the relations
\begin{eqnarray}
E=\frac {\ep} {1+\la_{DSR} \ep} \\
p=\frac {\pi} {1+\la_{DSR} \pi}
\end{eqnarray}
where $\la_{DSR}=E_{DSR}^{-1}$, $E,p$ are the physical/measured
energy/momentum and $\ep,\pi$ are called the ''pseudo-energy'' and
''pseudo-momentum'' respectively. $\ep,\pi$ transform under the
usual Lorentz transforms, which induce corresponding
transformations of $E$ and $p$~\cite{Judes:2002bw}. Similarly, the
$\ep,\pi$ for particles are conserved as energy and momentum
normally are for a scattering problem.\footnote{It may therefore
seem that DSR theories are ''empty'', in the sense that the new
definitions of E,p are merely mathematical manipulations without
any physical meaning, i.e. $\ep, \pi$ represent the true energy
and momentum that one would measure. For a discussion of this
point see~\cite{Liberati:2004ju,Kowalski-Glikman:2004qa}.  Since
this is not a theoretical paper, we simply make the DSR assumption
that $E,p$ are the measurable energy and momentum (or more
specifically for analysis of particle interactions, the energy and
momentum that are assigned to in and out states in a scattering
problem).} Given this set of rules, for any measured particle
momentum and energy, we can solve for $\ep,\pi$ and calculate
interaction thresholds, etc. The invariant dispersion relation for
the DSR2 boosts is given by
\begin{equation}
E^2-p^2=\frac {m^2(1-\lambda_{DSR} E^2)} {1-\lambda_{DSR} m^2}.
\end{equation}
This concludes our (brief) discussion of the basics of DSR.  For
further introductions to DSR and DSR phenomenology
see~\cite{Kowalski-Glikman:2004qa,Amelino-Camelia:2003ex,Amelino-Camelia:2002vy,
Daszkiewicz:2004xy}.  We discuss the threshold behavior of DSR
theories in section \ref{subsubsec:ThreshDSR}.

\subsection{Non-systematic dispersion} \label{subsec:nonsystematicdisp}
As mentioned in Sec. \ref{subsec:sysvsnonsys}, Lorentz violation
is only one possibility for a signal of quantum gravity.  Another
common idea about quantum gravity is that spacetime should have a
stochastic~\cite{Dowker:2003hb,Shiokawa:2000em} or
``foamy''~\cite{Ng:1993jb} structure at very small scales.
Combining these two ideas has lead a number of authors to the idea
of stochastic/non-systematic dispersion where the modifications to
the dispersion relation fluctuate over time. Such dispersion
modifications have been phenomenologically parameterized by three
numbers, the usual coefficient $f^{(n)}$ and exponent $n$ of Sec.
\ref{subsec:systematicdisp}, and a length scale $L$ which
determines the length over which the dispersion is roughly
constant. After a particle has travelled a distance $L$ a new
coefficient $f^{(n)}$ is chosen based upon some model dependent
probability distribution $P$ that reflects the underlying
stochasticity. Usually $P$ is assumed to be a gaussian about $0$,
such that the average energy of the particle is given by its
Lorentz invariant value. As well, $L$ is in general taken to be
the de Broglie wavelength of the particle in question. Note that
in these models $n$ is not required to be an integer, the most
common choices are
$n=5/2,n=8/3,n=3$~\cite{Ng:1999hm}.~\footnote{The underlying
approach that yields certain choices of $n$ has been sharply
criticized in~\cite{Baez:2002ra}.  However, the constraints on
these models are so poor that any $n$ is observationally
feasible.} The only existing constraints on non-systematic
dispersion come from threshold reactions (Sec.
\ref{subsubsec:Threshnonsystem}) and the phase coherence of light
(Sec. \ref{subsec:PhaseCoherence}).

\newpage

\section{Dynamical frameworks for Lorentz violation} \label{sec:DynamicalFrameworks}
\subsection{Effective field theory} \label{subsec:EFT}
The most conservative approach for a framework in which to test
Lorentz violation from quantum gravity is that of effective field
theory (EFT).  Both the standard model and relativity can be
considered EFT's and the EFT framework can easily incorporate
Lorentz violation via the introduction of extra tensors.
Furthermore, in many systems where the fundamental degrees of
freedom are qualitatively different than the low energy degrees of
freedom, EFT applies and give correct results up to some high
energy scale.  Hence following the usual guideline of starting
with known physics, EFT is an obvious place to start looking for
Lorentz violation.

\subsubsection{Renormalizable operators and the Standard Model Extension} \label{subsubsec:mSME}
The standard model is a renormalizable field theory containing
only mass dimension $\leq 4$ operators.  If we considered the
standard model plus Lorentz violating terms then we would expect a
tower of operators with increasing mass dimension. However,
without some custodial symmetry protecting the theory from Lorentz
violating dimension $\leq4$ operators, the lower dimension
operators will be more important than irrelevant higher mass
dimension operators (see Sec. \ref{subsubsec:InducedLowEnergyLV}
for details). Therefore the first place to look from an EFT
perspective is all possible renormalizable Lorentz violating terms
that can be added to the standard model.  In
~\cite{Colladay:1998fq} Colladay and Kostelecky derived just such
a theory in flat space  - the so-called (minimal) Standard Model
Extension (mSME).\footnote{In the literature the mSME is often
referred to as just the SME, although technically it was
introduced in~\cite{Colladay:1998fq} as a minimal subset of an
extension that involved non-renormalizable operators as well.}

One can classify the mSME terms by whether or not they are CPT odd
or even.  We first will show the terms with manifestly $SU(3)
\times SU(2) \times U(1)$ gauge invariance.  After that, we shall
give the coefficients in a more practical notation that exhibits
broken gauge invariance.

\subsubsection{Manifestly invariant form}
We deal with CPT odd terms first. The additional Lorentz violating
CPT odd operators for leptons are
\begin{equation} \label{eq:SMElepton}
-(a_L)_{\mu A B} \overline{L}_A \gamma^\mu L_B -(a_R)_{\mu A B}
\overline{R}_A \gamma^\mu R_B
\end{equation}
where $L_A$ is the left-handed lepton doublet $L_A=(\nu_A~
l_A)_L$,$R_A$ is the right singlet $(l_A)_R$ and $A,B$ are flavor
indices.  The coefficients $(a_{L,R})_{\mu A B}$ are constant
vectors that can mix flavor generations.\footnote{$(a_{L,R})_{\mu
A B}$ can be constant because the mSME deals with only Minkowski
space.  If one wishes to make the mSME diffeomorphism invariant,
these and other coefficients would be dynamical (section
\ref{subsubsec:Diffeomorphism}).} For quarks we have similarly
\begin{equation} \label{eq:SMEquark}
-(a_Q)_{\mu A B} \overline{Q}_A \gamma^\mu Q_B -(a_U)_{\mu A B}
\overline{U}_A \gamma^\mu U_B-(a_D)_{\mu A B} \overline{D}_A
\gamma^\mu D_B
\end{equation}
where $Q_A=(u_A~d_A)_L$, $U_A=(u_A)_R$, and $D_A=(d_A)_R$. In the
gauge sector we have
\begin{eqnarray} \label{eq:SMEgaugeodd}
(k_0)_\ka B^\ka+(k_1)_\ka \ep^{\ka\la\mu\nu} B_\la B_{\mu\nu} \nonumber \\
+ (k_2)_\ka \ep^{\ka\la\mu\nu} {\rm Tr} (W_\la W_{\mu\nu} + \frac
2 3 i g W_\la W_\mu W_\nu) \nonumber\\
+(k_3)_\ka \ep^{\ka\la\mu\nu} {\rm Tr} (G_\la G_{\mu\nu} + \frac 2
3 i g_3 G_\la G_\mu G_\nu).
\end{eqnarray}
Here $B_\mu, W_\mu, G_\mu$ are the U(1),SU(2),SU(3) gauge fields
and $B_{\mu \nu}, W_{\mu \nu}, G_{\mu \nu}$ are their respective
field strengths.  The $k_0$ term in (\ref{eq:SMEgaugeodd}) is
usually required to vanish as it makes the theory unstable.  The
remaining $a,k$ coefficients have mass dimension one.

The CPT even operators for leptons in the mSME are
\begin{equation}
\frac{1} {2} i (c_L)_{\mu\nu AB} \overline{L}_A \ga^{\mu} \lrDnu
L_B + \frac {1} {2} i (c_R)_{\mu\nu AB} \overline{R}_A \ga^{\mu}
\lrDnu R_B
\end{equation}
while we have for quarks
\begin{equation}
\frac{1} {2} i (c_Q)_{\mu\nu AB} \overline{Q}_A \ga^{\mu} \lrDnu
Q_B  + \frac {1} {2} i (c_U)_{\mu\nu AB} \overline{U}_A \ga^{\mu}
\lrDnu U_B
 + \frac {1} {2} i (c_D)_{\mu\nu AB} \overline{D}_A \ga^{\mu} \lrDnu
 D_B.
\end{equation}
For gauge fields the CPT even operators are
\begin{eqnarray}\label{eq:SMEgaugeeven}  -\frac 1
4 (k_B)_{\ka\la\mu\nu} B^{\ka\la}B^{\mu\nu} -\frac {1} {2}
(k_W)_{\ka\la\mu\nu} {\rm Tr} (W^{\ka\la}W^{\mu\nu})  -\frac {1}
{2} (k_G)_{\ka\la\mu\nu} {\rm Tr} (G^{\ka\la}G^{\mu\nu}).
\end{eqnarray}

The coefficients for all CPT even operators in the mSME are
dimensionless.  While the split of CPT even and odd operators in
the mSME correlates with even and odd mass dimension, we caution
the reader that this does not carry over to higher mass dimension
operators.  Finally, we will in general drop the subscripts $A,B$
when discussing various coefficients.  These terms without
subscripts are understood to be the flavor diagonal coefficients.

Besides the fermion and gauge field content, the mSME also has
Yukawa couplings between the fermion fields and the Higgs.  These
CPT even terms are
\begin{equation}
-\frac{1} {2} [(H_L)_{\mu \nu A B} \overline{L}_A \phi \sigma^{\mu
\nu} R_B  + (H_U)_{\mu \nu A B} \overline{Q}_A \phi^* \sigma^{\mu
\nu} U_B  + (H_D)_{\mu \nu A B} \overline{Q}_A \phi \sigma^{\mu
\nu} D_B] + h.c.
\end{equation}
Finally, there are also additional terms for the Higgs field
alone.  The CPT odd term is
\begin{equation}
i (k_\phi)^\mu \phi^\dagger D_\mu \phi + h.c.
\end{equation}
while the CPT even terms are
\begin{equation}
\frac {1} {2} [(k_{\phi \phi})^{\mu \nu} (D_\mu \phi)^\dagger
D_\nu \phi -(k_{\phi B})^{\mu \nu} \phi^\dagger \phi B_{\mu \nu}
-(k_{\phi W})^{\mu \nu} \phi^\dagger W_{\mu \nu} \phi] + h.c.
\end{equation}
This concludes the description of the mSME terms with manifest
gauge invariance.

\subsubsection{Practical form} \label{subsubsec:SMEPractical}
Tests of the mSME are done at low energies, where the SU(2) gauge
invariance has been broken.  It will be more useful to work in a
notation where individual fermions are broken out of the doublet
with their own Lorentz violating coefficients.  With gauge
breaking, the fermion Lorentz violating terms above give the
additional CPT odd terms
\begin{equation} \label{eq:LVQEDodd}
 - a_{\mu } \overline{\ps} \ga^{\mu} \ps
- b_{\mu} \overline{\ps} \ga_5 \ga^{\mu}\ps
\end{equation}
and the CPT even terms
\begin{equation}\label{eq:LVQEDeven}  - \frac {1} {2} H_{\mu\nu}
\overline{\ps} \si^{\mu\nu} \ps  + \frac {1} {2} i c_{\mu\nu}
\overline{\ps} \ga^{\mu} \lrDnu \ps  + \frac {1} {2} i d_{\mu\nu}
\overline{\ps} \ga_5 \ga^\mu \lrDnu \ps
\end{equation}
where the fermion spinor is denoted by $\ps$.  Each possible
particle species has its own set of coefficients. For a single
particle the $a_\mu$ term can be absorbed by making a field
redefinition $\ps \longrightarrow e^{-ia\cdot x} \ps$. However, in
multi-particle theories involving fermion interactions one cannot
remove $a_\mu$ for all fermions~\cite{Colladay:1996iz}.  However,
one can always eliminate \textit{one} of the $a_\mu$, i.e. only
the differences between $a_\mu$ for various particles are actually
observable.

As an aside, we note that there are additional dimension $\leq 4$
U(1) invariant terms for fermions that could be added to the mSME
once SU(2) gauge invariance is broken. These operators are
\begin{equation} \label{eq:SMEfermion}
\frac {1} {2} e_\nu \overline{\psi} \lrDnu \psi - \frac {1} {2}
f_\nu \overline{\psi} \gamma_5 \lrDnu \psi + \frac {i} {4} g_{\la
\mu \nu} \overline{\psi} \sigma^{\la \mu} \lrDnu \psi.
\end{equation}
These terms do not arise from gauge breaking of the renormalizable
mSME in the previous section.  However, they might arise from
non-renormalizable terms in an EFT expansion.  As such,
technically they should be constrained along with everything else.
However, since their origin can only be from higher dimension
operators they are expected to be much smaller than the terms that
come directly from the mSME.\footnote{Note that many of the
fermion terms in the mSME can also be found in the extended Dirac
equation framework (c.f.~\cite{Audretsch:1993ke,
Lammerzahl:1998qp}). Similarly, parts of the electromagnetic
sector were previously known (c.f.~\cite{Ni-chig}) in the context
of equivalence principle violations.}

Current tests of Lorentz invariance for gauge bosons directly
constrain only the electromagnetic sector.  The Lorentz violating
terms for electromagnetism are

\begin{equation}
\label{eq:LVQEDphoton} -\frac 1 4 (k_F)_{\ka\la\mu\nu}
F^{\ka\la}F^{\mu\nu} + \frac {1} {2} (k_{AF})^\ka
\ep_{\ka\la\mu\nu} A^\la F^{\mu\nu}
\end{equation}
where the $k_F$ term, is CPT even and the $k_{AF}$ term is CPT
odd.  The $k_{AF}$ term makes the theory unstable, so we assume it
is zero from here forward unless otherwise noted (see section
\ref{subsec:Birefringence}). Now that we have the requisite
notation to compare Lorentz violating effects directly with
observation we turn to the most common subset of the mSME, Lorentz
violating QED.

\subsubsection{Lorentz violating QED} \label{subsubsec:LVQED}
In many Lorentz violating tests, the relevant particles are
photons and electrons, making Lorentz violating QED the
appropriate theory.  The relevant Lorentz violating operators are
given by (\ref{eq:LVQEDodd},
\ref{eq:LVQEDeven},\ref{eq:LVQEDphoton}). The dispersion relation
for photons will be useful when deriving birefringence constraints
on $k_F$. If $k_F \ne 0$ spacetime acts as a anisotropic medium,
and different photon polarizations propagate at different speeds.
The two photon polarizations, labelled $\ep_\pm$, have the
dispersion relation~\cite{Kostelecky:2002hh}
\begin{equation}
E=(1+\rho\pm\si)|\overrightarrow{p}|
\end{equation}
where $\rho=\frac {1} {2} \tilde{k}_\al^\al$, $\si^2=\frac {1} {2}
(\tilde{k}_{\al\be})^2-\rho^2$, $\tilde{k}_{\al\be}=(k_F)_{\al\ga
\be \de}\hat{p}^{\ga}\hat{p}^{\de}$, and
$\hat{p}^\al=p^\al/|\overrightarrow{p}|$.  Strong limits can be
placed on this birefringent effect from astrophysical
sources~\cite{Kostelecky:2002hh}, as detailed in Sec.
\ref{subsec:Birefringence}.

A simplifying assumption that is often made is rotational
symmetry. With rotational symmetry all the Lorentz violating
tensors must be reducible to products of a vector field, which we
denote by $u^\alpha$, that describes the preferred frame. We will
normalize $u^\alpha$ to have components $(1,0,0,0)$ in the
preferred frame, placing constraints on the coefficients instead.
The rotationally invariant extra terms are
\begin{equation} \label{eq:LVQEDelectronrotinv}
- b u_{\mu} \overline{\ps} \ga_5 \ga^{\mu}\ps + \frac {1} {2} i c
u_\mu u_\nu \overline{\ps} \ga^{\mu} \lrDnu \ps  + \frac {1} {2} i
d u_\mu u_\nu \overline{\ps} \ga_5 \ga^\mu \lrDnu \ps
\end{equation}
for electrons and
\begin{equation}\label{eq:LVQEDphotonrotinv}
-\frac 1 4 (k_F){u_\ka \eta_{\la\mu} u_\nu} F^{\ka\la}F^{\mu\nu}
\end{equation}
for photons.  The high energy ($E_{Pl}\gg E \gg m$) dispersion
relations for the mSME will be necessary later.  To lowest order
in the Lorentz violating coefficients they are
\begin{eqnarray} \label{eq:SMErotinvdisp}
E^2=m^2+p^2+f^{(1)}_e p+f^{(2)}_ep^2\\
E^2=(1+ f^{(2)}_\ga ){p^2}
\end{eqnarray}
where, if $s=\pm1$ is the helicity state of the electron,
$f^{(1)}_e=-2bs,f^{(2)}_e=-(c-ds)$, and $f^{(2)}_\ga=k_F/2$. The
positron dispersion relation is the same as
(\ref{eq:SMErotinvdisp}) with the replacement $p\rightarrow -p$,
which will change only the $f^{(1)}_e$ term.

In the QED sector dimension five operators that give rise to $n=3$
type dispersion have also been investigated by~\cite{Myers:2003fd}
with the assumption of rotational symmetry.
\begin{equation}\label{eq:MPops}
\frac{\xi}{E_{Pl}}u^\mu
F_{\mu\al}(u\cdot\partial)(u_\nu\tilde{F}^{\nu
\al})+\frac{1}{E_{Pl}} u_\mu\bar{\psi}\gamma^\mu(\eta_L P_L+\eta_R
P_R)(u\cdot
\partial)^2\psi
\end{equation}
where $P_{L,R}$ are the usual left and right projection operators
$P_{L,R}=1/2(1 \pm \gamma^5)$ and $\tilde F^{\al\be}$ is the dual
of $F_{\al \be}$, $\tilde F^{\al\be}= \frac {1}{2} \ep^{\al \be
\ga \de}F_{\ga\de}$.  One should note that these operators violate
CPT.  Furthermore, they are not the only dimension five operators,
a mistake that has sometimes been made in the literature.  For
example, we could have $u_\al u_\be \bar{\ps} D^\al D^\be \ps$.
These other operators, however, do not give rise to $n=3$
dispersion as they are CPT even.

The birefringent dispersion relation for photons that results from
(\ref{eq:MPops}) is
\begin{equation} \label{eq:MPdispphot}
E^2=p^2\pm \frac {f^{(3)}_\ga E^3} {E_{Pl}}
\end{equation}
for right (+) and left (-) circularly polarized photons, where
$f^{(3)}_\ga = 2\xi$. Similarly, the high energy electron
dispersion is
\begin{equation} \label{eq:MPdispelec}
E^2=m^2 + p^2 + \frac {f^{(3)}_{e(R,L)} E^3} {E_{Pl}}
\end{equation}
where $f^{(3)}_{e(R,L)}=2 \et_{R,L}$.\footnote{These dispersion
relations also arise in some approaches to low energy dynamics
from loop quantum gravity~\cite{Alfaro:2001rb, Alfaro:1999wd,
Gambini:1998it}.  However, the ultimate low-energy status of
Lorentz invariance in loop quantum gravity is still far from clear
(c.f. ~\cite{Bojowald:2004bb, Livine:2004xy, Smolin:2004sx}).}

We note that since the dimension five operators violate CPT, they
give rise to different dispersions for positrons than electrons.
Whereas the coefficients for the positive and negative helicity
states of an electron are $2 \et_R$ and $2\et_L$, the
corresponding coefficients for a positron's positive and negative
helicity states are $-2\et_L$ and $-2\et_R$.  This will be
crucially important when deriving constraints on these operators
from photon decay.

\subsection{Non-commutative spacetime}
A common conjecture for the behavior of spacetime in quantum
gravity is that the algebra of spacetime coordinates is actually
noncommutative.  This idea has led to a large amount of research
in Lorentz violation and we would be remiss if we did not briefly
discuss Lorentz violation from non-commutativity. We will look at
only the most familiar form of spacetime non-commutativity,
``canonical'' non-commutativity, where the spacetime coordinates
acquire the commutation relation
\begin{equation} \label{eq:NC}
[x_\alpha,x_\beta]=i \frac {1} {\Lambda_{NC}^2}
\Theta_{\alpha\beta}.
\end{equation}
$\Theta_{\alpha\beta}$ is an $O(1)$ tensor that describes the
non-commutativity and $\Lambda_{NC}$ is the characteristic
noncommutative energy scale.  $\Lambda_{NC}$ is presumably near
the Planck scale if the non-commutativity comes from quantum
gravity.  However, in large extra dimension scenarios
$\Lambda_{NC}$ could be as low as 1 TeV.  For discussions of other
types of non-commutativity, including those that preserve Lorentz
invariance or lead to DSR-type theories
see~\cite{Kowalski-Glikman:2002jr, Morita:2002cv}. The
phenomenology of canonical non-commutativity as it relates to
particle physics can be found in~\cite{Hinchliffe:2002km,
Douglas:2001ba}.

The existence of $\Theta_{\alpha\beta}$ manifestly breaks Lorentz
invariance and hence the size of $\Lambda_{NC}$ is constrained by
tests of Lorentz violation.  However, in order to match a
non-commutative theory to low energy observations, we must have
the appropriate low energy theory, which implies that the infamous
UV/IR mixing problem of non-commutative field theory must be tamed
enough to create a well-defined low energy expansion. No general
method for doing this is known, although
supersymmetry~\cite{Matusis:2000jf} can perhaps do the
trick.\footnote{Other methods of removing UV/IR mixing exist, for
an example see~\cite{Vaidya:2003ew}.} If the UV/IR mixing is
present but regulated by a cutoff then the resulting field theory
can be re-expressed in terms of the
mSME~\cite{Anisimov:2001zc,Carroll:2001ws}.

In order to see how constraints come about, consider for the
moment non-commutative QED. The Seiberg-Witten
map~\cite{Seiberg:1999vs} can be used to express the
non-commutative fields in terms of ordinary gauge fields.  At
lowest order in $\Lambda_{NC}$ the effective action for low energy
is then
\begin{eqnarray}
 S &=& \frac 1 2 i \overline{\ps} \ga^\mu \lrD_\mu \ps - m
\overline{\ps} \ps - \frac 1 4 F_{\mu\nu} F^{\mu\nu}
\nonumber\\
&& - \frac 1 8 i q \frac {\th^{\al\be}} {\Lambda_{NC}^2}
F_{\al\be} \overline{\ps} \ga^\mu \lrD_\mu \ps + \frac 1 4 i
q\frac {\th^{\al\be}} {\Lambda_{NC}^2} F_{\al\mu} \overline{\ps}
\ga^\mu \lrD_\be \ps
\nonumber\\
&& + \frac 1 4 m q \frac {\th^{\al\be}} {\Lambda_{NC}^2}F_{\al\be}
\overline{\ps} \ps
\nonumber\\
&& - \frac 1 2 q \frac {\th^{\al\be}} {\Lambda_{NC}^2} F_{\al\mu}
F_{\be\nu} F^{\mu\nu} + \frac 1 8 q \frac {\th^{\al\be}}
{\Lambda_{NC}^2} F_{\al\be} F_{\mu\nu} F^{\mu\nu}. \label{oqed}
\end{eqnarray}
Direct constraints on the dimension six non-renormalizable
operators from cosmological birefringence and atomic clocks have
been considered in~\cite{Carroll:2001ws}.  A stronger bound of
$\Lambda_{NC}>5 \cdot 10^{14}$ GeV~\cite{Mocioiu:2000ip} on the
non-commutativity scale can be derived from clock comparison
experiments with Cs/Hg magnetometers~\cite{Berglund} (see section
\ref{subsec:clock}). Similarly, the possibility of constraints
from synchrotron radiation in astrophysical systems has been
analyzed in~\cite{Castorina:2002vs}.

Other strong constraints can be derived by noting that without a
custodial symmetry loop effects with the dimension six operators
will induce lower dimension operators. In~\cite{Anisimov:2001zc},
the authors calculated what dimension four operators would be
generated, assuming that the field theory has some cutoff scale
$\Lambda$. The dimension six operators induce dimension four
operators of the form $B(\Theta^2)^{\al\be} F_{\al \nu} F_\be^\nu$
and $A\Theta_{\al \be} \Theta_{\mu \nu} F^{\al \mu} F^{\be \nu}$,
where $A,B$ are dimensionless numbers that depend on
$\Lambda_{NC},{\Lambda}$. There are two different regimes of
behavior for $A,B$. If $\Lambda \gg \Lambda_{NC}$ then $A,B$ are
O(1) (up to loop factors and coupling coefficients), independent
of the scale $\Lambda_{NC}$.  Such strong Lorentz violation is
obviously ruled out by current experiment, implying that in this
perturbative approach such a limit is observationally not viable.
If instead one takes $\Lambda<<\Lambda_{NC}$ then $A,B \propto
\Lambda^2/\Lambda_{NC}^2$. The resulting field theory becomes a
subset of the standard model extension; specifically the new
operators have the form of the $(k_F)_{\al\be\ga\de} F^{\al \be}
F^{\ga\de}$ term in (\ref{eq:LVQEDphoton}).  It has been
argued~\cite{Carroll:2001ws} that any realistic non-commutative
theory must eventually reduce to part of the mSME.  The approach
of~\cite{Anisimov:2001zc} shows this is possible, although the
presence of such a low energy cutoff must be explained.

All of the above approaches use an expansion in $\Theta^{\al
\be},\Lambda_{NC}$ to get some low energy effective field theory.
In terms of Lorentz tests, the results are all based upon this EFT
expansion and not on the full non-commutative theory.  Therefore
we will restrict ourselves to discussing limits on various terms
in effective field theories rather than directly quoting limits on
the non-commutative scale. We leave it up to the reader to
translate this value into a constraint (if any) on $\Lambda_{NC}$
and or $\Lambda$.

\subsection{Symmetry and relevant/irrelevant Lorentz violating operators} \label{subsubsec:InducedLowEnergyLV}%
The above section illustrates a crucial issue in searches for
Lorentz violation that are motivated by quantum gravity: why is
Lorentz invariance such a good approximate symmetry at low
energies?  To illustrate the problem, let us consider the standard
assumption made in much of the work on Lorentz violation in
astrophysics - that there exist corrections to particle dispersion
relations of the form $f^{(n)}p^n E_{Pl}^{n-2}$ with $n \geq 3$
and $f^{(n)}$ of order one. Without any protective symmetry,
radiative corrections involving this term will generate dispersion
terms of the form $f^{(n)} p ^2 + E_{Pl} f^{(n)} p$. These terms
are obviously ruled out by low energy experiment.\footnote{While
we are primarily concerned in this section with dimensional
transmutation of higher dimension operators, Lorentz violating
renormalizable operators for one particle of course also yield
radiative corrections to other particle operators.  For a specific
example see~\cite{Andrianov:2001zj}.} Accordingly, the first place
to look for Lorentz violation is in terrestrial experiments using
the standard model extension rather than astrophysics with higher
dimension operators. However, no evidence for such violation has
been found. The absence of lower dimension operators implies that
either there is a fine tuning in the Lorentz violating
sector~\cite{Collins:2004bp}, some other symmetry is present that
protects the lower dimension operators, or Lorentz invariance is
an exact symmetry.

It is always possible that Lorentz violation is finely tuned -
there are other currently unexplained fine-tuning problems (such
as the cosmological constant) in particle physics. However, it
would be far preferable if there was some symmetry or partial
symmetry that could naturally suppress/forbid lower dimension
operators. For rotation invariance, a discrete remnant of the
original symmetry is enough. For example, hypercubic symmetry on a
lattice is enough to forbid dimension four rotation breaking
operators for scalars.\footnote{This can easily be seen by the
following argument.  Consider a kinetic term in Euclidean space
for a scalar field $\phi$ of the form $M^{ab} \partial_a \phi
\partial_b \phi$. In four dimensions $M^{ab}$ must be a dimensionless
tensor that has hypercubic symmetry.  The only such tensor is
$\delta^{ab}$ so rotation invariance is automatically preserved.
Interaction terms are by their very nature rotation invariant,
which implies that the entire action is invariant under the full
rotation group.} No physically meaningful equivalent construction
exists for the full Lorentz group however (see
~\cite{MooreLecture} for a further discussion of this point).  A
discrete symmetry that can forbid some of the possible lower
dimension operators is CPT. A number of the most observationally
constrained operators in the mSME are CPT violating, so imposing
CPT symmetry would explain why those operators are absent.
However, the CPT even operators in the mSME are also very tightly
bounded, so CPT cannot completely resolve the naturalness problem
either.

Supersymmetry is currently the only known symmetry (other than
Lorentz symmetry itself) that can protect Lorentz violating
operators of dimension four or less~\cite{GrootNibbelink:2004za,
Jain:2005as, Bolokhov:2005cj}, much as SUSY protects some lower
dimension operators in non-commutative field
theory~\cite{Matusis:2000jf}. If one imposes \textit{exact} SUSY
on a Lorentz violating theory, the first allowed operators are of
dimension five~\cite{GrootNibbelink:2004za}.  These dimension five
operators do not induce $n=3$ type dispersion like the operators
(\ref{eq:MPops}). Instead, in a rotationally invariant setting
they produce dispersion relations of the form
\begin{equation}
E^2=p^2+m^2(1+ f^{(1)}\frac{E} {E_{Pl}}+...).
\end{equation}
Such modifications are completely unobservable in astrophysical
processes, although high precision terrestrial experiments can
still probe them.  Dimension 6 SUSY operators in SQED also yield
dispersion relations that are untestable by high energy
astrophysics~\cite{Bolokhov:2005cj}.

Fortunately, we do not live in a SUSY world, so it may be that
upon SUSY breaking appropriate sized operators at each mass
dimension are generated. This question has recently been explored
in~\cite{Bolokhov:2005cj}. For CPT violating dimension five SUSY
operators in SQED, the authors find that SUSY breaking yields
dimension three operators of the form $\alpha m_s^2/M$, where
$m_s$ is the SUSY breaking scale, $M$ is the scale of Lorentz
violation and $\alpha$ is an O(1) coefficient. For $m_s$ as light
as it could be (around 100 GeV), spin polarized torsion balances
(see section \ref{subsec:balance}) are able to place limits on $M$
 between $10^5-10^{10} E_{Pl}$. It therefore is probable that these
operators are observationally unacceptable.  However, dimension
five SUSY operators are CPT violating, so a combination of CPT
invariance and SUSY would forbid Lorentz violating operators below
dimension six. The low energy dimension four operators induced by
SUSY breaking in the presence of dimension six operators would
then presumably be suppressed by  $m_s^2/M^2$. This is enough
suppression to be compatible with current experiment if $M$ is at
the Planck scale and $m_s \leq 1$ TeV.

Another method by which Lorentz violation can occur but might have
small dimension $\leq 4$ matter operators is via extra dimension
scenarios. For example, in~\cite{Burgess:2002tb} a braneworld
scenario was considered where 4-d Lorentz invariance was preserved
on the brane but broken in the bulk. The only particle which can
then directly see Lorentz violation is the graviton - the matter
fields, being trapped on the brane, can only feel the bulk Lorentz
violation through graviton loops. The induced dimension $\leq 4$
operators can be quite small, depending on the exact
extra-dimension scenario considered.  Note though that this
approach has been criticized in~\cite{Collins:2004bp}, whose
authors argue that significant Lorentz violation in the infrared
would still occur.

In summary, the current status of Lorentz violation in EFT is
mildly disconcerting for a phenomenologist (if one really wants to
believe in Lorentz violation). From an EFT point of view, without
custodial symmetries one would expect that we would have seen
signs of Lorentz violation by now. Imposing SUSY+CPT or a
braneworld scenario may fix this problem, but then we are left
with a model with more theoretical assumptions.  Furthermore a
SUSY+CPT model is unlikely to ever be testable with astrophysics
experiments and requires significant improvement in terrestrial
experiments to be seen~\cite{Bolokhov:2005cj}. Fortunately, since
this is a phenomenological review we can blithely ignore the above
considerations and simply classify and constrain all possible
operators at each mass dimension.  This is also the safest
approach. After all, we are searching for a possible signal from
the mysterious realm of quantum gravity and so must be careful
about overly restricting our models.

\subsection{Lorentz violation with gravity in EFT}\label{subsubsec:LVandgravity} %
The previous field theories dealt only with the possible Lorentz
violating terms that can be added to the matter sector.  Inclusion
of gravity into the mix yields a number of new phenomena.  Lorentz
violating theories with a preferred frame have been studied
extensively (c.f. ~\cite{Gasperini,Jacobson:2001yj, Will:2001mx}
and references therein), while an extension of the mSME into
Riemann-Cartan geometry has been performed
in~\cite{Kostelecky:2003fs}. Ghost condensate models, in which a
scalar field acquires a constant time derivative, thereby choosing
a preferred frame, were introduced in ~\cite{Arkani-Hamed:2003uy}.
Let us first look at the more generic case
of~\cite{Kostelecky:2003fs}.

In order to couple Lorentz violating coefficients to fermions, one
must work in the vierbein formalism (for a discussion
see~\cite{Wald:1984rg}). In Riemann-Cartan geometry the
gravitational degrees of freedom are the vierbein and spin
connection which give the Riemann and torsion tensors in
spacetime.  For the purposes of this review we will set the
torsion to zero and work strictly in Riemannian geometry; for the
complete Lorentz violating theory with torsion
see~\cite{Kostelecky:2003fs} (for more general reviews of torsion
in gravity see~\cite{Hehl:1976kj,Hammond:2002rm}). The low energy
action involving only second derivatives in the metric is given by
\begin{equation}
S=\frac {1} {16 \pi G} \int d^4x e[R-2\Lambda + s^{\al \be} R_{\al
\be} + t^{\al \be \ga \de} R_{\al \be \ga\de}]
\end{equation}
where $e$ is the determinant of the vierbein, $R,R_{\al \be},
R_{\al \be \ga \de}$ are the Ricci scalar, Ricci tensor, and
Riemann tensor respectively, and $\Lambda$ is the cosmological
constant. $G$ is the gravitational coupling constant, which can be
affected by Lorentz violation. Since there is no longer
translation invariance, in principle the Lorentz violating
coefficients $s^{\al \be},t^{\al \be \ga \de}$ vary with location,
so they also behave as spacetime varying couplings. $s^{\al
\be},t^{\al \be \ga \de}$ can furthermore be assumed to be
trace-free as the trace can be absorbed into $G$ and $\Lambda$.
There are then 19 degrees of freedom left.

The difficulty with this formulation is that it constitutes prior
geometry and generically leads to energy-momentum
non-conservation, similar to the bimetric model in section
\ref{subsubsec:Diffeomorphism}.  Again the matter stress tensor
will not be conserved unless very restrictive conditions are
placed on $s^{\al \be},  t^{\al \be \ga \de}$ (for example that
they are covariantly constant).  It is unclear whether or not such
restrictions can be consistently imposed in a complicated metric
as would describe our universe.

A more flexible approach is to presume that the Lorentz violating
coefficients are dynamical, as has been pursued
in~\cite{Gasperini,Jacobson:2001yj,Arkani-Hamed:2003uy,Kostelecky:1988zi,Moffat:2002nu}.
In this scenario, the matter stress tensor is automatically
conserved if all the fields are on-shell. The trade-off for this
is that the coefficients $s^{\al \be},t^{\al \be \ga \de}$ must be
promoted to the level of fields. In particularly they can have
their own kinetic terms. Not surprisingly, this rapidly leads to a
very complicated theory, as not only must $s^{\al \be},t^{\al \be
\ga \de}$ have kinetic terms, but they must also have potentials
that force them to be non-zero at low energies.  (If such
potentials were not present, then the vacuum state of the theory
would be Lorentz invariant.) For generic $s^{\al \be},t^{\al \be
\ga \de}$, the complete theory is not known, but a simpler theory
of a dynamical ``aether'', first looked at by~\cite{Gasperini} and
expanded on by~\cite{Jacobson:2001yj,Kostelecky:1988zi,
Eling:2004dk, Bertolami:2005bh} has been explored.

The aether models assume that all the Lorentz violation is
provided by a vector field $u^\al$.\footnote{The aether models
take as a starting point general relativity plus an external
vector field.  Recent work~\cite{Heinicke:2005bp} has shown that
there is an alternate formulation with the same Lorentz violating
consequences in terms of a non-zero shear field in metric-affine
gravity.} With this assumption, $s^{\al \be}$ can be written as
$u^\al u^\be$ and $t^{\al \be \ga \de}$ can always be reduced to
an $s^{\al \be}$ term due to the symmetries of the Riemann tensor.
The most generic action in D-dimensions that is quadratic in
fields is therefore

\begin{equation} \label{eq:aetheraction}
S=\frac{-1}{16\pi G } \int \!d^Dx \sqrt{-g} \Bigl(R+K^{\al
\be}{}{}_{\mu \nu} \nabla_\al u^\mu \nabla_\be u^\nu + V(u^\al
u_\al)\Bigr)
\end{equation}
where
\begin{equation} \label{eq:aethercoeffs}
    K^{\al\be}{}{}_{\mu\nu}=c_1 g^{\al\be} g_{\mu\nu} + c_2 \delta^a_\mu
    \delta^\be_\nu
    + c_3 \delta^\al_\nu \delta^\be_\mu + c_4 u^\al u^\be g_{\mu\nu}.
    \label{K}
\end{equation}
where the $s^{\al \be}$ has been integrated by parts and replaced
with the $c_1,c_3$ terms.  The coefficients $c_{1,2,3,4}$ are
dimensionless constants, $R$ is the Ricci scalar, and the
potential $V(u^\al u_\al)$ is some function that enforces a
non-zero value for $u^a$ at low energies.  With a proper scaling
of coefficients and $V$ this value can be chosen to be unit at low
energies. The model of (\ref{eq:aetheraction}) still allows for
numerous possibilities.  Besides the obvious choice of which
coefficients are actually present, $u^\al$ can be either
space-like or time-like and in extra dimension scenarios can point
in one of the four uncompactified dimensions or the $D-4$
compactified ones.

At low energies $u^\al$ acquires an expectation value, $\bar
{u}^\al$, and there will be excitations $\delta u^\al$ about this
value.  Generically, there will be a single massive excitation and
three massless ones.  It has been argued in~\cite{Elliott:2005va}
that the theory suffers stability problems unless $V$ is of the
form $\lambda(u^\al u_\al-1)$ where $\lambda$ is a Lagrange
multiplier.  The theory is also ghost free with this potential and
the further assumption that $c_1+c_4<0$~\cite{Graesser:2005bg}.
Assuming these conditions, aether theories possess a set of
coupled aether-metric modes which act as new gravitational degrees
of freedom that can be searched for with gravitational wave
interferometers or by determining energy loss rates from inspiral
systems like the binary pulsar. The same scenario generically
happens for any tensor field that acquires a VEV dynamically (see
section \ref{subsec:GravitationalWaves}), which implies that
Lorentz violation can be constrained by the gravitational sector
as well as by direct matter couplings.

The aether models use a vector field to describe a preferred
frame.  Ghost condensate gives a more specific model involving a
scalar field.  In this scenario the scalar field $\phi$ has a
Lagrangian of the form $P(X)$, where $X=\partial_\al \phi
\partial^\al \phi$.  $P(X)$ is a polynomial in $X$ with a minimum
at some value $X=m$, i.e. $\phi$ acquires a constant velocity at
its minimum.  In a cosmological setting, Hubble friction drives
the field to this minimum, hence there is a global preferred frame
determined by the velocity of $\phi$.  This theory gives rise to
the same Lorentz violating effects of aether theories, such as
\v{C}erenkov radiation and spin dependent
forces~\cite{Arkani-Hamed:2004ar}.  In general systems that give
constraints on the coefficients of the aether theory are likely to
also yield constraints on the size of the velocity $m$.

\newpage

\section{Terrestrial constraints on Lorentz violation} \label{sec:ExperimentsEarth}
Having laid out the necessary theoretical background, we now
discuss the various experiments and observations that give the
best limits on Lorentz violation.
\subsection{Penning traps} \label{subsec:penning}
A Penning trap is a combination of static magnetic and electric
fields that can keep a charged particle localized within the trap
for extremely long periods of time (for a review of Penning traps
see~\cite{Brown:1985rh}).  A trapped particle moves in a number of
different ways.  The two motions relevant for Lorentz violation
tests are the cyclotron motion in the magnetic field and Larmor
precession due to the spin. The ratio of the precession frequency
$\omega_s$ to the cyclotron frequency $\omega_c$ is given by
\begin{equation}
\omega_s/\omega_c= g/2
\end{equation}
where $g$ is the g-factor of the charged particle.  The energy
levels for a spin $1/2$ particle are given by $E_n^s=n \omega_c +
s \omega_s$ where n is an integer and $s=\pm 1/2$.  For electrons
and positrons, where $g \approx 2$, the state $n,s=-1/2$ is almost
degenerate with the state $n-1,s=+1/2$. The degeneracy breaking is
solely due to the anomalous magnetic moment of the electron and is
usually denoted by $\omega_a=\omega_s-\omega_c$. By introducing a
small oscillating magnetic field into the trap one can induce
transitions between these almost degenerate energy states and very
sensitively determine the value of $\omega_a$.

The primary use of measurements of $\omega_a$ is that they
directly give a very accurate value of $g-2$.  However, due to
their precision, these measurements also provide good tests of CPT
and Lorentz invariance. In the mSME, the only framework that has
been applied to Penning trap experiments, the $g$ factor for
electrons and positrons receive no corrections at lowest order.
However, the frequencies $\omega_a,\omega_c$ both receive
corrections~\cite{Bluhm:1997qb}. At lowest order in the Lorentz
violating coefficients these corrections are (with the trap's
magnetic field in the z-direction)
\begin{eqnarray} \label{eq:penningfreqs}
\omega^{e^-}_c=(1-c^e_{00}-c^e_{XX}-c^e_{YY}) \omega^{e,0}_c \\
\omega^{e^\mp}_a=\omega^{e,0}_a \mp 2 b^e_Z+2 d^e_{Z0} m_e +
2H^e_{XY}
\end{eqnarray}
expressed in a non-rotating frame.  The unmodified frequencies are
denoted by $\omega^{e,0}_{c,a}$ and the Lorentz violating
parameters are various components of the general set given in
(\ref{eq:LVQEDodd}) and (\ref{eq:LVQEDeven}).

The functional form of (\ref{eq:penningfreqs}) immediately makes
clear that there are two ways to test for Lorentz violation. The
first is to look for instantaneous CPT violation between electrons
and positrons which occurs if the $b_Z$ parameter is non-zero. The
observational bound on the difference between $\omega_a$ for
electrons and positrons is $|\omega_a^+-\omega_a^-|< 2.4 \cdot
10^{-21} m_e$~\cite{Dehmelt:1999jh}.  This leads to a bound on
$b_Z$ of order $b_Z \leq 10^{-21} m_e$.  The second approach is to
track $\omega_{a,c}$ over time, looking for sidereal variations as
the orientation of the experimental apparatus changes with respect
to the background Lorentz violating tensors.  This approach has
been used in~\cite{Mittleman:1999it} to place a bound on the
diurnal variation of the anomaly frequency of $\Delta
\omega^{e^-}_{a} \leq 1.6 \cdot 10^{-21} m_e$, which limits a
particular combination of components of $b_\mu, c_{\mu \nu},
d_{\mu \nu} \, H_{\mu \nu}$ at this level. Finally , we note that
similar techniques have been used to measure CPT violations for
proton/anti-proton and hydrogen ion
systems~\cite{Gabrielse:1999kc}. By measuring the cyclotron
frequency over time, bounds on the cyclotron frequency variation
(\ref{eq:penningfreqs}) for the anti-proton have established a
limit at the level of $10^{-26}$ on components of $c^{p^-}_{\mu
\nu}$.

\subsection{Clock comparison experiments} \label{subsec:clock}
The classic clock comparison experiments are those of
Hughes~\cite{Hughes} and Drever~\cite{Drever} and their basic
approach is still used today.  Two ``clocks'', usually two atomic
transition frequencies, are co-located at some point in space.  As
the clocks move they pick out different components of the Lorentz
violating tensors in the mSME, yielding a sidereal drift between
the two clocks.  The difference between clock frequencies can be
measured over long periods, yielding extremely high precision
limits on the amount of drift and hence the parameters in the
mSME.\footnote{Clock comparison experiments have also been used to
place bounds on Lorentz violation in a conjectured low energy
state for loop quantum gravity~\cite{Sudarsky:2002ue}.} Note that
this approach is only possible if the clocks are made of different
materials or have different orientations.

The best overall limit is in the neutron sector of the mSME and
comes from a $^3He/^{129}Xe$ maser system~\cite{Bear:2000cd,
Bear2}. In this setup, both noble gases are co-located.  The gases
are placed into a population inverted state by collisions with a
pumped vapor of rubidium.  In a magnetic field of 1.5 Gauss, each
gas acts as a maser at frequencies of 4.9 kHz and 1.7 kHz for He
and Xe, respectively.  The Xe emission is used as a magnetometer
to stabilize the magnetic field while the He emission frequency is
tracked over time, looking for sidereal variation. At lowest order
in Lorentz violating couplings, the Lorentz violating effect for
each gas is that of a single valence neutron so this experiment is
sensitive only to neutron parameters in the mSME.  The magnitude
of the sidereal variation $\Delta f_J$ is given by
\begin{equation}
2 \pi |\Delta f_J| = |-3.5 \tilde{b}_J +0.012 (\tilde{d}_J
-\tilde{g}_{D,J})|
\end{equation}
where J stands for the X,Y components of the Lorentz violating
tensors in a non-rotating frame that are orthogonal to the earth's
rotation axis. All parameters are understood to be the ones for
the neutron sector of the mSME. The coefficients $\tilde{b},
\tilde{d},\tilde{g}$ are related to the mSME coefficients of
section \ref{subsubsec:SMEPractical} by~\cite{Kostelecky:1999mr}
\begin{eqnarray} \label{eq:tildes}
\tilde{b}_J=b_J-m d_{J0} + \frac {1} {2} m \ep_{JKL} g_{KL0} -
\frac {1} {2} \ep_{JKL} H_{KL} \\
\tilde{d}_J=m(d_{0J} + d_{J0}) - \frac {1} {4}(2 m d_{J0} +
\ep_{JKL} H_{KL})\\
\tilde{g}_{D,J}= m \ep_{JKL}(g_{K0L} + \frac {1} {2} g_{KL0})-b_J.
\end{eqnarray}
Here $m$ is the neutron mass and $\ep_{IJK}$ is the 3-dimensional
antisymmetric tensor.  Barring conspiratorial cancellations among
the coefficients, the bound on
$\tilde{b}_\perp=\sqrt{\tilde{b}_X^2+\tilde{b}_Y^2}$ is $6.4 \pm
5.4 \cdot 10^{-32}$ GeV, which is the strongest clock comparison
limit on mSME parameters.  Similarly, one can derive bounds on
$\tilde{d}_\perp, \tilde{g}_{D,\perp}$ that are two to three
orders of magnitude lower.  Hence certain components of these
coefficients are bounded at the level of $10^{-28}$ GeV.  A
continuation of this experiment has recently been able to directly
constrain boost violation at the level of $10^{-27}$
GeV~\cite{Cane:2003wp} (sidereal variations look at rotation
invariance). Besides the bounds above, other clock comparison
experiments ~\cite{Kostelecky:1999mr} are able to establish the
following bounds on other coefficients in the neutron sector of
the mSME:
\begin{eqnarray}
|\tilde{c}_{Q,J}|=|m(c_{JZ}+c_{ZJ})| < 10^{-25}~GeV \\ \nonumber%
|\tilde{c}_-|=|m(c_{XX}-c_{YY})| <10^{-27}~GeV\\ \nonumber %
|\tilde{c}_{XY}|=|m(c_{XY}+c_{YX})|<10^{-27}~GeV.
\end{eqnarray}

A constraint of the dimension five operators of (\ref{eq:MPops})
for neutrons was recently derived in~\cite{Bertolami:2004bf} using
limits on the spatial variation of the hyperfine nuclear spin
transition in $Be^+$ as a function of the angle between the spin
axis and an external magnetic field~\cite{Bollinger}. Assuming the
reference frame of the earth is not aligned with the four vector
$u^\alpha$, the extra terms in (\ref{eq:MPops}) generically
introduce a small orientation dependent potential into the
non-relativistic Schrodinger equation for any particle. For
$Be^+$, the nuclear spin can be thought of as being carried by a
single neutron, so this experiment limits the neutron Lorentz
violating coefficients. This extra potential for the neutron leads
to anisotropy of the hyperfine transition frequency, which can be
bounded by experiment.  The limits are roughly $|\eta_1|< 6 \times
10^{-3}, |\eta_2|<3$ if $u^\alpha$ is timelike and coincides with
the rest frame of the CMBR.  If $u^\alpha$ is spacelike one has
$|\eta_1|< 2 \times 10^{-8}, |\eta_2|<10^{-8}$. If $u^\alpha$ is
lightlike both coefficients are bounded at the $10^{-8}$ level.
Note that all these bounds are approximate, as they depend on the
spatial orientation of the experiment with respect to spatial
components of $u^\alpha$ in the lab frame. The authors
of~\cite{Bertolami:2004bf} have assumed that the orientation is
not special.

The above constraints applies solely to the neutron sector. Other
clock comparison experiments have been performed that yield
constraints on the proton
sector~\cite{Chupp,Prestage:1985zm,Lamoreaux:1986xz, Berglund,
Phillips:2000dr} in the mSME. The best proton limit, on the
$\tilde{b}_\perp$ parameter, is $|\tilde{b}_\perp|< 2 \cdot
10^{-27}$ GeV~\cite{Phillips:2000dr}, with corresponding limits on
$\tilde{d}_\perp, \tilde{g}_{D,\perp}$ of order $10^{-25}$ GeV.
Similar bounds have been estimated~\cite{Kostelecky:1999mr} from
the experiment of Berglund et. al.~\cite{Berglund} using the
Schmidt model~\cite{Schmidt} for nuclear structure, where an
individual nucleon is assumed to carry the entire nuclear angular
momentum. The experiments of Chupp~\cite{Chupp},
Prestage~\cite{Prestage:1985zm}, and
Lamoreaux~\cite{Lamoreaux:1986xz} are insensitive to proton
coefficients in this model so no proton bounds have yet been
established from these experiments.  As noted
in~\cite{Kostelecky:1999mr}, proton bounds would be derivable with
a more detailed model of nuclear structure.

\subsection{Cavity experiments} \label{subsec:Cavity}
From the Michelson-Morley experiments onward, interferometry has
been an excellent method of testing relativity. Modern cavity
experiments extend on the ideas of interferometry and provide very
precise tests on the bounds of certain photon parameters.  The
main technique of a cavity experiment is to detect the variation
of the resonance frequency of the cavity as its orientation
changes with respect to a stationary frequency standard.  In this
sense, it is similar to a clock comparison experiment.  However,
since one of the clocks involves photons, cavity experiments
constrain the electromagnetic sector of the mSME as well.

The analysis of cavity experiments is much easier if we make a
field redefinition of the electromagnetic sector of the
mSME~\cite{Kostelecky:2002hh}. In analogy to the theory of
dielectrics, we define two new fields, $\vec{D}, \vec{H}$ by
\begin{equation}
\left(%
\begin{array}{c}
  \vec{D} \\
  \vec{H} \\
\end{array}%
\right) =\left(%
\begin{array}{cc}
  1 + \kappa_{DE} & \kappa_{DB} \\
  \kappa_{HE} & 1 + \kappa_{HB} \\
\end{array}%
\right)
\left(%
\begin{array}{c}
  \vec{E} \\
  \vec{B} \\
\end{array}%
\right).
\end{equation}
The $\kappa$ coefficients are related to the mSME coefficients by
\begin{eqnarray}
(\kappa_{DE})^{jk}=-2(k_F)^{0j0k} \\
(\kappa_{HB})^{jk}=\frac {1} {2} \ep^{jkq} \ep^{krs}
(k_F)^{pqrs}\\
(\kappa_{DB})^{jk}=-(\kappa_{HE})^{jk}= \ep^{kpq} (k_F)^{0jpq}.
\end{eqnarray}
With this choice of fields, the modified Maxwell equations from
the mSME take the suggestive form
\begin{eqnarray} \label{eq:MaxwelDH}
\vec{\nabla} \times \vec{H} -\partial_0 \vec{D}= 0 \\
\vec{\nabla} \cdot \vec{D}=0\\
\vec{\nabla} \times \vec{E} +\partial_0 \vec{B}= 0 \\
\vec{\nabla} \cdot \vec{B}=0.
\end{eqnarray}
This redefinition shows that the Lorentz violating background
tensor $(k_F)_{\mu \nu \al \be}$ can be thought of as a dielectric
medium with no charge or current density.  Hence we can apply much
of our intuition about the behavior of fields inside a dielectric
to construct tests of Lorentz violation. Note that since $\vec{H}$
and $\vec{D}$ depend on the components of $(k_F)_{\mu \nu \al
\be}$ the properties of the dielectric are orientation dependent.

Constraints from cavity experiments are not on the $\ka$
parameters themselves, but rather on the linear combinations
\begin{eqnarray}
\tilde{\ka}_{tr}=\frac {1} {3} (\ka_{DE})^{ll}\\
(\tilde{\kappa}_{e+})^{jk}=\frac {1} {2} (\ka_{DE}+\ka_{HB})^{jk}\\
(\tilde{\kappa}_{e-})^{jk}=\frac {1} {2} (\ka_{DE}-\ka_{HB})^{jk}-
 \de^{jk}\tilde{\ka}_{tr} \\
(\tilde{\kappa}_{o+})^{jk}=\frac {1} {2} (\ka_{DB}+\ka_{HE})^{jk}\\
(\tilde{\kappa}_{o-})^{jk}=\frac {1} {2} (\ka_{DB}-\ka_{HE})^{jk}.
\end{eqnarray}
$\tilde{\ka}_{tr}, \tilde{\ka}_{e+}, \tilde{\ka}_{e-}$ are all
parity even while $\tilde{\ka}_{o+}$ and $\tilde{\ka}_{o-}$ are
parity odd.  The usefulness of this parameterization can be seen
if we rewrite the Lagrangian in these
parameters~\cite{Kostelecky:2002hh},
\begin{equation}
L=\frac {1} {2} [(1+\tilde{\ka}_{tr}) \vec{E}^2 -
(1-\tilde{\ka}_{tr})\vec{B}^2] + \frac {1} {2} \vec{E} \cdot
[\tilde{\kappa}_{e+}+\tilde{\kappa}_{e-}] \cdot \vec{E}-\frac {1}
{2} \vec{B} \cdot [\tilde{\kappa}_{e+}-\tilde{\kappa}_{e-}] \cdot
\vec{B} + \vec{E} \cdot [\tilde{\kappa}_{o+}+\tilde{\kappa}_{o-}]
\cdot \vec{B}.
\end{equation}
From this expression it is easy to see that $\tilde{\ka}_{tr}$
corresponds to a rotationally invariant shift in the speed of
light. It can be shown that $\tilde{\kappa}_{e-}$ and
$\tilde{\kappa}_{o+}$ also yield a shift in the speed of light,
although in a direction dependent manner.  The coefficients
$\tilde{\kappa}_{e+}$ and $\tilde{\kappa}_{o-}$ control
birefringence.  Cavity experiments yield limits on
$\tilde{\kappa}_{e-}$ and $\tilde{\kappa}_{o+}$ while
birefringence (section \ref{subsec:Birefringence}) bounds
$\tilde{\kappa}_{e+}$ and $\tilde{\kappa}_{o-}$.

The most straightforward way to constrain Lorentz violation with
cavity resonators is to study the resonant frequency of a cavity.
Since we have a cavity filled with an orientation dependent
dielectric the resonant frequency will also vary with orientation.
The resonant frequency of a cavity is
\begin{equation}
f_r=\frac {m c} {2 n L}
\end{equation}
where $m$ is the mode number, $c$ is the speed of light, $n$ is
the index of refraction (including Lorentz violation) of any
medium in the cavity, and $L$ is the length of the cavity.  $f_r$
can be sensitive to Lorentz violating effects through $c, n, L$.
Depending on the construction of the cavity some effects can
dominate over others. For example, in sapphire cavities the change
in $L$ due to Lorentz violation is negligible compared to the
change in $c$. This allows one to isolate the electromagnetic
sector.

In general, all cavities are sensitive to the the photon $\kappa$
parameters.  In contrast to sapphire, for certain materials the
strain induced on the cavity by Lorentz violation is large. This
allows sensitivity to the electron parameters $c_{\mu \nu}$ at a
level equivalent to the photon parameters. Furthermore, by using a
cavity with a medium, the dependence of $f_r$ on $n$ gives
additional electron sensitivity~\cite{Mueller:2004zp}.

The complete bounds on the mSME coefficients for cavity
experiments are given in~\cite{Amelino-Camelia:2005qa,
Braxmaier:2001wu,Muller:2004mb,Mueller:2004zp,Wolf:2004gg,Lipa:2003mh,Antonini:2005yb,
Stanwix:2005yv}. The strongest bounds are displayed in table
\ref{tb:cavity}. Roughly, the components of $\tilde{\kappa}_{e-}$
and $c_{\mu \nu}$ are bounded at $O(10^{-15})$ while
$\tilde{\kappa}_{o+}$ is bounded at $O(10^{-11})$.  The $10^4$
difference arises as $\tilde{\kappa}_{o+}$ enters constraints
suppressed by the boost factor of the earth relative to the solar
``rest'' frame where the coefficients are taken to be constant.

\begin{table} [htb]
  \centering

\begin{tabular}{|l|r|}
\hline
   Parameter & Value ($\times 10^{-15}$)\\ \hline
    $c^e_{XY}$ & $0.76 \pm 0.35$\\ [2pt]
    $c^e_{YZ}$ & $0.21 \pm 0.46$\\[2pt]
    $c^e_{XZ}$ & $-0.16 \pm 0.63$\\[2pt]
    $c_{XX}-c_{YY}$ & $1.15 \pm 0.64$\\[2pt]
    $|c_{XX}+c_{YY} - 2 c_{ZZ} - 0.25 \tilde{\kappa}_{e^-}^{ZZ}|$ &
    $10^3$\\[2pt]
    \hline
    $\tilde{\kappa}_{e-}^{XY}$ & $-0.63 \pm 0.43$\\[2pt]
    $\tilde{\kappa}_{e-}^{YZ}$ & $-0.45 \pm 0.37$\\[2pt]
    $\tilde{\kappa}_{e-}^{XZ}$ & $0.19 \pm 0.37$ \\[2pt]
    $\tilde{\kappa}_{e-}^{XX} - \tilde{\kappa}_{e-}^{YY}$ & $-1.3 \pm
    0.9$\\[2pt]
    $\tilde{\kappa}_{e-}^{ZZ}$ & $-20 \pm 2$\\[2pt]
    \hline
    $\tilde{\kappa}_{o+}^{XY}$ & $(0.20 \pm 0.21) \times 10^4$\\[2pt]
    $\tilde{\kappa}_{o+}^{YZ}$ & $(0.44 \pm 0.46) \times 10^4$\\[2pt]
    $\tilde{\kappa}_{o+}^{XZ}$ & $(-0.91 \pm 0.46) \times 10^4$\\[2pt]
   \hline
\end{tabular}
  \caption{Cavity limits on $c_{\mu \nu},\tilde{\kappa}_{e-}, \tilde{\kappa}_{o+}$ (Taken from~\cite{Amelino-Camelia:2005qa,Mueller:2004zp, Antonini:2005yb, Stanwix:2005yv}).
  Components are in a sun centered equatorial frame.  Error bars are $1 \sigma$. The non-zero value of
$\tilde{\kappa}_{e-}^{ZZ}$ is argued by the authors to be due to
systematics in the
experiment~\cite{Antonini:2005yb}.}\label{tb:cavity}
\end{table}

\subsection{Spin polarized torsion balances} \label{subsec:balance}
Clock comparison experiments constrain the $\tilde{b}_J$ parameter
for protons and neutrons.  Spin polarized torsion balances are
able to place comparable limits on the electron sector of the
mSME~\cite{Bluhm:1999ev}. The best limits on $\tilde{b}_i$ (where
i is the spatial direction, including that parallel to the earth's
rotation axis) for the electron come from two balances, one in
Washington~\cite{Kostelecky:1999dx, Balance} and one in
Taiwan~\cite{Balance2}. We detail the Washington experiment for
pedagogical purposes - the two approaches are similar. In the
Washington experiment a two different types of magnets (SmCo and
Alnico) are arranged in an octagonal shape. Four SmCo magnets are
on one side of the octagon and four Alnico magnets are on the
other.  The magnetization of both types of magnets is set to be
equal and in the angular direction around the octagon.  This
minimizes any magnetic interactions. However, with equal
magnetization the net electron spin of the SmCo and Alnico magnets
differs as the SmCo magnets have a large contribution to their
overall magnetization from orbital angular momentum of Sm ions.
Therefore the octagonal pattern of magnets has an overall spin
polarization in the octagon's plane.

A stack of four of these octagons are suspended from a torsion
fiber in a vacuum chamber.  The magnets give an estimated net spin
polarization $\overrightarrow{\sigma}$ equivalent to approximately
$10^{23}$ aligned electron spins.   The whole apparatus is then
mounted on a turntable. As the turntable rotates a bound on
Lorentz violation is obtained in the following manner. Lorentz
violation in the mSME gives rise to an interaction potential for
non-relativistic electrons of the form $V=\tilde{b}_i \sigma^i$,
where i stands for direction and $\sigma^i$ is the electron
magnetic moment.  As the turntable rotates, since $\tilde{b}$
points in some fixed direction in space, the interaction produces
a torque on the torsion balance. The magnet apparatus therefore
twists on the torsion fiber by an amount given by
\begin{equation}
\Theta=V_H ~\sin(\phi_0-\omega t)/ \kappa
\end{equation}
where $V_H$ is the horizontal component of $V$, $\omega$ is the
frequency of rotation, $\phi _0$ is an initial phase due to
orientation, and $\kappa$ is the torsion constant. Since $\kappa$
and $\omega$ are known, a measurement of $\Theta$ will give the
magnitude of $V_H$.  Since $\sigma^i$ is also known, $V_H$ gives a
limit on the size of $\tilde{b}_i$. The absence of any extra twist
limits all components of $|\tilde{b}|$ for the electron to be less
than $10^{-28}$ GeV. The Taiwan experiment uses a different
material ($Dy_6 Fe_23$)~\cite{Balance2}.  The bounds from this
experiment are of order $10^{-29}$ GeV for the components of
$\tilde{b}_i$ perpendicular to the spin axis and $10^{-28}$ GeV
for the parallel component.

To conclude this section, we note that the torsion balance
experiments are actually sensitive enough to also constrain the
dimension 5 operators in (\ref{eq:MPops}). Assuming that all lower
dimension operators are absent, the constraint on the dimension
five operators is $|\eta_R-\eta_L|<4$~\cite{Myers:2003fd}.

\subsection{Neutral mesons} \label{subsec:Mesons}
Mesons have long been used to probe CPT violation in the standard
model.  In the framework of the mSME, CPT violation also implies
Lorentz violation.  Let us focus on kaon tests, where most of the
work has been done.  The approach for the other mesons is
similar~\cite{Kostelecky:1998kz,Ackerstaff:1997vd}. The relevant
parameter for CPT and Lorentz violation in neutral kaon systems is
$a_\mu$ for the down and strange quarks (since $K=d\overline{s}$).
As we mentioned previously, one of the $a_\mu$ can always be
absorbed by a field redefinition. Therefore only the difference
between the quark $a_\mu$'s, $\Delta a_\mu=r_d a_\mu^d - r_s
a_\mu^s$ controls the amount of CPT violation and is physically
measurable. Here $r_{d,s}$ are coefficients that allow for effects
due to the quark bound state~\cite{Kostelecky:1994rn}.

A generic kaon state $\Psi_K$ is a linear combination of the
strong eigenstates, $K^0, \overline{K}^0$.  If we write $\Psi_K$
in two component form the time evolution of the $\Psi_K$
wavefunction is given by a Schrodinger equation:
\begin{equation}
i \partial_t \bigg{(}\begin{array}{c}
  K^0 \\
  \overline{K}^0 \\
\end{array}\bigg{)} = H \bigg{(} \begin{array}{c}
  K^0 \\
  \overline{K}^0 \\
\end{array} \bigg{)}
\end{equation}
where the Hamiltonian $H$ is a $2 \times 2$ complex matrix. $H$
can be decomposed into real and imaginary parts, $H=M - i\Gamma$.
$M$ and $\Gamma$ are hermitian matrices usually called the mass
matrix and decay matrix respectively. The eigenstates of $H$ are
the physically propagating states, which are the familiar short
and long decay states $K_S$ and $K_L$.  CPT violation only occurs
when the diagonal components of $H$ are not equal~\cite{LeeWu}. In
the mSME, the lowest order contribution to the diagonal components
of $H$ occurs in the mass matrix $M$, contributions to $\Gamma$
are higher order~\cite{Kostelecky:1994rn}. Hence the relevant
observable for this type of CPT violation in the kaon system is
the $K^0, \overline{K}^0$ mass difference, $\Delta_K$ =
$|m_{K^0}-m_{\overline{K}^0}|/m_{K^0}$.~\footnote{For a more
thorough discussion of CPT (and CP) tests, see for example
~\cite{Lee-Franzini:1997fs}.}

In the mSME the deviation $\Delta_K$ is (as usual) orientation
dependent.  In terms of $\Delta a_\mu$, we
have~\cite{Kostelecky:1999bm}
\begin{equation}
\Delta_K \approx |\frac {\beta^\mu \Delta a_\mu} {m_{K^0}}|
\end{equation}
where $\beta^\mu$ is the four-velocity of the kaon in the
observer's frame.  The mass difference $\Delta_K$ has been
extremely well measured by experiments such as
KTeV~\cite{Nguyen:2001tg} or FNAL E773 at
Fermilab~\cite{Schwingenheuer:1995uf}.  By looking for sidereal
variations or other orientation effects one can derive bounds on
each component of $a_\mu$.  The best current bounds to not quite
achieve this but rather constrain a combination of parameters. A
linear combination of $\Delta a_0,\Delta a_z$ is bounded at the
level of $10^{-20}$ GeV~\cite{Kostelecky:1997mh} and a combination
of $\Delta a_x, \Delta a_y$ is constrained at the $10^{-21}$
GeV~\cite{Nguyen:2001tg} level.

\subsection{Doppler shift of lithium}
If Lorentz invariance is violated then the transformation laws for
clocks with relative velocity will be different from the usual
time dilation. The RMS framework of section \ref{subsec:RMS}
provides a convenient parameterization of how the Doppler shift
can deviate from its standard relativistic form.  Comparisons of
oscillator frequencies under boosts therefore can constrain the
$\al_{RMS}$ parameter in the RMS framework.  The best test to date
comes from spectroscopy of lithium ions in a storage
ring~\cite{Saathoff}. In this experiment, $^7Li^+$ ions are
trapped in a storage ring at a velocity of $0.064c$.  The
transition frequencies of the boosted ions are then measured and
compared to the transition frequencies at rest, providing a bound
on the deviation from the special relativistic Doppler shift of
$|\al_{RMS}|<2 \cdot 10^{-7}$ in the RMS framework. Recently, the
results of~\cite{Saathoff} have been reinterpreted in the context
of the mSME.  For the electron/proton sector the approximate
bounds are~\cite{Lane:2005jv}
\begin{eqnarray} \nonumber
|c^p_{XX} + c^p_{YY}-2c^p_{ZZ}| \leq 10^{-11}\\ \nonumber
|c^p_{TJ} + c^p_{JT}| \leq 10^{-8}\\ \nonumber %
|c^e_{XX} + c^e_{YY}-2c^e_{ZZ}| \leq 10^{-5} \\
|c^e_{TJ} + c^e_{JT}| \leq 10^{-2}
\end{eqnarray}
where $J=X,Y,Z$ in a heliocentric frame.  In the photon sector,
the limit $\tilde{\ka}_{tr} \leq O(10^{-5})$ can also be set from
this experiment~\cite{Tobar:2004vi}.

\subsection{Muon experiments}
Muon experiments provide another window into the lepton sector of
the mSME.  As discussed in section
\ref{subsubsec:InducedLowEnergyLV}, if the mSME coefficients are
to be small then there must be some small energy scale suppressing
the Lorentz violating coefficients.  There are only a few
available small scales, namely particle masses or a symmetry
breaking scale.  If we assume the scale is particle mass, then
muon based experiments would have a signal at least $10^2$ larger
than equivalent electron experiments due to the larger mass of the
muon.  The trade-off, of course, is that muons are unstable so
experiments are intrinsically more difficult.

There are two primary experiments that give constraints on the
muon sector.  First, spin transitions in muonium ($\mu^+ e^-$)
have been used to place a bound on $\tilde{b}_J$ for the muon (see
(\ref{eq:tildes}) for the definition of
$\tilde{b}_J$)~\cite{Hughes:2001yk}. Even though muonium is a
muon-electron system, the muon sector of the mSME can be isolated
by placing the muonium in a strong magnetic field and looking for
a particular frequency resonance that corresponds to muon spin
flips. The sidereal variation of this transition frequency is then
tracked yielding a limit on $\tilde{b}_J$ of
\begin{equation}
|\tilde{b}_J|<5 \times 10^{-22} m_\mu
\end{equation}
where $J=X,Y$ in a non-rotating frame with $Z$ oriented along the
earth's spin axis.

The second muon experiment that yields strong limits is the
$\mu^-/\mu^+$ g-2 experiment~\cite{Bluhm:1999dx,
Bailey:1978mn,Carey:1999dd}. In this experiment relativistic
$\mu^-$ (or $\mu^+$) are injected into a storage ring and allowed
to decay. The deposit rate of the decay products along the
detector is sensitive to the evolution of the spin of the muon,
which in turn is a function of $g-2$ for the muon.  Lorentz
violation changes this evolution equation, and therefore this type
of $g-2$ experiments can bound the mSME.  As in the case of the
$g-2$ experiments in section \ref{subsec:penning}, two types of
bounds can be placed from the muon $g-2$ experiment.  The first is
a direct comparison between the $g-2$ factors for $\mu^-$ and
$\mu^+$, which limits the CPT violating coefficient $b_Z<10^{-22}$
GeV.  Furthermore, an analysis of sidereal variations involving
only one of the $\mu^-/\mu^+$ at the current sensitivity in
~\cite{Carey:1999dd} could bound the $\tilde{b}_J$ coefficient at
the level of $10^{-25}$ GeV~\cite{Bluhm:1999dx}.

\subsection{Constraints on the Higgs sector}
Since the constraints on various parameters of the mSME are so
tight, one can derive interesting indirect constraints on
unmeasured sectors by considering loop effects.  Such an approach
has been recently taken in~\cite{Anderson:2004qi}, where loop
corrections to mSME coefficients from Lorentz violation in the
Higgs sector are considered.  Such an approach could be used with
any particle, but since the Higgs is an observationally hidden
sector such an analysis is more important as direct tests are
unlikely any time soon.  There are four parameters in the Higgs
sector of the mSME (see section \ref{subsubsec:mSME}).

Constraints on the antisymmetric part of $(k_{\phi \phi})^{\mu
\nu}$, which we denote $(k_{\phi \phi}^A)^{\mu \nu}$, and
$(k_{\phi B})^{\mu \nu},(k_{\phi W})^{\mu \nu}$ come from the
birefringence constraints on photon propagation (see section
\ref{subsec:Birefringence}). Here the loop corrections to the
photon propagator induce a non-zero $(k_F)_{\al \be \mu \nu}$,
which can be directly constrained. This yields a constraint on all
three coefficients of order $10^{-16}$.  A bound $(k_{\phi
\phi}^s)^{\mu \nu}<10^{-13}$ can be derived from the cyclotron
frequencies of hydrogen ions and anti-protons.  Bounds on the CPT
violating term $(k_\phi)^\mu$ come from both the spin polarized
torsion balance experiments and the noble gas maser.  The torsion
experiments bound the $t,z$ components (where z is parallel to the
earth's rotation axis) at the level of $10^{-27}$ GeV and the
transverse components at $10^{-25}$ GeV. The He/Xe maser system
gives a better, although less clean, bound on the transverse
components of order $10^{-31}$ GeV.

\newpage

\section{Astrophysical constraints on Lorentz violation} \label{sec:ExperimentsAstro}
\subsection{Relevance of astrophysical observations}
Terrestrial experiments are invariably concerned with low energy
processes. They are therefore best suited for looking at the mSME,
which involves lower dimension operators.  Astrophysics is more
suited for directly constraining higher dimension operators as the
Lorentz violating effects scale with energy.  As mentioned in
section \ref{subsubsec:InducedLowEnergyLV}, the existence of
Lorentz violating higher dimensional operators would generically
generate lower dimension ones.  At the level of sensitivity of
astrophysical tests, the size of the corresponding lower dimension
operators should give signals in terrestrial experiments. Hence if
a signal is seen in astrophysics for Lorentz violation, one must
then explain why Lorentz invariance passes all the low energy
tests. As mentioned in section \ref{subsubsec:InducedLowEnergyLV},
exact SUSY, which is the only known mechanism to completely
protect lower dimension operators, yields dispersion modifications
(the primary method used in astrophysics) that are unobservable.
In summary there is currently no ``natural''  and complete way
that astrophysics might observe Lorentz violation but terrestrial
experiments confirm Lorentz invariance.  That said, physics is
often surprising, and it is therefore still important to check for
Lorentz violating signals in all possible observational areas.

\subsection{Time of flight} \label{subsec:TimeOfFlight}
The simplest astrophysical observations that provide interesting
constraints on Planck scale Lorentz violation are time of flight
measurements of photons from distant
sources~\cite{Amelino-Camelia:1997gz, Ellis:1999sd, Ellis:2002in}.
This is also one of two processes (the other being birefringence)
that can be directly applied to kinematic models. With a modified
dispersion relation of the form (\ref{eq:rotinvdisp}) and the
assumption that the velocity is given by $v=\partial E/
\partial p$\footnote{In $\kappa$-Minkowski space there is currently some
debate as to whether the standard relation for group velocity is
correct~\cite{Amelino-Camelia:2002tc,Kowalski-Glikman:2004qa}.
Until this is resolved, $v=\partial E/
\partial p$ remains an assumption that might be modified in a DSR context. It
obviously holds in field theoretic approaches to Lorentz
violation.} , the velocity of a photon is given by
\begin{equation}
v_\ga=1+\frac {(n-1)f^{(n)}_\ga E^{n-2}} {2 E_{Pl}^{n-2}}.
\end{equation}
If $n \neq 2$ the velocity is a function of energy, and the time
of arrival difference $\De T$ between two photons at different
energies travelling over a time T is
\begin{equation}
\De T=\De v T = \frac {(n-1)f^{(n)}_\ga (E_1^{n-2} - E_2^{n-2})}
{2 E_{Pl}^{n-2}} T
\end{equation}
where $E_{1,2}$ are the photon energies.  The large time $T$ plays
the role of an amplifier in this process, compensating for the
small ratio $E/E_{Pl}$.\footnote{This is the first example of a
significant constraint on terms in particle dispersion/effective
field theory that are Planck suppressed, which would naively seem
impossible.  The key feature of this reaction is the interplay
between the long travel time and the large Planck energy.  In
general any experiment that is sensitive to Planck suppressed
operators is either extremely precise (as in terrestrial tests of
the mSME) or has some sort of ``amplifier''. An amplifier is some
other scale (such as travel time or particle mass) which combines
with the Planck scale to magnify the effect.} For $n=1$ there are
much better low energy constraints, while for $n=4$ the
constraints are far too weak to be useful. Hence we shall
concentrate on $n=3$ type dispersion, where this constraint has
been most often applied.

The best limits~\cite{Biller:1998hg} are provided by observations
of rapid flares from Markarian 421, a blazar at a redshift of
approximately $z=0.03$, although a number of other objects give
comparable results~\cite{Schaefer:1998zg, Boggs:2003kx}.  The most
rapid flare from Markarian 421 showed a strong correlation of flux
at 1 and 2 TeV on a timescale of 280 seconds.  If we assume that
the flare was emitted from the same event at the source, the time
of arrival delay between 1 and 2 TeV photons must be less than 280
s. Combining all these factors yields the limit $|f^{(3)}|<128$.

A possible problem with the above bound is that in a single
emission event it is not known if the photons of different
energies are produced simultaneously. If different energies are
emitted at different times that might mask a LV signal. One way
around this is to look for correlations between time delay and
redshift, which has been done for a set of GRB's
in~\cite{Ellis:2002in}. Since time of flight delay is a
propagation effect that increases over time, a survey of GRB's at
different redshifts can separate this from intrinsic source
effects. This enables constraints to be imposed (or LV to be
observed) despite uncertainty regarding source effects.  The
current data from GRB's limit $f^{(3)}$ to be less than
$O(10^3)$~\cite{Ellis:2002in}.  Therefore significant
observational progress must be made in order to reach O(1) bounds
on $f^{(3)}$. Improvements on this limit might come from
observations of gamma ray bursts with new instruments such as
GLAST, however concerns have been raised that source effects may
severely impair this approach as well~\cite{Piran:2004qe,
Ellis:1999sd}.  Higher order dispersion corrections seem unlikely
to ever be probed with time of flight measurements.

The limit $|f^{(3)}|<128$ can be easily applied to the EFT
operators in (\ref{eq:MPops}).  From (\ref{eq:MPdispphot}) we
trivially see that the constraint on $\xi$ is $|\xi|<64$, again
comparing the 2 TeV peak to the 1 TeV. It might seem that we can
get a better constraint by demanding the time delay between 2 TeV
right handed and left handed photons is less than 280 seconds.
However, the polarization of the flare is unknown, so it is
possible (although perhaps unlikely) that only one polarization is
being produced. If one can show that both polarizations are
present, then one can further improve this constraint.  However,
the time of flight constraints are much weaker than other
constraints that can be derived on the operators in
(\ref{eq:MPops}) from birefringence so this line of research would
not be fruitful.

DSR theories may also predict a time of flight signal, where the
speed of light is effectively given by the group velocity of an
$n=3$ type dispersion relation.\footnote{There currently seems to
be some disagreement about this.  For example,
in~\cite{Amelino-Camelia:2003ex, Smolin:2005cz} an energy
dependent speed of light is argued for, whereas
in~\cite{Kowalski-Glikman:2004qa} no such modification is found.}
If there is such a frequency dependence, it is not expected that
DSR also yields birefringence as in the EFT case.  An $n=3$ type
dispersion for photons without birefringence would hence be a
strong signal for DSR or something similar. Coupled with the fact
that DSR does not affect threshold reactions or exhibit sidereal
effects, time of flight analyses provide the only currently
realistic probe of DSR theories. Unfortunately, since the
invariant energy scale is usually taken to be the Planck energy,
time of flight constraints are still one to two orders of
magnitude below what is needed to constrain/probe DSR.

As an aside, note that the actual measurement of the dependence of
the speed of light with frequency in a telescope such as
GLAST~\cite{GLAST} has a few subtleties in a DSR framework.  Let
us make the (unrealistic) assumption that the situation is as good
as it could possibly be experimentally: there is a short, high
energy gamma ray burst from some astrophysical source where all
the photons are emitted from the same point at the same time.  The
expected observational signal is then a correlation between the
photon time of arrival and energy. The time of arrival is fairly
straightforward to measure, but the reconstruction of the initial
photon energy is not so easy.  GLAST measures the initial photon
energy by calorimetry - the photon goes through a conversion foil
and converts to an electron-positron pair.  The pair then enters a
calorimeter, which measures the energy by scintillation. The
initial particle energy is then only known by reconstruction from
many events.  Energy reconstruction requires addition of the
multitude of low energy signals back into the single high energy
incoming photon.  Usually this addition in energy is linear (with
corrections due to systematics/experimental error). However, if we
take the DSR energy summation rules as currently postulated the
energies of the low energy events add non-linearly, leading to a
modified high energy signal.  One might guess that since the
initial particle energy is well below the Planck scale the
non-linear corrections make little difference to the energy
reconstruction.  However, to concretely answer such a question the
multi-particle sector of DSR must be properly understood (for a
discussion of the problems with multi-particle states in DSR
see~\cite{Kowalski-Glikman:2004qa}).

Finally, while photons are the most commonly used particle in time
of flight tests other particles may also be employed.  For
example, it has been proposed in~\cite{Choubey:2002bh} that
neutrino emission from GRB's may also be used to set limits on
$n=3$ dispersion.  Observed neutrino energies can be much higher
than the TeV scale used for photon measurements, hence one expects
that any time delay is greatly magnified.  Neutrino time delay
might therefore be a very precise probe of even $n>3$ dispersion
corrections. Of course, first an identifiable GRB neutrino flux
must be detected, which has not happened yet~\cite{Ahrens:2003pv}.
Assuming that a flux is seen and able to be correlated on the sky
with a GRB, one must still disentangle the signal.  In a DSR
scenario, where time delay scales uniformly with energy this is
not problematic, at least theoretically. However, in an EFT
scenario there can be independent coefficients for each helicity,
thereby possibly masking an energy dependent signal. For $n=3$
this complication is irrelevant if one assumes that all the
neutrinos are left-handed (as would be expected if produced from a
standard model interaction) as only $f^{(3)}_{\nu L}$ would then
apply.  For $n>3$ the possible operators are not yet known, so it
is not clear what bounds would be set by limits on neutrino time
of flight delays.

\subsection{Birefringence} \label{subsec:Birefringence}
A constraint related to time of flight is birefringence.  The
dimension five operators in (\ref{eq:MPops}) as well as certain
operators in the mSME induce birefringence - different speeds for
different photon polarizations
(\ref{eq:MPdispphot})\footnote{Gravitational birefringence has
also been studied extensively in the context of non-metric
theories of gravitation, which also exhibit Lorentz violation. See
for example~\cite{Solanki:1998ik, Gabriel:1991kf,Preuss:2004pp}
for discussions of these theories and the parallels with the
mSME.} A number of distant astrophysical objects exhibit strong
linear polarization in various low energy bands (see for example
the sources
 in ~\cite{Kostelecky:2001mb,Gleiser:2001rm}). Recently linear
polarization at high energies from a GRB has been
reported~\cite{Coburn:2003pi}, though this claim has been
challenged~\cite{Rutledge:2003wa,Wigger:2004by}. Lorentz violating
birefringence can erase linear polarization as a wave propagates,
hence measurements of polarization constrain the relevant
operators.

The logic is as follows. We assume for simplicity the framework of
Sec. \ref{subsubsec:LVQED} and rotation invariance; the
corresponding analysis for the general mSME case can be found
in~\cite{Kostelecky:2001mb}. At the source, assume the emitted
radiation is completely linearly polarized, which will provide the
most conservative constraint. To evolve the wave, we must first
decompose the linear polarization into the propagating circularly
polarized states,
\begin{equation}
A^\al=\ep_L^{\al} e^{-i k_L \cdot x} +\ep_R^{\al} e^{-i k_R \cdot
x}.
\end{equation}
If we choose coordinates such that the wave is travelling in the
z-direction with initial polarization (0,1,0,0) then
$\ep_L=(0,1,-i,0)$ and $\ep_R=(0,1,i,0)$.  Rearranging slightly we
have
\begin{equation} \label{eq:polarizationrot}
A^\al=e^{-i (\om_L t - k_z z)} (\ep_L^{\al}  + \ep_R^{\al} e^{-i
(\om_R- \om_L) t})
\end{equation}
which describes a wave with a rotating polarization vector.  Hence
in the presence of birefringence a linearly polarized wave rotates
its direction of polarization during propagation.  This fact alone
has been used to constrain the $k_{AF}$ term in the mSME to the
level of $10^{-42}$ GeV by analyzing the plane of polarization of
distant galaxies~\cite{Carroll:1989vb}.

A variation on this constraint can be derived by considering
birefringence when the difference $\Delta \om = \om_R-\om_L$ is a
function of $k_z$. A realistic polarization measurement is an
aggregate of the polarization of received photons in a narrow
energy band.  If there is significant power across the entire
band, then a polarized signal must have the polarization direction
at the top nearly the same as the direction at the bottom.  If the
birefringence effect is energy dependent, however, the
polarization vectors across the band rotate differently with
energy.  This causes polarization ``diffusion'' as the photons
propagate.  Given enough time the spread in angle of the
polarization vectors becomes comparable to $2 \pi$ and the initial
linear polarization is lost. Measurement of linear polarization
from distant sources therefore constrains the size of this effect
and hence the Lorentz violating coefficients.  We can easily
estimate the constraint from this effect by looking at when the
polarization at two different energies (representing the top and
bottom of some experimental band) is orthogonal, i.e
$A_{(E_T)}^\al A_{(E_B)_\al}=0$. Using (\ref{eq:polarizationrot})
for the polarization gives
\begin{equation}
|\Delta \om_{(E_T)} - \Delta \om_{(E_B)} |t \approx \pi.
\end{equation}

Three main results have been derived using this approach.
Birefringence has been applied to the mSME
in~\cite{Kostelecky:2001mb,Kostelecky:2002hh}.  Here, the ten
independent components of the two coefficients
$\tilde{\kappa}_{e+}$ and $\tilde{\kappa}_{o-}$ (see section
\ref{subsec:Cavity}) that control birefringence are expressed in
terms of a ten-dimensional vector $k^a$~\cite{Kostelecky:2002hh}.
The actual bound, calculated from the observed polarization of
sixteen astrophysical objects, is $|k^a| \leq 10^{-32}$.
\footnote{This severe limit on birefringence provides an
interesting limitation to allowable spacetime metrics in the
approach of Hehl and others~\cite{Hehl:2004yk}.  In this approach,
linear constitutive relations for electromagnetism are postulated
as fundamental and the metric is derived from the constitutive
relation.  A lack of birefringence implies that the metric must be
Riemannian in this approach~\cite{Lammerzahl:2004ww}.}  A similar
energy band was used to constrain $\xi$ in (\ref{eq:MPops}) to be
$|\xi|<O(10^{-4})$~\cite{Gleiser:2001rm}. Recently, the reported
polarization of GRB021206~\cite{Coburn:2003pi} was used to
constrain $\xi$ to $|\xi|<O(10^{-14})$~\cite{Jacobson:2003bn}, but
since the polarization claim is
uncertain~\cite{Rutledge:2003wa,Wigger:2004by} such a figure
cannot be treated as an actual constraint.

\subsection{Threshold constraints}
\label{subsec:ThreshIntro} %
We now turn our attention from astrophysical tests involving a
single particle species to threshold reactions, which often
involve many particle types. Before delving into the calculational
details of energy thresholds with Lorentz violation, we give a
pedagogical example that shows why particle decay processes (which
involve rates) give constraints that are only functions of
reaction threshold energies. Consider photon decay, $\gamma
\rightarrow e^+ e^-$ (see section \ref{subsubsec:QEDParticles} for
details). In ordinary Lorentz invariant physics the photon is
stable to this decay process. What forbids this reaction is solely
energy/momentum conservation - two timelike four-momenta (the
outgoing pair) cannot add up to the null four momentum of the
photon. If, however, we break Lorentz invariance and assume a
photon obeys a dispersion relation of the form
\begin{equation} \label{eq:kinphotondisp}
\omega^2=k^2 + f^{(3)}_\ga \frac {k^3} {E_{Pl}}
\end{equation}
while electrons/positrons have their usual Lorentz invariant
dispersion, then it is possible to satisfy energy conservation
equation if $f^{(3)}_\ga>0$ (to see this intuitively, note that
the extra term at high energies acts as a large effective mass for
a photon). Therefore a photon can decay to an electron positron
pair.

This type of reaction is called a \textit{threshold reaction} as
it can happen only above some threshold energy $\omega_{th} \sim
(m_e^2 E_{Pl}/f^{(3)}_\ga)^{1/3}$ where $m_e$ is the electron
mass. The threshold energy is translated into a constraint on
$f^{(3)}_\ga$ in the following manner. We see 50 TeV photons from
the Crab nebula~\cite{Tanimori:1997cq}, hence this reaction must
not occur for photons up to this energy as they travel to us from
the Crab. If the decay rate is high enough, one could demand that
$\omega_{th}$ is above 50 TeV, constraining $f^{(3)}_\ga$ and
limiting this type of Lorentz violation. For O(1) $f^{(3)}_\ga$,
$\omega_{th} \sim 10$ TeV, and so we can get a slightly better
than O(1) constraint on $f^{(3)}_\ga$ from 50 TeV
photons~\cite{Jacobson:2001tu}. If, however, the rate is very
small then even though a photon is above threshold it could still
reach us from the Crab. Using the Lorentz invariant expression for
the matrix element $\mathcal{M}$ (i.e. just looking at the
kinematical aspect of Lorentz violation) one finds that as
$\omega$ increases above $\omega_{th}$ the rate very rapidly
becomes proportional to $f^{(3)}_\ga \omega^2/E_{Pl}$. If a 50 TeV
photon is above threshold the decay time is then approximately
$10^{-11}/f^{(3)}_\ga$ seconds.  The travel time of a photon from
the Crab is $\sim 10^{11}$ seconds.  Hence if a photon is at all
above threshold it will decay almost instantly relative to the
observationally required lifetime. Therefore we can neglect the
actual rate and derive constraints simply by requiring that the
threshold itself is above 50 TeV.

It has been argued that technically, threshold constraints can't
truly be applicable to a kinematic model where just modified
dispersion is postulated and the dynamics/matrix elements are not
known. This isn't actually a concern for most threshold
constraints. For example, if we wish to constrain $f^{(3)}_\ga$ at
O(1) by photon decay then we can do so as long as $\mathcal{M}$ is
within 11 orders of magnitude of its Lorentz invariant value
(since the decay rate goes as $|\mathcal{M}|^2$). Hence for rapid
reactions, even an enormous change in the dynamics is irrelevant
for deriving a kinematic constraint. Since kinematic estimates of
reaction rates are usually fairly accurate (for an example see
~\cite{Lehnert:2004hq, Lehnert:2004be}) one can derive constraints
using only kinematic models.  In general, under the assumption
that the dynamics is not drastically different from that of
Lorentz invariant effective field theory, one can effectively
apply particle reaction constraints to kinematic theories since
the decay times are extremely short above threshold.

There are a few exceptions where the rate is important, as the
decay time is closer to the travel time of the observed particle.
Any type of reaction involving a weakly interacting particle such
as a neutrino or graviton will be far more sensitive to changes in
the rate. For these particles, the decay time of observed
particles can be comparable to their travel time. As well, any
process involving scattering, such as the GZK reaction
($p+\gamma_{CMBR} \longrightarrow p+ \pi^0$) or photon
annihilation ($2\gamma \longrightarrow e^+ + e^-$) is more
susceptible to changes in $\mathcal{M}$ as the interaction time is
again closer to the particle travel time.  Even for scattering
reactions, however, $\mathcal{M}$ would need to change
significantly to have any effect. Finally, $\mathcal{M}$ is
important in reactions like ($\gamma \longrightarrow 3\gamma$),
which are not observed in nature but do not have
thresholds~\cite{Jacobson:2002hd,
Kostelecky:2002ue,Adam:2002rg,Adam:2003yd, Gelmini:2005gy}. In
these situations, the small reaction rate is what may prevent the
reaction from happening on the relevant timescales. For all of
these cases, kinematics only models should be applied with extreme
care.  We now turn to the calculation of threshold constraints
assuming EFT.

\subsection{Particle threshold interactions in EFT} \label{subsec:ThresholdsEFT} %
When Lorentz invariance is broken there are a number of changes
that can occur with threshold reactions.  These changes include
shifting existing reaction thresholds in energy, adding additional
thresholds to existing reactions, introducing new reactions
entirely, and changing the kinematic configuration at
threshold~\cite{Coleman:1997xq,Gonzalez-Mestres:1995xe,Jacobson:2002hd,
Lehnert:2003ue}. By demanding that the energy of these thresholds
is inside or outside a certain range (so as to be compatible with
observation) one can derive stringent constraints on Lorentz
violation.

In this section we will describe various threshold phenomena
introduced by Lorentz violation in EFT and the constraints that
result from high energy astrophysics. Thresholds in other models
are discussed in section \ref{subsec:ThresholdsOther}. We will use
rotationally invariant QED as the prime example when analyzing new
threshold behavior. The same methodology can easily be transferred
to other particles and interactions.  A diagram of the necessary
elements for threshold constraints and the appropriate sections of
this review is shown in Figure \ref{fig:flow}

\epubtkImage{}{
\begin{figure}[htbp]
 \def\epsfsize#1#2{0.75#1}
  \centerline{\epsfbox{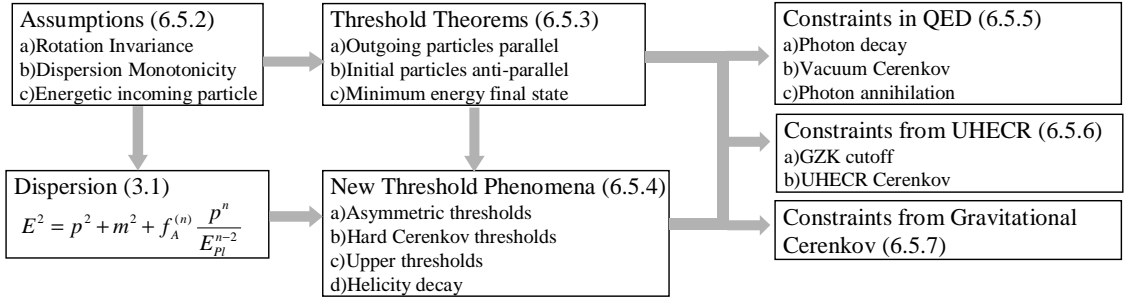}}
  \caption{Elements involved in threshold constraints.}
  \label{fig:flow}
\end{figure}}

Thresholds are determined by energy-momentum conservation. Since
we are working in straight EFT in Minkowski space, translational
invariance implies that the usual conservation laws hold, i.e.
$p^{A}_\al+p^{B}_\al+...=p^{C}_\al+p^{D}_\al+...$, where $p_\al$
is the four momentum of the various particles $A,B,C,D...$.  Since
this just involves particle dispersion, we can neglect the
underlying EFT for the general derivations of thresholds and
threshold theorems.  EFT comes back into the picture when we need
to determine a) the actual dispersion relations that occur in a
physical system to establish constraints and b) matrix elements
for actual reaction rates (c.f. ~\cite{Lehnert:2004be}).

Threshold constraints have been looked at for reactions which have
the same interaction vertices as in Lorentz invariant physics. The
reaction rate is therefore suppressed only by gauge couplings and
phase space. $n>2$ dispersion requires higher mass dimension
operators, and these operators will generically give rise to new
interactions when the derivatives are made gauge covariant.
However, the effective coupling for such interactions is the same
size as the Lorentz violation and hence is presumably very small.
These reactions are therefore suppressed relative to the Lorentz
invariant coupling and can most likely be ignored, although no
detailed study has been done.

\subsubsection{Required particle energy for ``Planck scale'' constraints}

We now give another simple example of constraints from a threshold
reaction to illustrate the required energy scales for constraints
on Planck scale Lorentz violation. The key concept for
understanding how threshold reactions are useful is that, as we
briefly saw for the photon decay reaction in section
\ref{subsec:ThreshIntro}, particle thresholds are determined by
particle mass, which is a small number that can offset the large
Planck energy. To see this in more detail, let us consider the
vacuum \v{C}erenkov effect, $A \rightarrow A + \gamma$, where $A$
is some massive charged particle.  In usual Lorentz invariant
physics, this reaction does not happen due to energy-momentum
conservation. However, consider now a Lorentz violating dispersion
relation for $A$ of the form
\begin{equation}
E^2=p^2+m^2+f_A^{(n)} \frac {p^n} {E_{Pl}^{n-2}}
\end{equation}
with $f_A^{(n)}>0$.  For simplicity in this pedagogical example we
shall not change the photon dispersion relation $\omega=k$.
\v{C}erenkov radiation usually occurs when the speed of the source
particle exceeds the speed of light in a medium.  The same
analysis can be applied in this case, although for more general
Lorentz violation there are other scenarios where \v{C}erenkov
radiation occurs even though the speed condition is not met (see
below)~\cite{Jacobson:2002hd}.  The group velocity of A,
$v=dE/dp$, is equal to one at a momentum
\begin{equation} \label{eq:pthreshceren}
p_{th}=\left[ \frac {m^2 E_{Pl}^{n-2}} {(n-1)f_A^{(n)}}
\right]^{1/n}
\end{equation}
and so we see that the threshold momenta can actually be far below
the Planck energy, as it is controlled by the particle mass as
well.  For example, electrons would be unstable with $n=3$ and
$f_e^{(3)}$ of O(1) at 10 TeV, well below the maximum electron
energies in astrophysical systems.  We can rewrite
(\ref{eq:pthreshceren}) for $f_A^{(n)}$ and see that the expected
constraint from the stability of $A$ at some momentum $p_{obs}$ is
\begin{equation}
f_A^{(n)} \leq \frac {m^2 E_{Pl}^{n-2}} {(n-1) p_{obs}^n}.
\end{equation}
Therefore constraints can be much less than order one with
particle energies much less than $E_{Pl}$. The orders of magnitude
of constraints on $f_A^{(n)}$ estimated from the threshold
equation alone (i.e. we have neglected the possibility that the
matrix elements are small) for various particles are given in
Table \ref{tab:ceren}~\cite{Jacobson:2002hd}.

\begin{table} [htb]
  \centering
  \caption{Orders of magnitude of vacuum \v{C}erenkov constraint for various particles} \label{tab:ceren}

\begin{tabular}{|c|c|c|c|}
  \hline
  \raisebox{0pt}[13pt][7pt]{Particle} & $\nu$ & $e^-$ & $p^+$ \\
  \hline \hline
  \raisebox{0pt}[13pt][7pt]{mass} & $\leq 1$ eV & $0.511$ MeV & $938$ GeV \\
  \hline
  \raisebox{0pt}[13pt][7pt]{$p_{obs}$} & 100 TeV & 50 TeV & $10^{20}$ eV \\
  \hline
  \raisebox{0pt}[13pt][2pt]{$n=2$} & $10^{-28}$ & $10^{-15}$ & $10^{-22}$ \\
  \raisebox{0pt}[13pt][2pt]{$n=3$} & $10^{-14}$ & $10^{-2}$ & $10^{-14}$ \\
  \raisebox{0pt}[13pt][2pt]{$n=4$} & $1$ & $10^{12}$ & $10^{-6}$ \\
  \hline
\end{tabular}
\end{table}

For neutrinos, $p_{obs}$ comes from AMANDA
data~\cite{Geenen:2003gg}.  The $p_{obs}$ for electrons comes from
the expected energy of the electrons responsible for the creation
of $\sim 50$ TeV gamma rays via inverse Compton
scattering~\cite{Koyama:1995rr,Tanimori:1997cq} in the Crab
nebula.  For protons, the $p_{obs}$ is from AGASA
data~\cite{Takeda:1998ps}.

We include the neutrino, even though it is neutral, since
neutrinos still have a non-vanishing interaction amplitude with
photons.  We shall talk more about neutrinos in section
\ref{subsec:Neutrinos}.  The neutrino energies in this table are
those currently observed; if future neutrino observatories see PeV
neutrinos (as expected) then the constraints will increase
dramatically.

This example is overly simplified, as we have ignored Lorentz
violation for the photon.  However, the main point remains valid
with more complicated forms of Lorentz violation:  constraints can
be derived with current data that are much less than O(1) even for
$n=4$ Lorentz violation.  We now turn to a discussion of the
necessary steps for deriving threshold constraints, as well as the
constraints themselves for more general models.

\subsubsection{Assumptions} \label{subsubsec:Assumptions}
One must make a number of assumptions before one can analyze
Lorentz violating thresholds in a rigorous manner.\\
\\
\textbf{Rotation Invariance.}  Almost all work on thresholds to
date has made the assumption that rotational invariance holds.  If
this invariance is broken, then our threshold theorems and results
do not necessarily hold.  For threshold discussions, we will
assume that the underlying EFT is rotationally invariant and use
the notation
$p=|\overrightarrow{p}|$.\\
\\
\textbf{Monotonicity.}  We will assume that the dispersion
relations for all particles is monotonically increasing.  This is
the case for the mSME with small Lorentz violating coefficients if
we work in a concordant frame.  Mass dimension $>4$ operators
generate dispersion relations of the form
\begin{equation}
E^2=m^2+p^2+f^{(n)} \frac {p^n} {E_{Pl}^{n-2}}
\end{equation}
which do not satisfy this condition at momentum near the Planck
scale if $f^{(n)}<0$.  The turnover momentum $p_{TO}$ where the
dispersion relation is no longer monotonically increasing is
$p_{TO}=(-2/(nf^{(n)}))^{1/(n-2)} E_{Pl}$.  The highest energy
particles known to propagate ar the trans-GZK cosmic rays with
energy $10^{-8} E_{Pl}$.  Hence unless $f^{(n)} \gg 1$, $p_{TO}$
is much higher than any relevant observational energy and we can
make
the assumption of monotonicity without loss of generality.\\
\\
\textbf{High energy incoming particle.}  If there is a
multi-particle in state, we will assume that one of the particles
is much more energetic than all the others.  This is the
observational situation in reactions such as photon-photon
scattering or pion production by cosmic rays scattering off the
cosmic microwave background (the GZK reaction, see section
\ref{subsubsec:GZK}).

\subsubsection{Threshold theorems} \label{subsubsec:ThresholdThms}
Eventually, any threshold analysis must solve for the threshold
energy of a particular reaction.  To do this, we must first know
the appropriate kinematic configuration that applies at a
threshold.  Of use will be a set of threshold theorems that hold
in the presence of Lorentz violation, which we state below.
Variations on these theorems were derived in~\cite{Coleman:1998ti}
for single particle decays with n=2 type dispersion
and~\cite{Mattingly:2002ba} for two in-two out particle
interactions with general dispersion. Here we state the more general versions.\\
\\
\textbf{Theorem 1:} The configuration at a threshold for a
particle
with momentum $p_1$ is the minimum energy configuration of all other particles that conserves momentum.\\
\\
\textbf{Theorem 2:} At a threshold all outgoing momenta are
parallel to $p_1$ and all other incoming momentum are anti-parallel.\\

\subsubsection{New threshold phenomena}\label{subsubsec:newthreshphenomena}%
\textbf{Asymmetric thresholds}\\
\\
Asymmetric thresholds are thresholds where two outgoing particles
with equal masses have unequal momenta.  This cannot occur in
Lorentz invariant reactions. Asymmetric thresholds occur because
the minimum energy configuration is not necessarily the symmetric
configuration.  To see this, let us analyze photon decay, where we
have one incoming photon with momentum $p_{in}$ and an
electron/positron pair with momenta $q_1,q_2$.  We will assume our
Lorentz violating coefficients are such that the electron and
positron have identical dispersion. \footnote{This can easily be
done by considering a CPT invariant EFT or choosing the
appropriate helicity states/coefficients for a CPT violating EFT.}

Imagine that the dispersion coefficients  $f^{(n)}$ for the
electron and positron are negative and such that the
electron/positron dispersion is given by the solid curve in Figure
~\ref{fig:asymm}. We define the energy $E_{symm}$ to be the energy
when both particles have the same momentum $q_1=q_2=p_{in}/2$.
This is not the minimum energy configuration, however, if the
curvature of the dispersion relation ($\partial^2 E/\partial p^2$)
at $p_{in}/2$ is negative. If we add a momentum $\Delta q$ to
$q_2$ and $-\Delta q$ to $q_1$ then we change the total energy by
$\Delta E=\Delta E_2- \Delta E_1$. Since the curvature is
negative,$\Delta E_1 > \Delta E_2$ and therefore $\Delta E<0$. The
symmetric configuration is not the minimum energy configuration
and is not the appropriate configuration to use for a threshold
analysis for all $p_{in}$.
\epubtkImage{}{
\begin{figure}[htbp]
 \def\epsfsize#1#2{0.75#1}
  \centerline{\epsfbox{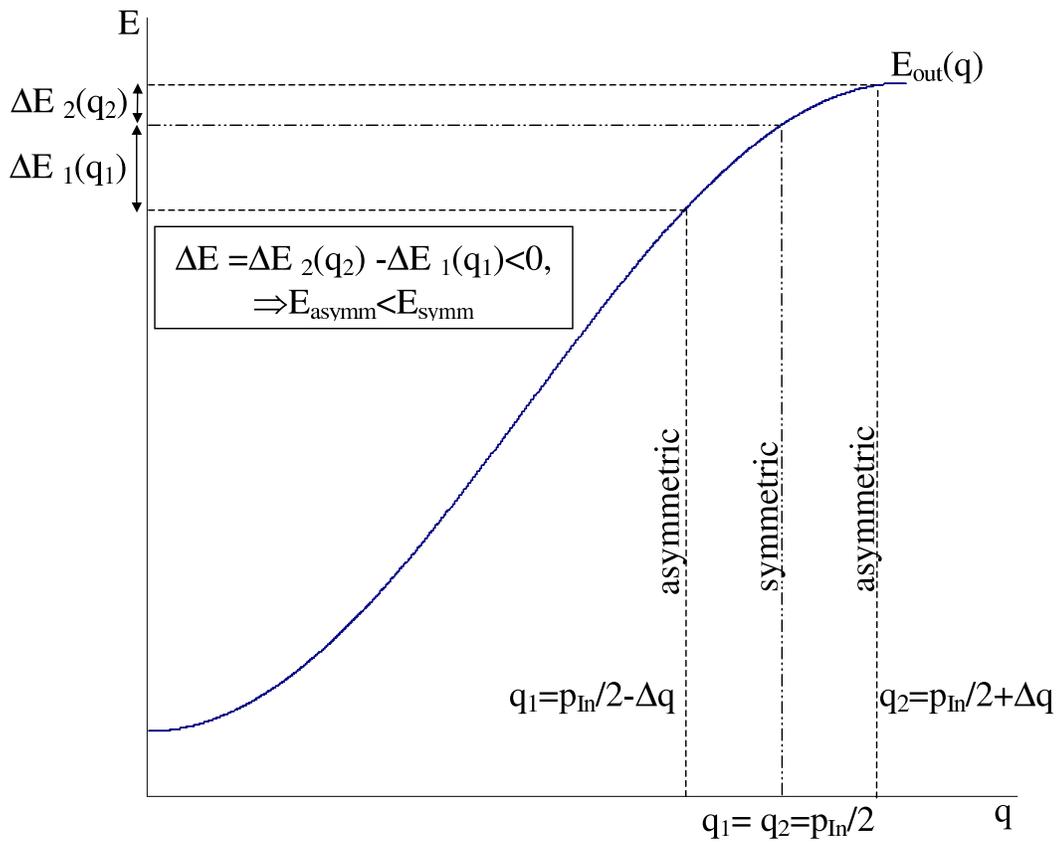}}
  \caption{Total outgoing particle energy in symmetric and
  asymmetric configurations.}
  \label{fig:asymm}
\end{figure}}
Note that part of the dispersion curve in Figure \ref{fig:asymm}
has positive curvature, as must be the case if at low energies we
have the usual Lorentz invariant massive particle dispersion.  If
we were considering the constraints derivable when $p_{in}/2$ is
small and in the positive curvature region, then the symmetric
configuration \textit{would} be the applicable one.  In general
when it is appropriate to use asymmetric thresholds or symmetric
ones depends heavily on the algebraic form of the outgoing
particle Lorentz violation and the energy that the threshold must
be above.  The only general statement that can be made is that
asymmetric thresholds are not relevant when the outgoing particles
have $n=2$ type dispersion modifications (either positive or
negative) or for strictly positive coefficients at any $n$. For
further examples of the intricacies of asymmetric thresholds,
see~\cite{Jacobson:2002hd,Konopka:2002tt}.\\
\\
\textbf{Hard \v{C}erenkov thresholds}\\
\\
Related to the existence of asymmetric thresholds is the hard
\v{C}erenkov threshold, which also occurs only when $n>2$ with
negative coefficients.  However, in this case both the outgoing
and incoming particles must have negative coefficients.  To
illustrate the hard \v{C}erenkov threshold, we consider photon
emission from a high energy electron, which is the rotated diagram
of the photon decay reaction.  In Lorentz invariant physics,
electrons emit soft \v{C}erenkov radiation when their group
velocity $\partial E/\partial p$ exceeds the phase velocity
$\om/k$ of the electromagnetic vacuum modes in a medium. This type
of \v{C}erenkov emission also occurs in Lorentz violating physics
when the group velocity of the electrons exceeds the low energy
speed of light in vacuum. The velocity condition does not apply to
hard \v{C}erenkov emission, however, so to understand the
difference we need to describe both types in terms of
energy-momentum conservation.

Let us quickly remind ourselves where the velocity condition comes
from.  The energy conservation equation (imposing momentum
conservation) can be written as
\begin{equation}
\om(k)=E(p)-E(p-k).
\end{equation}
Dividing both sides by $k$ and taking the soft photon limit
$k\rightarrow 0$ we have
\begin{equation} \label{eq:softcerenk}
 \lim_{k \rightarrow 0} \frac {\om(k)} {k} =\lim_{k \rightarrow
0} \frac {E(p)-E(p-k)} {k}=\frac {\partial E} {\partial p}.
\end{equation}
Equation (\ref{eq:softcerenk}) makes clear that the velocity
condition is only applicable for soft photon emission.   Hard
photon emission can occur even when the velocity condition is
never satisfied if the photon energy-momentum vector is spacelike
with $n>2$ dispersion. As an example, consider an unmodified
electron and a photon dispersion of the form
$\om^2=k^2-k^3/E_{Pl}$.  The energy conservation equation in the
threshold configuration is
\begin{equation}
p+\frac{m^2} {2p}=p-k + \frac {m^2} {2(p-k)} + k -\frac {k^2}
{2E_{Pl}}
\end{equation}
where $p$ is the incoming electron momentum.  Introducing the
variable $x=k/p$ and rearranging we have
\begin{equation}
\frac{m^2 E_{Pl}} {p^3} =x(1-x).
\end{equation}
Since all particles are parallel at threshold, $x$ must be between
0 and 1.  The maximum value of the right hand side is $1/4$, and
so we see that we can solve the conservation equation if $p>(4m^2
E_{Pl})^{1/3}$, which is approximately 23 TeV.  At threshold,
$x=1/2$ so this corresponds to emission of a hard photon with an
energy of 11.5 TeV.\\
\\
\textbf{Upper thresholds}\\
\\
Upper thresholds do not occur in Lorentz invariant physics. It is
easy to see that they are possible with Lorentz violation,
however.  In figure \ref{fig:threshupper} the region \textbf{R} in
energy space spanned by $E_{Out}(X_k,p_1)$ is bounded below, since
each individual dispersion relation is bounded below. However, if
one can adjust the dispersion $E_1(p_1)$ freely, as would be the
case if the incoming particle was a unique species in the
reaction, then one can choose Lorentz violating coefficients such
that $E_1(p_1)$ moves in and out of \textbf{R}.
\epubtkImage{}{
\begin{figure}[htbp]
  \def\epsfsize#1#2{#1}
  \centerline{\epsfbox{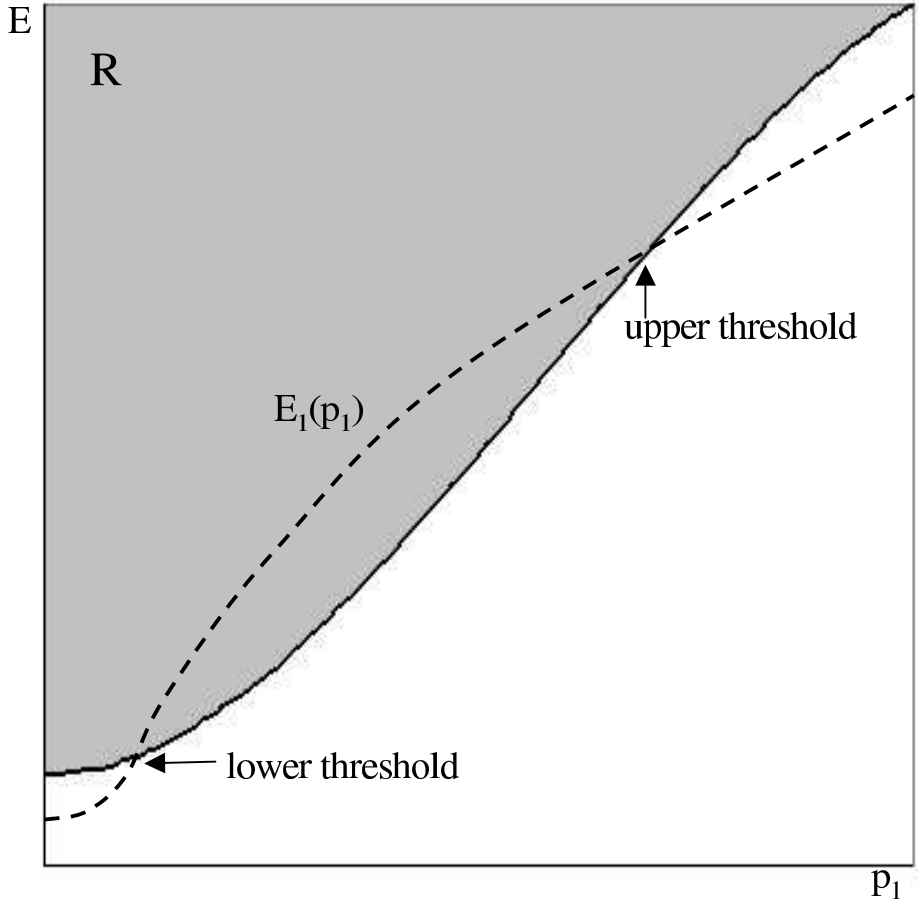}}
  \caption{An example of an upper and lower threshold. \textbf{R} is the region spanned
  by all $X_k$ and $E_1(p_1)$ is the energy of the incoming
  particle.  Where $E_1(p_1)$ enters and leaves \textbf{R} are
  lower and upper thresholds, respectively.}
 \label{fig:threshupper}
\end{figure}}

As a concrete example consider photon decay, $\gamma
\longrightarrow e^+ + e^-$, with unmodified photon dispersion and
an electron/positron dispersion relation of
\begin{equation}
E^2=p^2+m^2- \frac {7m^{2/3}} {2E_{Pl}^{2/3}} p^2 + \frac {p^3}
{E_{Pl}}
\end{equation}
chosen strictly for algebraic convenience. This dispersion
relation has positive curvature everywhere implying that the the
electron and positron have equal momenta at threshold.  The energy
conservation equation, where the photon has momentum $k$ is then
\begin{equation}
k=2(\frac {k} {2} + \frac {m^2} {k} - \frac {7m^{2/3}}
{8E_{Pl}^{2/3}} k + \frac {k^3} {8 E_{Pl}})
\end{equation}
which reduces to
\begin{equation} \label{eq:upperthresh1}
8m^2-\frac {7m^{2/3}} {E_{Pl}^{2/3}}k^2 + \frac {k^3} {E_{Pl}}=0.
\end{equation}
(\ref{eq:upperthresh1}) has two positive real roots, at $k=(4\pm
2\sqrt{2}) m^{2/3}E_{Pl}^{1/3}$, corresponding to a lower and
upper threshold at 14 and 82 TeV, respectively.  Such a threshold
structure would produce a deficit in the observed photon spectrum
in this energy band. \footnote{Currently, the observed photon
spectrum only extends to 50 TeV.  Hence this type of lower/upper
threshold structure is observationally indistinguishable from a
simple lower threshold with current data.}  Very little currently
exists in the literature on the observational possibilities of
upper thresholds.  A complicated lower/upper threshold structure
has been applied to the trans-GZK cosmic ray events, with the
lower threshold mimicking the GZK-cutoff at $5 \cdot 10^{19}$ GeV
and the upper entering below the highest events at $3 \cdot
10^{20}$ GeV~\cite{Jacobson:2002hd}.  The region of parameter
space where such a scenario might happen is extremely small, however.\\
\\
\textbf{Helicity decay}\\
\\
In previous work on the \v{C}erenkov effect based on EFT it has
been assumed that left and right handed fermions have the same
dispersion.  As we have seen, however, this need not be the case.
When the fermion dispersion is helicity dependent the phenomenon
of helicity decay occurs.  One of the helicities is unstable and
will decay into the other as a particle propagates, emitting some
sort of radiation depending on the exact process considered.
Helicity decay has no threshold in the traditional sense; the
reaction happens at all energies.  However, below a certain energy
the phase space is highly suppressed, so we have an effective
threshold that practically speaking is indistinguishable from a
real threshold. As an example, consider the reaction
$e_L\longrightarrow e_R + \gamma$, with an unmodified photon
dispersion and the electron dispersion relation
\begin{eqnarray}
E_R^2=m^2+p^2\\
E_L^2=m^2+p^2+f^{(4)}_{eL} p^2
\end{eqnarray}
for right and left-handed electrons.  Furthermore, assume that
$f^{(4)}_{eL}>0$. The opening up of the phase space can be seen by
looking at the minimum and maximum values of the longitudinal
photon momentum.  The energy conservation equation is
\begin{equation} \label{eq:helicitydecaycons}
p+\frac {m^2} {2p} + \frac {f^{(4)}_{eL}} {2} p = p-k + \frac
{m^2} {2(p-k)} + |k|
\end{equation}
where $p$ is the incoming momentum and $k$ is the outgoing photon
momentum.  We have assumed that the transverse momentum is zero,
which gives us the minimum and maximum values of $k$.  $k$ is
assumed to be less than $p$; one can check a posteriori that this
assumption is valid. It can be negative, however, which is
different from a threshold calculation where all momenta are
necessarily parallel.  Solving (\ref{eq:helicitydecaycons}) for
$k_{min},k_{max}$ yields to lowest order in $m,f^{(4)}_{eL}$
\begin{eqnarray} \label{eq:helicitydecayk}
k_{min}=-\frac {f_{eL}^{(4)} p} {4}\\
\nonumber k_{max}=p \frac {f_{eL}^{(4)} p^2} {m^2+f_{eL}^{(4)}
p^2}.
\end{eqnarray}
From (\ref{eq:helicitydecayk}) it is clear that when
$p^2<<m^2/f^{(4)}_{eL}$ the phase space is highly suppressed,
while for $p^2<<m^2/f^{(4)}_{eL}$ the phase space in $k$ becomes
of order $p$. The momentum $p_{th}=(m^2/f^{(4)}_{eL})^{1/2}$ acts
as an effective threshold, where the reaction is strongly
suppressed below this energy. Constraints from helicity decay in
the current literature~\cite{Jacobson:2005bg} are complicated and
not particularly useful.  Hence we shall not describe them here,
instead focussing our attention on the strict \v{C}erenkov effect
when the incoming and outgoing particle have the same helicity.
For an in depth discussion of helicity decay constraints
see~\cite{Jacobson:2005bg}.

\subsubsection{Threshold constraints in QED} \label{subsubsec:QEDParticles}%
With the general phenomenology of thresholds in hand, we now turn
to the actual observational constraints from threshold reactions
in Lorentz violating QED.  We will continue to work in a
rotationally invariant setting.  Only the briefest listing of the
constraints is provided here; for a more detailed analysis
see~\cite{Jacobson:2002hd, Jacobson:2003bn, Jacobson:2005bg}. Most
constraints in the literature have been placed by demanding that
the threshold for an unwanted reaction is above some observed
particle energy. As mentioned previously, a necessary step in this
analysis is to show that the travel times of the observed
particles are much longer than the reaction time above threshold.
A calculation of this for the vacuum \v{C}erenkov has been done
for QED with dimension four Lorentz violating operators
in~\cite{Moore:2001bv}. More generally, a simple calculation shows
that the energy loss rate above threshold from the vacuum
\v{C}erenkov effect rapidly begins to scale as $e^2 A E^n /
E_{Pl}^{n-2}$ where $A$ is a coefficient that depends on the
coefficients of the Lorentz violating terms in the EFT. Similarly,
the photon decay rate is $e^2 A E^{n-1} / E_{Pl}^{n-2}$.  In both
cases the reaction times for high energy particles are roughly
$(e^2A)^{-1} E_{Pl}^{n-2}/E^{n-1}$, which is far shorter than the
required lifetimes for electrons and photons in astrophysical
systems for $n=2,3$.\footnote{For $n=4$ no QED particles reach
energies high enough to provide constraints. The only particles of
the required energy are ultra-high energy cosmic rays or
neutrinos. Assuming the cosmic rays are protons, the corresponding
reaction time for \v{C}erenkov emission is $10^{-17}$ sec.} The
lifetime of a high energy particle in QED above threshold is
therefore short enough that we can establish constraints simply by
looking at threshold conditions.

\paragraph{Photon decay}
Lorentz violating terms can be chosen such that photons become
unstable to decay into electron-positron
pairs~\cite{Jacobson:2001tu}.  We observe 50 TeV photons from the
Crab nebula. There must exist then at least one stable photon
polarization. The thresholds for $n=2,3$ dispersion have been
calculated in~\cite{Jacobson:2002hd}. Demanding that these
thresholds are above 50 TeV yields the following best constraints.

For $n=2$ with CPT preserved we have $f^{(2)}_\ga - f^{(2)}_e \leq
4 m^2/p_{th}^2=4 \cdot 10^{-16}$~\cite{Jacobson:2002hd}.  If we
set $d=0$ in (\ref{eq:SMErotinvdisp}) so that there is no helicity
dependence, this translates to the constraint $k_F/2+c \leq 4
\cdot 10^{-16}$. If $d\neq 0$ then both helicities of
electrons/positrons must satisfy this bound since the photon has a
decay channel into every possible combination of electron/positron
helicity. The corresponding limit is $k_F/2+(c\pm d) \leq 4 \cdot
10^{-16}$.

For $n=3$ the situation is a little more complicated, as we must
deal with photon and electron helicity dependence, positron
dispersion, and the possibility of asymmetric thresholds. The 50
TeV Crab photon polarizations are unknown, so only the region of
parameter space in which both polarizations decay can be excluded.
We can simplify the problem dramatically by noting that the
birefringence constraint on $\xi$ in (\ref{eq:MPops}) is $|\xi|
\leq 10^{-4}$~\cite{Gleiser:2001rm}. The level of constraints from
threshold reactions at 50 TeV is around
$10^{-2}$~\cite{Jacobson:2001tu, Konopka:2002tt}. Since the
birefringence constraint is so much stronger than threshold
constraints, we can effectively set $\xi=0$ and derive the photon
decay constraint in the region allowed by birefringence.

With this assumption, we can derive a strong constraint on both
$\eta_R$ and $\eta_L$ by considering the individual decay channels
$\ga \rightarrow e^-_R + e^+_L$ and $\ga \rightarrow e^-_L+e^+_R$,
where $L,R$ stand for the helicity.  For brevity, we shall
concentrate on $\ga \rightarrow e^-_R + e^+_L$, the other choice
is similar.  The choice of a right-handed electron and left-handed
positron implies that both particle's dispersion relations are
functions of only $f^{(3)}_{eR}$ and hence $\et_R$ (see Sec.
\ref{subsubsec:LVQED}).  The matrix element can be shown to be
large enough for this combination of helicities that constraints
can be derived by simply looking at the threshold. Imposing the
threshold configuration and momentum conservation and substituting
in the appropriate dispersion relations the energy conservation
equation becomes
\begin{equation}
k=p+\frac{m^2} {2p} + \frac{f^{(3)}_{eR}} {2 E_{Pl}} p^2 + k-p +
\frac {m^2} {2(k-p)}-\frac{f^{(3)}_{eR}} {2 E_{Pl}} (k-p)^2
\end{equation}
where $k$ is the incoming photon momentum and $p$ is the outgoing
electron momentum. Cancelling the lowest order terms and
introducing the variable $z=2 p/k -1$ this can be rewritten as
\begin{equation} \label{eq:photdecayk}
k^3=\frac {4 m^2 E_{Pl}} {f^{(3)}_{eR} (-z+z^3)}.
\end{equation}
To find the minimum energy configuration we must minimize the
right hand side of (\ref{eq:photdecayk}) with respect to $z$
(keeping the right hand side positive). We note that since the
range of $z$ is between -1 and 1, the right hand side of
(\ref{eq:photdecayk}) can be positive for both positive and
negative $f^{(3)}_{eR}$, which implies that the bound will be two
sided.

As an aside, it may seem odd that photon decay happens at all when
the outgoing particles have opposite dispersion modifications
since the net effect on the total outgoing energy might seem to
cancel. However, this is only the case if both particles have the
same momenta.  We can always choose to place more of the incoming
momentum into the outgoing particle with a negative coefficient,
thereby allowing the process to occur. This reasoning also
explains why the bound is two sided, as the threshold
configuration gives more momentum to whichever particle has a
negative coefficient.

Returning to the calculation of the threshold, minimizing
(\ref{eq:photdecayk}),  we find that the threshold momentum is
\begin{equation}
k_{thresh}=\left|\frac {m^2 E_{Pl}} {f^{(3)}_{eR}} 6
\sqrt{3}\right|^{1/3}.
\end{equation}
The absolute value here appears because we find the minimum
positive value of (\ref{eq:photdecayk}).  Placing $k_{thresh}$ at
50 TeV yields the constraint $|f^{(3)}_{eR}|<0.25$ and hence
$|\et_R|<0.125$.  The same procedure applies in the opposite
choice of outgoing particle helicity, so $\et_L$ obeys this bound
as well.

\paragraph{Vacuum \v{C}erenkov}
The 50 TeV photons observed from the Crab nebula are believed to
be produced via inverse Compton (IC) scattering of charged
particles off the ambient soft photon background.\footnote{The
theoretical model for the non-thermal emission of the Crab nebula
is called the synchrotron-self-Compton (SSC) model. For recent
fits of data to the SSC spectrum see~\cite{Aharonian:2004gb}.} If
one further assumes that the charged particles are electrons, it
can then be inferred that 50 TeV electrons must propagate.
However, only one of the electron helicities may be propagating,
so we can only constrain one of the helicities.

For $n=2$ the constraint is $f^{(2)}_e - f^{(2)}_\ga \leq
m^2/p_{th}^2= 10^{-16}$~\cite{Jacobson:2002hd}, where $f^{(2)}_e$
is the coefficient for one of the electron helicities.  In terms
of the mSME parameters this condition can be written as
$c-|d|-k_F/2\leq 10^{-16}$. For $n=3$ an added complication
arises.  If we consider just electrons as the source of the 50 TeV
photons, then we have that either $f^{(3)}_{eL}$ or $f^{(3)}_{eR}$
must satisfy
\begin{eqnarray}
{\rm a)} \quad f^{(3)}_e&<&\displaystyle{ 0.012 } \qquad \qquad
\quad \:\:\: \mbox{for $f^{(3)}_e>0$ and
$f^{(3)}_\ga\geq-3f^{(3)}_e$}, \label{eq:cerconda}
\\ \nonumber
{\rm b)} \quad
f^{(3)}_\ga&>&\displaystyle{f^{(3)}_e-0.048-2\sqrt{\left(0.024\right)^2-0.048
f^{(3)}_e}} \quad \;\mbox{for $f^{(3)}_\ga<-3f^{(3)}_e<0$ or
$f^{(3)}_\ga<f^{(3)}_e \leq 0$} \label{eq:cercondb}
\end{eqnarray}
and the translation to $\xi$ and $\eta_{R,L}$ is as before. Note
that for the range of $\xi$ allowed by birefringence, the relevant
constraint is $\eta_R<0.012$ \textit{or} $\eta_L<0.012$.

A major difficulty with the above constraint is that positrons may
also be producing some of the 50 TeV photons from the Crab nebula.
Since positrons have opposite dispersion coefficients in the $n=3$
case, there is always a charged particle able to satisfy the
\v{C}erenkov constraint. Hence by itself, this IC \v{C}erenkov
constraint can always be satisfied in the Crab and gives no limits
at all. However, as we shall see in section
\ref{subsec:Synchrotron} the vacuum \v{C}erenkov constraint can be
combined with the synchrotron constraint to give an actual
two-sided bound.

\paragraph{Photon annihilation}
The high energy photon spectrum (above 10 TeV) from astrophysical
sources such as Markarian 501 and 421 has been observed to show
signs of absorption due to scattering off the IR background. While
this process occurs in Lorentz invariant physics, the amount of
absorption is affected by Lorentz violation.  The resulting
constraint is not nearly as clear cut as in the photon decay and
\v{C}erenkov cases, as the spectrum of the background IR photons
and the source spectrum are both important, neither of which is
entirely known. Various authors have argued for different
constraints on the $n=3$ dispersion relation, based upon how far
the threshold can move in the IR background.  The constraints vary
from O(1) to O(10). However, none of the analyses take into
account the EFT requirement for $n=3$ that opposite photon
polarization have opposite Lorentz violating terms.  Such an
effect would cause one polarization to be absorbed more strongly
than in the Lorentz invariant case and the other polarization to
be absorbed less strongly. The net result of such a situation is
currently unknown, although current data from blazars suggest that
both polarizations must be absorbed to some
degree~\cite{Stecker:2003wm}. Since even at best the constraint is
not competitive with other constraints, and since there is so much
uncertainty about the situation, we will not treat this constraint
in any more detail. For discussions
see~\cite{Jacobson:2002hd,Amelino-Camelia:2002dx}.

\subsubsection{The GZK Cutoff and Ultra-high energy cosmic rays} \label{subsubsec:GZK}
\paragraph{The GZK Cutoff}
Ultra-high energy cosmic rays (UHECR), if they are protons, will
interact strongly with the cosmic microwave background and produce
pions, $p+\ga \longrightarrow p + \pi^0$, losing energy in the
process. As the energy of a proton increases, the GZK reaction can
happen with lower and lower energy CMBR photons.  At very high
energies ($5 \cdot 10^{19}$ eV), the interaction length (a
function of the power spectrum of interacting background photons
coupled with the reaction cross section) becomes of order 50 Mpc.
Since cosmic ray sources are probably at further distances than
this, the spectrum of high energy protons should show a cutoff
around $5 \cdot 10^{19}$ eV~\cite{Greisen:1966jv,
Zatsepin:1966jv}. A number of experiments have looked for the GZK
cutoff, with conflicting results.  AGASA found trans-GZK events
inconsistent with the GZK cutoff at $2.5
\sigma$~\cite{DeMarco:2003ig} while Hi-Res has found evidence for
the GZK cutoff (although at a lower confidence level, for a
discussion see~\cite{Stecker:2003wm}). New experiments such as
AUGER~\cite{AUGER} may resolve this issue in the next few years.
Since Lorentz violation shifts the location of the GZK cutoff,
significant information about Lorentz violation (even for $n=4$
type dispersion) can be gleaned from the UHECR spectrum.  If the
cutoff is seen then Lorentz violation will be severely
constrained, while if no cutoff or a shifted cutoff is seen then
this might be a positive signal.

For the purposes of this review, we will \textit{assume} that the
GZK cutoff has been observed and describe the constraints that
follow. We can estimate their size by noting that in the Lorentz
invariant case the conservation equation can be written as
\begin{equation}
(p+k)^2=(m_\pi+m_p)^2
\end{equation}
as the outgoing particles are at rest at threshold.  Here $p$ is
the UHECR proton 4-momentum and $k$ is the soft photon 4=momentum.
At threshold the incoming particles are anti-parallel, which gives
a threshold energy for the GZK reaction of
\begin{equation}
E_{GZK}\simeq 3 \cdot 10^{20} GeV \cdot \bigg{(} \frac {\om_0}
{2.7K} \bigg{)}
\end{equation}
where $\om_0$ is the energy of the CMBR photon.  The actual GZK
cutoff occurs at $5 \cdot 10^{19}$ eV due to the tail of the CMBR
spectrum and the particular shape of the cross section (the
$\Delta$ resonance).  From this heuristic threshold analysis,
however, it is clear that Lorentz violation can become important
when the modification to the dispersion relation is of the same
order of magnitude as the proton mass.  For $n=2$ dispersion, a
constraint of $f^{(2)}_\pi - f^{(2)}_p < O(10^{-23})$ was derived
in ~\cite{Coleman:1998ti, Coleman:1998en,Alfaro:2005rs}.  The case
of $n=3$ dispersion with $f^{(3)}_\pi = f^{(3)}_p$ was studied
in~\cite{Gonzalez-Mestres:1995xe,Gonzalez-Mestres:2000rn,Gonzalez-Mestres:1996zv,Bertolami:2000qa,
Bertolami:1999dc,
Bertolami:1999da,Amelino-Camelia:2000zs,Amelino-Camelia:2001qf,Amelino-Camelia:2000ev,Kifune:1999ex,Aloisio:2000cm,Konopka:2002tt,
Stecker:2004xm, Alfaro:2002ya}, while the possibility of
$f^{(3,4)}_\pi \neq f^{(3,4)}_p$ was studied
in~\cite{Jacobson:2002hd}. A simple
constraint~\cite{Jacobson:2002hd} can be summarized as follows. If
we demand that the GZK cutoff is between $2 \cdot 10^{19}$ eV and
$7 \cdot 10^{19}$ eV then for $f^{(3)}_\pi = f^{(3)}_p$ we have
$|f^{(3)}_p|<O(10^{-14})$.  If $f^{(3)}_\pi \neq f^{(3)}_p$ then
there is a wedge shaped region in the parameter space that is
allowed~\cite{Jacobson:2002hd}.

The numerical values of these constraints should not be taken too
literally. While the order of magnitude is correct, simply moving
the value of the threshold for the proton that interacts with a
CMBR photon at some energy does not give accurate numbers. GZK
protons can interact with any photon in the CMBR distribution
above a certain energy. Modifying the threshold modifies the phase
space for a reaction with all these photons in the region to
varying degrees, which must be folded in to the overall reaction
rate. Before truly accurate constraints can be calculated from the
GZK cutoff a more detailed analysis to recompute the rate in a
Lorentz violating EFT considering the particulars of the
background photon distribution and $\Delta$-resonance must be
done.  However, the order of magnitude of the constraints above is
roughly correct. Since they are so strong, the actual numeric
coefficient is not particularly important.\footnote{This would
change, of course, if a shift in the cutoff was recorded, as then
the detailed dependence of the cutoff location on the Lorentz
violating physics would be important.}

Another difficulty with constraints using the GZK cutoff is the
assumption that the source spectrum follows the same power law
distribution as at lower energies.  It may seem that proposing a
deviation from the power law source spectrum at that energy would
be a conspiracy and considered unlikely.  However, this is not
quite correct. A constraint on $f^{(n)}$ will, by the arguments
above, be such that the Lorentz violating terms are important only
near the GZK energy - below this energy we have the usual Lorentz
invariant physics. However, such new terms could then also
strongly affect the source spectrum only near the GZK energy.
Hence the GZK cutoff could vanish or be shifted due to source
effects as well. Unfortunately, we have little idea as to the
mechanism that generates the highest energy cosmic rays, so we
cannot say how Lorentz violation might affect their generation. In
summary, while constraints from the position of the GZK cutoff are
impressive and useful, their actual values should be taken with a
grain of salt since a number of unaccounted for effects may be
tangled up in the GZK cutoff.

\paragraph{UHECR \v{C}erenkov}
A complimentary constraint to the GZK analysis can be derived by
recognizing that $10^{19}-10^{20}$ eV protons reach us - a vacuum
\v{C}erenkov effect must be forbidden up to the highest observed
UHECR energy~\cite{Coleman:1998ti,Jacobson:2002hd,Gagnon:2004xh}.
The direct limits from photon emission, treating a $5 \cdot
10^{19}$ eV proton as a single constituent are $f^{(2)}_p -
f^{(2)}_\gamma<4 \cdot 10^{-22}$~\cite{Jacobson:2002hd,
Coleman:1997xq,Gagnon:2004xh} for $n=2$,\footnote{The bounds
quoted in~\cite{Coleman:1997xq,Gagnon:2004xh} vary from this bound
by roughly an order of magnitude, as the proton energy is taken to
be $10^{20}$ eV rather than right at the GZK cutoff.} $f^{(3)}_p -
f^{(3)}_\gamma<O(10^{-14})$ for $n=3$~\cite{Jacobson:2002hd}, and
$f^{(4)}_p - f^{(4)}_\gamma<O(10^{-5})$ for
$n=4$~\cite{Jacobson:2002hd}.  Equivalent bounds on Lorentz
violation in a conjectured low energy limit of loop quantum
gravity have also been derived using UHECR
\v{C}erenkov~\cite{Lambiase:2003bq}.

\v{C}erenkov emission for UHECR has been used most extensively
in~\cite{Gagnon:2004xh}, where two-sided limits on Lorentz
violating dimension 4,5, and 6 operators for a number of particles
are derived.  The argument is as follows. If we view a UHECR
proton as actually a collection of constituent partons (i.e.
quarks, gauge fields, etc.) then the dispersion correction should
be a function of the corrections for the component partons. By
evaluating the parton distribution function~\footnote{This
approach requires that one can evaluate the parton distribution
functions up to UHECR energies without errors from the uncertainty
in low energy parton distribution functions or effects from new
physics becoming appreciable.  We caution that such effects can
adjust the constraints, as the authors of~\cite{Gagnon:2004xh}
note.} for protons and other particles at high energies, one can
get two sided bounds by considering multiple reactions, in the
same way one obtains two sided bounds in QED. As a simple example,
consider only dimension four rotationally invariant operators
(i.e. $n=2$ dispersion) and assume that all bosons propagate with
speed one while all fermions have a maximum speed of $1-\ep$. Let
us take the case $\ep<0$.  A proton is about half fermion and half
gauge boson, while a photon is 80 percent gauge boson and 20
percent fermion. The net effect, therefore, is that a proton
travels faster than a photon and hence \v{C}erenkov radiates.
Demanding that a $10^{20}$ eV proton not radiate yields the bound
$\ep>-10^{-23}$, similar to the standard \v{C}erenkov bound above.

If instead $\ep>0$, then $e^+e^-$ pair emission becomes possible
as electrons and positrons are 85 percent fermion and 15 percent
gauge boson.  Pair emission would also reduce the UHECR energy, so
one can demand that this reaction is forbidden as well.  This
yields the bound $\ep<10^{-23}$.  Combined with the above bound we
have $|\ep|<10^{-23}$, which is a strong two sided bound. The
parton approach yields two-sided bounds on dimension six operators
of order $|f^{(4)}|<O(10^{-2})$ for all constituent particles,
depending on the assumptions made about equal parton dispersion
corrections.  Bounds on the coefficients of CPT violating
dimension five operators are of the order
$10^{-15}$.\footnote{Dimension five CPT violating operators yield
helicity dependent dispersion.  To derive bounds on these
operators~\cite{Gagnon:2004xh} assumes that the coefficients are
roughly equal.} For the exact constraints and assumptions,
see~\cite{Gagnon:2004xh}. Note that if one treated electrons,
positrons, and protons as the fundamental constituents with only
$n=2$ dispersion and assigned each a common speed $1-\ep$ one
would obtain no constraints. Therefore the parton model is more
powerful. However, for higher dimension operators that yield
energy dependent dispersion, simply assigning electrons and
protons equal coefficients $f^{(n)}$ does yield comparable
constraints. Finally, we comment that~\cite{Gagnon:2004xh} does
not explicitly include possible effects such as SUSY that would
change the parton distribution functions at high energy.

\subsubsection{Gravitational \v{C}erenkov} \label{subsubsec:GravitonCerenkov}
High energy particles travelling faster than the speed of graviton
modes will emit graviton \v{C}erenkov radiation. The authors
of~\cite{Moore:2001bv} have analyzed the emission of gravitons
from a high energy particle with $n=2$ type dispersion and find
the rate to be
\begin{equation} \label{eq:dedetGrav}
\frac{dE} {dt}=\frac {G p^4} {3} (c_p-1)^2
\end{equation}
where $c_p$ is the speed of the particle and $G$ is Newton's
constant.  We have normalized the speed of gravity to be one.  The
corresponding constraint from the observation of high energy
cosmic rays is $c_p-1\leq 2 \cdot 10^{-15}$.  This bound assumes
that the cosmic rays are protons, uses the highest record energy
$3 \cdot 10^{20}$ eV, and assumes that the protons have travelled
over at least 10 kpc. Furthermore, the bound assumes that all the
cosmic ray protons travel at the same velocity, which is not the
case if CPT is violated or $d \neq 0$ in the mSME.

The corresponding bounds for $n=3,4$ type dispersion are not
known, but one can easily estimate their size.   The particle
speed is approximately $1+f^{(n)} (E/E_{Pl})^{n-2}$.  For a proton
at an energy of $10^{20}$ eV ($10^{-8} E_{Pl}$) the constraint on
the coefficient $f^{(3)}$ is then of O($10^{-7}$). Note though,
that in this case only one of the UHECR protons must satisfy this
bound due to helicity dependence. Similarly, the $n=4$ bound is of
O(10).

Equation (\ref{eq:dedetGrav}) only considers the effects of
Lorentz violation in the matter sector which give rise to a
difference in speeds, neglecting the effect of Lorentz violation
in the gravitational sector. Specifically, the analysis couples
matter only to the two standard graviton polarizations. However,
as we shall see in Sec. \ref{subsec:GravitationalWaves},
consistent Lorentz violation with gravity can introduce new
gravitational polarizations with different speeds.  In the aether
theory (section \ref{subsubsec:LVandgravity}) there are three new
modes, corresponding to the three new degrees of freedom
introduced by the constrained aether vector. The corresponding
\v{C}erenkov constraint from possible emission of these new modes
has recently been analyzed in~\cite{Elliott:2005va}. Demanding
that high energy cosmic rays not emit these extra modes and
assuming no significant Lorentz violation for cosmic rays yields
the bounds
\begin{eqnarray} \nonumber
-c_1-c_3 < 1 \times 10^{-15} \\ \nonumber %
\frac {(c_1+c_3)^2
(c_1^2 + 3 c_1 c_3 -2c_4)} {c_1^2} < 1.4 \times 10^{-31}\\
\nonumber %
\frac{(c_3 - c_4)^2} {|c_1+c_4|} < 1 \times 10^{-30}\\
\frac {c_4-c_2-c_3} {c_1} < 3 \times 10^{-19}
\end{eqnarray}
on the coefficents in (\ref{eq:aethercoeffs}).  The next to last
bound requires that $(c_4-c_2-c_3)/c_1 > 10^{-22}$.  If, as the
authors of~\cite{Elliott:2005va} argue, no gravity-aether mode can
be superluminal, then these bounds imply that every coefficient is
generically bounded by $|c_i| < 10^{-15}$. There is, however, a
special case given by $c_3=-c_1, c_4=0, c_2=c_1/(1-2c_1)$ where
all the modes propagate at exactly the speed of light and hence
avoid this bound.

\subsection{Threshold reactions in other models}
\label{subsec:ThresholdsOther}
\subsubsection{Thresholds and DSR} \label{subsubsec:ThreshDSR}
Doubly special relativity modifies not only particle dispersion
relation but also the form of the energy conservation equations.
The situation is therefore very different from that in EFT. The
first difference between DSR and EFT is that DSR evades all of the
photon decay and vacuum \v{C}erenkov constraints that give strong
limits on EFT Lorentz violation. Since there is no EFT type
description of particles and fields in a DSR framework, one has no
dynamics and cannot calculate reaction rates.  However, one still
can use the DSR conservation laws to analyze the threshold
kinematics. By using the pseudo-momentum and and energy $\pi, \ep$
one can show that if a reaction does not occur in ordinary Lorentz
invariant physics, it does not occur in DSR~\cite{Heyman:2003hs}.
Physically, this is obvious.  If the vacuum \v{C}erenkov effect
for, say. electrons began to occur at some energy $E_{th}$, in a
different reference frame the reaction would occur at some other
energy $E_{th}'$, as the threshold energy is not an invariant.
Therefore frames could be distinguished by labelling them
according to the energy when the vacuum \v{C}erenkov effect for
electrons begins to occur.  This violates the equivalence of all
inertial frames that is postulated in DSR theories.

A signal of DSR in threshold reactions would be a shift of the
threshold energies for reactions that do occur, such as the GZK
reaction or $\gamma$-ray annihilation off the infrared
background~\cite{Amelino-Camelia:2003ex}.  However, the actual
shift of threshold energies due to DSR is negligible at the level
of sensitivity we have with astrophysical
observations~\cite{Amelino-Camelia:2003ex}. Hence DSR cannot be
ruled out or confirmed by any threshold type analysis we currently
have.  The observational signature of DSR would therefore be a
possible energy dependence of the speed of light (section
\ref{subsec:TimeOfFlight}) without any appreciable change in
particle thresholds~\cite{Smolin:2005cz}.

\subsubsection{Thresholds and non-systematic dispersion} \label{subsubsec:Threshnonsystem}
Similar to DSR, the lack of dynamics in the non-systematic
dispersion framework of Sec. \ref{subsec:nonsystematicdisp} makes
it more problematic to set bounds on the parameters $f^{(n)}$.
In~\cite{Jankiewicz:2003sm, Aloisio:2000cm,Aloisio:2002ed,
Amelino-Camelia:2002ws}, the authors assume that the net effect of
spacetime foam can be derived by considering energy conservation
and non-systematic dispersions at a point.  There is a difficulty
with this, which we shall address, but for now let us assume that
this approach is correct.

As an example of the consequences of non-systematic dispersion let
us consider the analysis of the GZK reaction
in~\cite{Aloisio:2002ed}. The authors consider $n=3$ nonsystematic
dispersion relations with normally distributed coefficients
$f^{(3)}_{p,\pi}$ that can take either sign and have a variance of
O(1).  Looking solely at the kinematical threshold condition, they
find that all cosmic ray protons would undergo photo-pion
production at energies above $10^{15}$ eV.  This is perhaps
expected, as the energy scale that an $n=3$ term becomes important
is $E \approx (m^2E_{Pl}/f^{(3)})^{1/3} \approx 10^{15}$ eV for
$f^{(3)}$ of O(1).  There is a large region of the
$f^{(3)}_{p,\pi}$ parameter space that is susceptible to the
vacuum \v{C}erenkov effect with pion
emission~\cite{Jacobson:2002hd} and hence a significant amount of
the time the random coefficients will fall in this region of
parameter space.  If an ultra-high energy proton can emit a pion
without scattering off of the CMBR, then certainly it can scatter
as well, which implies that GZK reaction is also accessible.  This
same type of argument can be rapidly extended to $n=4$ dispersion,
yielding a cutoff in the spectrum at $10^{18}$ eV.  The $n=4$
cutoff could easily be pushed above GZK energies if the
coefficients had a variance slightly less than O(1). In short,
since we see high energy cosmic rays at energies of $10^{20}$ eV,
the results of~\cite{Jankiewicz:2003sm,
Aloisio:2000cm,Aloisio:2002ed, Amelino-Camelia:2002ws} imply that
we could not have $n=3$ non-systematic dispersion unless the
coefficients are $\ll O(1)$, while for $n=4$ the coefficients
would only have to be an order of magnitude or two below O(1).

We now return to a possible problem with this type of analysis,
which has been raised in~\cite{Basu:2005kt}.  Performing threshold
analyses on non-systematic dispersion assumes that energy-momentum
conservation can be applied with a single fluctuation (i.e. the
reaction effectively happens at a point). It further assumes that
the matrix element is roughly unchanged. In GZK or \v{C}erenkov
reactions, however, one of the outgoing particles is much softer
than the incoming particle. In this situation the interaction
region is much larger than the de Broglie wavelength of the high
energy incoming particle, which means that many dispersion
fluctuations will occur during the interaction.  The amplitude of
low energy emission in regular quantum field theory changes
dramatically in this situation (e.g. Bremsstrahlung with a rapidly
wiggling source) as opposed to the case in which there is only one
fluctuation (e.g. the \v{C}erenkov effect). The above approach,
modified conservation plus unchanged matrix element/rate when the
reaction is allowed, is not correct when a low energy particle is
involved. If the outgoing particle has an energy comparable to the
incoming particle, then it may be possible to avoid this problem.
However, in this case the reverse reaction is also kinematically
possible with a different fluctuation of the same order of
magnitude, so it is unclear what the net effect on the spectrum
should be. Note, finally, that these arguments only concern the
rate of decay - the conclusion that high energy particles would
decay in this framework is unchanged.

\subsection{Synchrotron radiation} \label{subsec:Synchrotron}
Ref.~\cite{Jacobson:2002ye} noticed that the synchrotron emission
from the Crab nebula (and other astrophysical objects) is very
sensitive to $n=3$ modified dispersion relations. The logic behind
the constraint in~\cite{Jacobson:2002ye} is as follows.  The Crab
nebula emits electromagnetic radiation from radio to multi-TeV
frequencies.  As noted in section \ref{subsubsec:QEDParticles} the
spectrum is well fit over almost the entire frequency range by the
synchrotron self-Compton model. If this model is correct, as seems
highly probable, then the observed radiation from the Crab at 100
MeV is due to synchrotron emission from high energy electrons
and/or positrons.  The authors of~\cite{Jacobson:2002ye}, in the
context of effective field theory, argued that the maximum
frequency $\omega_c$ of synchrotron radiation in the presence of
Lorentz violation is given by
\begin{equation} \label{eq:synchmax}
\omega_c=\frac {3} {2} \frac {eB \gamma^3} {E}
\end{equation}
where e is the charge, $B$ is the magnetic field, and $\gamma,E$
are the gamma factor and energy of the source particle. The
derivation of (\ref{eq:synchmax}) was challenged
in~\cite{Castorina:2004rc}.  More detailed
calculations~\cite{Montemayor:2004mh,
Ellis:2003sd,Montemayor:2005ka} show that in the case of the Crab
nebula (\ref{eq:synchmax}) is correct, although the argument
of~\cite{Jacobson:2002ye} does not necessarily hold in
general\footnote{For a discussion of synchrotron radiation in
other models with Planck suppressed dispersion corrections
see~\cite{Gonzalez:2005qa}.}. Assuming the source particles are
electrons, if $f^{(3)}_e<0$ then there is a maximum electron
velocity for any energy particle and hence a maximum possible
value for $\omega_c$. The maximum frequency must be above 100 MeV
in the Crab, which leads to a constraint of either $f^{(3)}_{e,R}$
or $f^{(3)}_{e,L}
>-7 \cdot 10^{-8}$, i.e. at least one of the electron parameters
must be above this value.

The analysis of~\cite{Jacobson:2002ye} does not take into account
the possibility that the high energy synchrotron emission could be
due to positrons, which may also be generated near the pulsar.
This is an important possibility, since in the EFT that gives rise
to $f^{(3)}$ terms for electrons, (\ref{eq:MPops}), the positron
has an opposite dispersion modification.  Hence there is always
some charged particle in the Crab with a dispersion modification
that evades the synchrotron constraint.  The possibility that
there are two different populations, one electrons and one
positrons, that contribute to the overall spectrum would be a
departure from the synchrotron-self Compton model, which
presupposes only one population of particles injected into the
nebula.  However, such a possibility cannot be ruled out without
more detailed modelling of the Crab nebula and a better
understanding of how the initial injection spectrum of particles
from the pulsar is produced.\footnote{It has recently been
suggested that Lorentz violation might play a role in the
formation of the pulsar itself, specifically in large pulsar
anomalous velocities~\cite{Lambiase:2005iz}.}

The possible importance of positrons in the Crab implies that the
synchrotron constraint is always satisfied when considered by
itself.  However, the synchrotron constraint can be combined with
\v{C}erenkov constraints to create a two sided
bound~\cite{Jacobson:2005bg}. Essentially, the SSC model is such
that whatever species of particle is producing the synchrotron
spectrum must also be responsible for the inverse Compton
spectrum.  Hence at least one helicity of electron or positron
must satisfy \textit{both} the vacuum \v{C}erenkov and synchrotron
constraints.  This is not automatically satisfied in EFT and
constitutes a true constraint.  For more discussion
see~\cite{Jacobson:2005bg}. The combined synchrotron, threshold,
birefringence, and time of flight constraints are displayed in
Figure \ref{fig:QEDn3} (taken from ~\cite{Jacobson:2005bg}). In
this plot $\eta_\pm$ is equal to $f^{(3)}_{e(R,L)}$ (not to be
confused with the $\eta_{R,L}$ in (\ref{eq:MPops}) which differs
by a factor of two).

\epubtkImage{}{
\begin{figure}[htbp]
\def\epsfsize#1#2{#1}
\centerline{\epsfbox{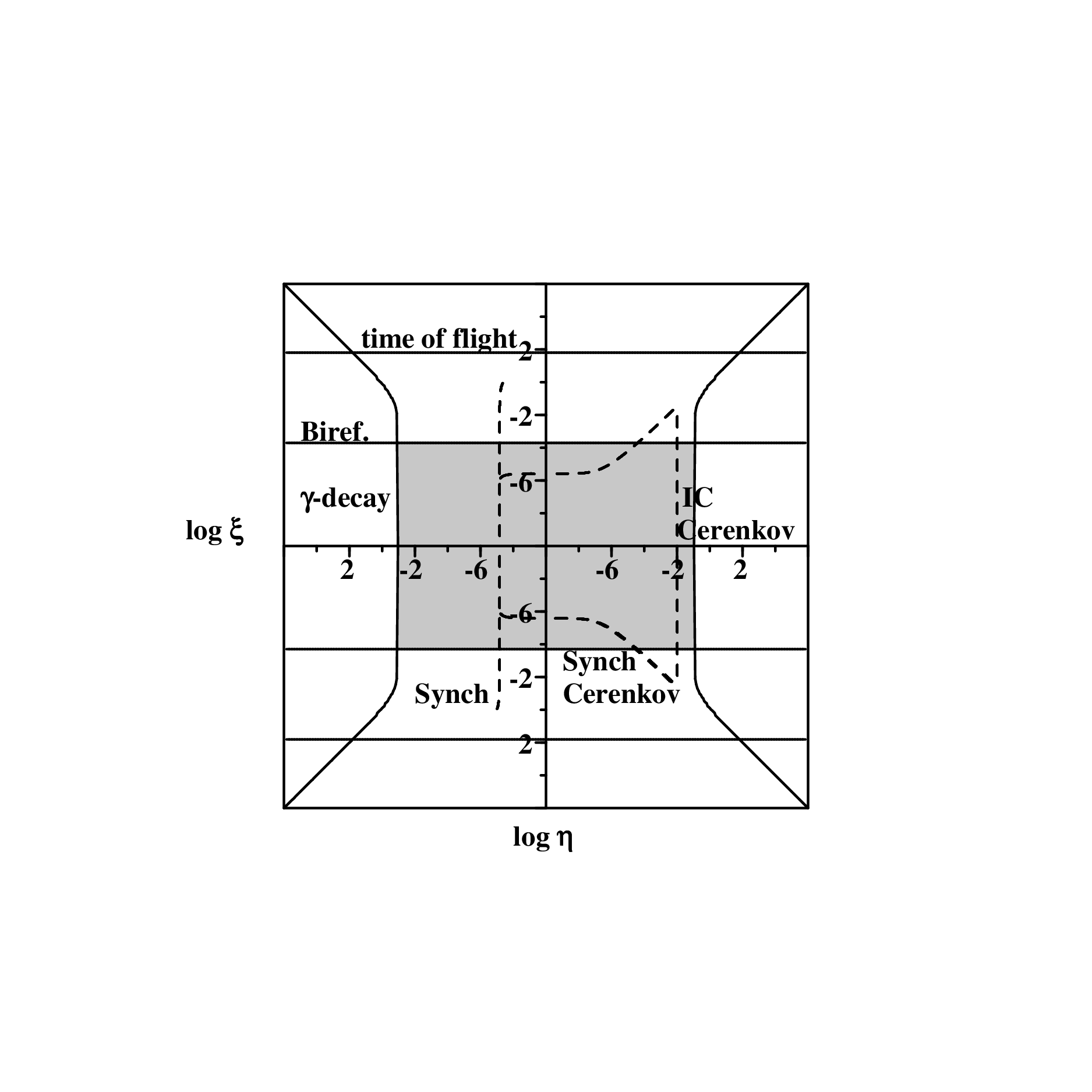}} \caption{Constraints on LV in QED
at $n=3$ on a log-log plot. For negative parameters minus the
logarithm of the absolute value is plotted, and region of width
$10^{-10}$ is excised around each axis. The constraints in solid
lines apply to $\xi$ and both $\eta_\pm$, and are symmetric about
both the $\xi$ and the $\eta$ axis. At least one of the two pairs
$(\eta_\pm,\xi)$ must lie within the union of the dashed
bell-shaped region and its reflection about the $\xi$ axis.
Intersecting lines are truncated where they cross.}
  \label{fig:QEDn3}
\end{figure}
}

Finally, note that if the effective field theory is CPT conserving
then positrons/electrons have the same dispersion relation.  So,
for $n=2$ dispersion the 100 MeV synchrotron radiation from the
Crab yields a parallel constraint of roughly
$f^{(2)}_{e,e^+}>-10^{-22}$ for at least one helicity of
electron/positron.

\subsection{Neutrinos} \label{subsec:Neutrinos}
Neutrinos can provide excellent probes of Lorentz violation, as
their mass is much smaller than any other known particle.  To see
this consider the modified dispersion framework.  For an electron
with $n=3,4$ dispersion the energies at which Lorentz violation
can become appreciable are at $10, 10^5$ TeV respectively.
However, for a neutrino with mass even at 1 eV the corresponding
energies are only 1 GeV for $n=3$ and 1 TeV for $n=4$, well within
the realm of accelerator physics. The most sensitive tests of
Lorentz violation in the neutrino sector come from neutrino
oscillation experiments, which we now describe. For a more
comprehensive overview of neutrino mixing, see for
example~\cite{Eidelman:2004wy, Kayser:2001ki}.

\subsubsection{Neutrino oscillations}
Lorentz violating effects in the neutrino sector have been
considered by many authors~\cite{Coleman:1997xq,
Kostelecky:2004hg, Kostelecky:2003xn, Kostelecky:2003cr,
Christian:2004xb, Blasone:2003wf,
Amelino-Camelia:2003nj,Glashow:2004im, Lambiase:2003sq,
Choubey:2003ke}. To illustrate how Lorentz violation affects
neutrino propagation, we consider the simplest case, where the
limiting speeds for mass eigenstates of the neutrino are
different, i.e. neutrinos have dispersions
\begin{equation}
E^2=(1+f^{(2)}_{\nu_i}) p^2+ m_i^2
\end{equation}
where $i$ denotes the energy eigenstate.  In this case, the energy
eigenstates are also the mass eigenstates (this is not necessarily
the case with general Lorentz violation).  This is a special case
of the neutrino sector of the mSME if we assume that
$c_{\nu}^{00}$ is flavor diagonal and is the only non-zero term.
For relativistic neutrinos, we can expand the energy to be
\begin{equation}
E=(1+\frac{f^{(2)}_{\nu_i}} {2}) p + \frac {m_i^2} {2p}.
\end{equation}
Now consider a neutrino produced via a particle reaction in a
definite flavor eigenstate $I$ with energy E.  We denote the
amplitude for this neutrino to be in a particular energy
eigenstate $i$ by the matrix $U_{Ii}$ where $\sum_i U_{Ji}^\dag
U_{Ii}=\delta_{IJ}$. The amplitude for the neutrino to be observed
in another flavor eigenstate $J$ at some distance $L,T$ from the
source is then
\begin{equation}
A_{IJ}=\sum_i U_{Ji}^\dag e^{-i(ET-pL)} U_{Ii}\approx \sum_i
U_{Ji} ^\dag e^{-i(\frac{f^{(2)}_{\nu_i}} {2} E + \frac {m_i^2}
{2E}) L } U_{Ii}
\end{equation}
for relativistic neutrinos.  If we define an ``effective mass''
$N_i$ as
\begin{equation}
N_i^2=m_i^2 + f^{(2)}_{\nu_i} E^2
\end{equation}
then the probability $P_{IJ}=|A_{IJ}|^2$ can be written as,
\begin{equation} \label{eq:oscprob}
P_{IJ}=\delta_{IJ}-\sum_{i,j>i}  4F_{IJij} sin^2 \bigg{(}\frac
{\delta N_{ij}^2 L} {4E} \bigg{)} + 2 G_{IJij} sin \bigg{(}\frac
{\delta N_{ij}^2 L} {2E} \bigg{)}
\end{equation}
where $\delta N_{ij}= N_i^2-N_j^2$ and $F_{IJij}$, $G_{IJij}$ are
functions of the $U$ matrices.

We can immediately see from (\ref{eq:oscprob}) that Lorentz
violation can have a number of consequences for standard neutrino
oscillation experiments.  The first is simply that neutrino
oscillation still occurs even if the mass is zero.  In fact, some
authors have proposed that Lorentz violation could be partly
responsible for the observed
oscillations~\cite{Kostelecky:2003xn}.  Oscillations due to the
type of Lorentz violation above vary as
$EL$~\cite{Kostelecky:2003xn}. Current data support neutrino
oscillations that vary as a function of
$L/E$~\cite{Ashie:2004mr,Giacomelli:2004cw} so it seems unlikely
that Lorentz violation could be the sole source of neutrino
oscillations.  It is possible, however, that Lorentz violation may
explain some of the current problems in neutrino physics by giving
a contribution in addition to the mass term. For example it has
been proposed in~\cite{Kostelecky:2004hg} that Lorentz violation
might explain the LSND (Liquid Scintillator Neutrino Detector)
anomaly~\cite{Athanassopoulos:1997pv}.\footnote{There are other
attempts to explain the LSND results with CPT
violation~\cite{Murayama:2000hm, Barenboim:2001ac,
Barenboim:2002hx}.  However, these CPT violating models are not
directly correlated with Lorentz violation as they may involve
non-local field theories.  It is also seems likely that the LSND
result is simply incorrect.}, which is an excess of
$\overline{\nu}_\mu \rightarrow \overline{\nu}_\mu$ events that
cannot be reconciled with other neutrino
experiments~\cite{Gonzalez-Garcia:2002dz}. We note that the above
model for Lorentz violating effects in neutrino oscillations is
perhaps the simplest case. In the neutrino sector of the mSME
there can be more complicated energy dependence, directional
dependence, and new oscillations that do not occur in the standard
model.  For a discussion of these various possibilities
see~\cite{Kostelecky:2003cr}.

The difference in speeds between electron and muon neutrinos was
bounded in~\cite{Coleman:1998ti} to be $|f^{(2)}_{\nu_e}-
f^{(2)}_{\nu_\mu}|<10^{-22}$.  Oscillation data from Super
Kamiokande have improved this bound to
$O(10^{-24})$~\cite{Fogli:1999fs}.  Current neutrino oscillation
experiments are projected to improve on this by three orders of
magnitude, giving limits on maximal speed differences of order
$10^{-25}$~\cite{Glashow:2004im}. For comparison, the time of
flight measurements from supernova 1987A constrain
$|f^{(2)}_{\nu_i}-f^{(2)}_\gamma|
<10^{-8}$~\cite{Stodolsky:1987vd}.  Neutrino oscillations are
sensitive enough to directly probe non-renormalizable Lorentz
violating terms.  In~\cite{Brustein:2001ik} current neutrino
oscillation experiments are shown to yield bounds on dimension
five operators stringent enough that the energy scale suppressing
the operator must be a few orders of magnitude above the Planck
energy.  Such operators are therefore very unlikely in the
neutrino sector. Ultra-high energy neutrinos, when observed, will
provide further information about neutrino Lorentz violation. For
example, flavor oscillations of ultra-high energy neutrinos at
$10^{21} eV$ propagating over cosmic distances would be able to
probe Lorentz violating dispersion suppressed by seven powers of
$E_{Planck}$~\cite{Christian:2004xb} (or more if the energies are
even higher).

Additionally, neutrino Lorentz violation can modify the energy
thresholds for reactions involving neutrinos, which can have
consequences for the expected flux of ultra-high energy neutrinos
for detectors such as ICECUBE. The expected flux of ultra-high
energy neutrinos is bounded above by the Bahcall-Waxman
bound~\cite{Waxman:1998yy} if the neutrinos are produced in active
galactic nuclei or gamma ray bursters.  It has been
shown~\cite{Amelino-Camelia:2003nj} that Lorentz violation can in
fact raise (or lower) this bound significantly. A higher than
expected ultra-high energy neutrino flux therefore could be a
signal of Lorentz violation.

\subsubsection{Neutrino \v{C}erenkov effect}
\label{subsubsec:NeutrinoCerenkov}

Finally, neutrinos can also undergo a vacuum \v{C}erenkov effect.
Even though a neutrino is neutral there is a non-zero matrix
element for interaction with a photon as well as a graviton.
Graviton emission is very strongly suppressed and unlikely to give
any useful constraints.  The matrix element for photon emission,
while small, is still larger than that for graviton emission and
hence the photon \v{C}erenkov effect is more promising. The
photon-neutrino matrix element can be split into two channels, a
charge radius term and a magnetic moment term. The charge radius
interaction is suppressed by the $W$ mass, leading to a reaction
rate too low for current neutrino observatories such as AMANDA to
constrain $n=3,4$ Lorentz violation. However, the rate from the
charge radius interaction scales strongly with energy, and it has
been estimated~\cite{Jacobson:2002hd} that atmospheric PeV
neutrinos may provide good constraints on $n=3$ Lorentz violation.
The magnetic moment interaction has not yet been conclusively
analyzed, so possible constraints from the magnetic moment
interaction are unknown.  In Lorentz invariant physics, the
magnetic moment term is suppressed by the small neutrino mass, so
energy loss rates are likely small. However, it should be noted
that some Lorentz violating terms in an effective field theory
give rise to effective masses that scale with energy. These might
be much larger than the usual neutrino mass at high energies,
yielding a large neutrino magnetic moment.

\subsection{Phase coherence of light} \label{subsec:PhaseCoherence}
An interesting and less well known method of constraining
non-systematic Lorentz violation is looking at Airy rings
(interference fringes) from distant astrophysical objects.  In
order for an interference pattern from an astrophysical source to
be observed, the photons reaching the detector must be in phase
across the detector surface.  However, if the dispersion relation
is fluctuating, then the phase velocity $v_\phi=\om/k$ is also
changing.  If the fluctuations are uncorrelated then an initially
in phase collections of photons will lose phase coherence as they
propagate.  Uncorrelated fluctuations are reasonable, since for
most of their propagation time, photons that strike across a
telescope mirror are separated by macroscopic distances.
Observation of Airy rings implies that photons are in phase and
hence limits the fluctuations in the dispersion
relation~\cite{Lieu:2003ee,Ragazzoni:2003tn,Ng:2003ag}. The
aggregate phase fluctuation is given by~\cite{Ng:2003ag}
\begin{equation}
\Delta \phi=2 \pi f^{(n)} \frac {L_{Pl}^{n-2} D^{3-n}} {\la}
\end{equation}
where $L_{Pl}$ is the Planck length, D is the distance to the
source, and $\la$ is the wavelength of the observed light. This
technique was originally applied by~\cite{Lieu:2003ee}, but the
magnitude of the aggregate phase shift was overestimated.
PKS1413+135, a galaxy at a distance of 1.2 Gpc, shows Airy rings
at a wavelength of $1.6 \mu m$.  Demanding that the overall phase
shift is less than $2 \pi$ yields O(1) constraints for $n=5/2$ and
constraints of order $10^9$ for $n=8/3$.  Hence this type of
constraint is only able to minimally constrain Lorentz violating
non-systematic models.  In principle the frequency of light used
for the measurement can be increased, however, in which case this
type of constraint will improve.  However,Ref.~\cite{Coule:2003td}
has argued that other effects mask the loss of phase coherence
from quantum gravity, making even this approach uncertain.

\newpage

\section{Gravitational observations} \label{sec:GravitationalObservations}
So far we have restricted ourselves to Lorentz violating tests
involving matter fields.  It is also possible that Lorentz
violation might manifest itself in the gravitational sector. There
are three obvious areas where the consequences of such Lorentz
violation might manifest itself: gravitational waves, cosmology,
and post-Newtonian corrections to weak field general relativity.

\subsection{Gravitational waves} \label{subsec:GravitationalWaves}
In the presence of dynamical Lorentz violation, where the entire
action is diffeomorphism invariant, one generically expects new
gravitational wave polarizations~\footnote{There are exceptions,
for example see~\cite{Jackiw:2003pm}.  Here gravity is modified by
a Chern-Simons form, yet there are still only two gravitational
wave polarizations. The only modification is that the intensity of
the polarizations differs from what would be expected in general
relativity.}. The reason is simple. Any dynamical Lorentz
violating tensor field must have kinetic terms involving
derivatives of the form $\nabla_\mu U^{\al \be ...}$ where $U^{\al
\be ...}$ is the Lorentz violating tensor. Furthermore, $U$ must
take a non-zero vacuum expectation value if it violates Lorentz
invariance. At linear order in the perturbations $h_{\al \be},
u^{\al \be...}$ (where $g_{\al \be}=\eta_{\al \be} + h_{\al \be},
U^{\al \be...} = <U^{\al \be...}> + u^{\al \be...}$), the
connection terms in the covariant derivative are also first order,
for example $\partial_\al h_{\be \ga} <U^{\be \de ...}>$.  Upon
varying the linearized metric, these terms contribute to the
graviton equations of motion. The extra terms in the graviton
equations give rise to new solutions. Since the potential that
forces $U$ to take a non-zero vacuum expectation value must
involve the metric, variations in $U$ are usually coupled to
metric variations, implying that the new graviton modes mix with
excitations of the Lorentz violating tensor fields.

There is a large literature on gravitational wave polarizations in
theories of gravity other than general relativity.  For a thorough
discussion, see~\cite{Will:2001mx} and references therein.  Many
of the models with preferred frame effects are similar to the
types of theories that give rise to dynamical Lorentz violation.
For example, the vector-tensor theories of Will, Hellings and
Nordtvedt~\cite{WillNord1,WillNord2, NordHellings} have many
similarities to the aether theory of section
\ref{subsubsec:LVandgravity}.  The aether model's wave spectrum
has been calculated in~\cite{Jacobson:2004ts, Gripaios:2004ms} and
limits from the absence of \v{C}erenkov emission of these modes by
cosmic rays has been studied in~\cite{Elliott:2005va} (see section
\ref{subsubsec:GravitonCerenkov}). Other consequences of dynamical
Lorentz violation in Riemann-Cartan spacetimes have been examined
in~\cite{Bluhm:2004ep}.

Unfortunately, few constraints currently exist on dynamical
Lorentz violation from gravitational wave observations as the
spectrum is only part of the story. Currently, the expected rate
of production of these modes from astrophysical sources as a
function of the coefficients in the Lagrangian is unknown.
However, both the energy loss from inspiral systems due to
gravitational radiation and gravitational wave observatories such
as LIGO and LISA should produce strict bounds on the possibility
of dynamical Lorentz violating fields.\footnote{It has also been
proposed that laser interferometry may eventually be capable of
direct tests of Planck suppressed LV
dispersion~\cite{Amelino-Camelia:2003zf}.} We note that aether
type theories seem to be free of certain obvious problems such as
a van Dam-Veltman-Zakharov type
discontinuity~\cite{Gripaios:2004ms}. The theories can therefore
be made arbitrarily close to GR by tuning the coefficients to be
near zero.

\subsection{Cosmology} \label{subsec:Cosmology}
Cosmology also provides a way to test Lorentz violation.  The most
obvious connection is via inflation.  If the number of e-foldings
of inflation is high enough, then the density fluctuations
responsible for the observed cosmic microwave background (CMB)
spectrum have a size shorter than the Planck scale before
inflation.  It might therefore be possible for trans-Planckian
physics/quantum gravity to influence the currently observed CMB
spectrum. If Lorentz violation is present at or near the Planck
scale (as is implicit in models that use a modified dispersion
relation at high energies~\cite{Martin:2002kt}), then the
microwave background may still carry an imprint.\footnote{The
B-mode polarization of the CMB might also carry an imprint of
Lorentz violation due to modifications in the gravitational
sector~\cite{Lim:2004js}.} A number of authors have addressed the
possible signatures of trans-Planckian physics in the CMB, for a
sampling see
~\cite{Danielsson:2004mf,Martin:2000xs,Martin:2003kp,Starobinsky:2001kn,Easther:2001fi,Kaloper:2002uj,Brandenberger:2002hs,Niemeyer:2002kh}
and references therein. While the possibility of such constraints
is obviously appealing, the CMB imprint (if any) of
trans-Planckian physics, much less Lorentz violation, is model
dependent and currently the subject of much debate.\footnote{The
above approach presumes inflation and speculates about the low
energy signature of Lorentz violating physics.  Lorentz violation
can also be a component in the so-called variable speed of light
(VSL) cosmologies (for a review see~\cite{Magueijo:2003gj}) which
are a possible alternative to inflation.  Some bounds on VSL
theories are known from Lorentz symmetry tests, but in these cases
the VSL model can be equivalently expressed in one of the
frameworks of this review.}  In short, although such cosmological
explorations are interesting and may provide an eventual method
for ultra-high energy tests of Lorentz invariance, for the
purposes of this review we forego any more discussion on this
approach.

A simple low energy method to limit the coefficients in the aether
model (\ref{eq:aetheraction}) that is less fraught with
ambiguities has been explored by Carroll and
Lim~\cite{Carroll:2004ai}. They consider a simplified version of
the model (\ref{eq:aetheraction}) without the $c_4$ term and
choose the potential $V(u_\al u^\al)$ to be of the form $\lambda
(u^\al u_\al-a^2)$ where $\lambda$ is a Lagrange multiplier.
Without loss of generality, we can rescale the coefficients $c_i$
in (\ref{eq:aetheraction}) to set $a^2=1$.  In the Newtonian limit
Carroll and Lim find that Newton's constant as measured on earth
is rescaled to be
\begin{equation}
G_N^{obs}=\frac {2G} {2 - c_1}.
\end{equation}
In comparison, the effective cosmological Newton's constant is
calculated to be
\begin{equation}
G_{cosmo}^{obs}=\frac {2G} {2 -  (c_1+3c_2+c_3)}.
\end{equation}

The difference between the cosmological and Newtonian regimes
implies that we have to adjust our measured Newton's constant
before we insert it into cosmological evolution equations.  Such
an adjustment modifies the rate of expansion.  A change in the
expansion rate modifies big bang nucleosynthesis and changes the
ratio of the primordial abundance of $^4He$ to $H$.  By comparing
this effect with observed nucleosynthesis limits, Carroll and Lim
are able to constrain the size of $c_1,c_2,c_3$.  In addition to
the nucleosynthesis constraint, the authors impose restrictions on
the choice of coefficients such that in the preferred frame
characterized by $\bar{u}^\al$ the perturbations $\delta u^\al$
have a positive definite Hamiltonian, are non-tachyonic, and
propagate subluminally.  With these assumptions Carroll and Lim
find the following constraint
\begin{equation}
0<(14c_1+21c_2+7c_3 + 7 c_4)<2
\end{equation}
where the $c_4$ dependence has been included for completeness.

\subsection{PPN Parameters} \label{subsec:PPN}
Preferred frame effects, as might be expected from Lorentz
violating theories, are nicely summarized in the parameterized
post-Newtonian formalism, otherwise known as PPN (for a
description, see~\cite{Will:2001mx} or~\cite{Will:1993ns}).  The
simplest setting in which the PPN parameters might be different
than GR is in the static, spherically symmetric case. For static,
spherically symmetric solutions in vector-tensor models the only
PPN parameters that do not vanish are the
Eddington-Robertson-Schiff (ERS) parameters $\gamma$ and $\beta$.
For GR, $\beta=\gamma=1$. The ERS parameters for the general
Hellings-Nordvedt vector-tensor theory~\cite{NordHellings} are not
necessarily unity~\cite{Will:1993ns}, so one might expect that the
constrained aether model also has non-trivial ERS parameters.
However, it turns out that the constrained aether model with the
Lagrange multiplier potential also has $\beta=\gamma=1$ for
generic choices of the coefficients~\cite{Eling:2003rd}.
Therefore, at this point there is no method by which the ERS
parameters can be used to constrain Lorentz violating theories.
The ERS parameters for more complicated theories with higher rank
Lorentz violating tensors are largely unknown.

If we move away from spherical symmetry then more PPN parameters
become important, in particular $\al_1,\al_2,\al_3$ which give
preferred frame effects.  In~\cite{Graesser:2005bg} $\al_2$ has
been calculated to be
\begin{eqnarray} \al_2 &=& -\frac{1}{2c_1(c_1+c_2+c_3)}  \left[2 c_1^3 + 4c^2_3(c_2 +c_3)
 +c^2_1(3 c_2+5 c_3  +3 c_4) \right. \\ & & \left. + c_1( (6c_3 -c_4)(c_3 + c_4) +c_2(6 c_3 +c_4))
\right].
\end{eqnarray}
The observational limit on $\al_2$ is $|\al_2|<4 \cdot
10^{-7}$~\cite{Will:2001mx}.  Barring cancellations this
translates to a very strong bound of order $10^{-7}$ on the
coefficients $c_i$ in the aether action.

\newpage
\newpage

\section{Conclusions and prospects} \label{sec:Conclusions}
As we have seen, over the last decade or two a tremendous amount
of progress has been made in tests of Lorentz invariance.
Currently, we have no experimental evidence that Lorentz symmetry
is not an exact symmetry in nature.  The only not fully understood
experiments where Lorentz violation might play a role is in the
(possible) absence of the GZK cutoff and the LSND anomaly. New
experiments such as AUGER, a cosmic ray telescope, and
MiniBooNE~\cite{Boone}, a neutrino oscillation experiment
specifically designed to test the LSND result, may resolve the
experimental status of both systems and allow us to determine if
Lorentz violation plays a role.

Terrestrial experiments will continue to improve. Cold
anti-hydrogen can now be produced in enough
quantities~\cite{Gabrielse:2002xc, Amoretti:2004ci} for
hydrogen/anti-hydrogen spectroscopy to be performed.  The
frequency of various atomic transitions (1S-2S, 2S-nd, etc.) can
be observationally determined with enough precision to improve
bounds on various mSME parameters~\cite{Bluhm:1998rk,
Shore:2004sh}. Spectroscopy of hydrogen-deuterium molecules might
lead to limits on electron mSME parameters an order of magnitude
better than current cavity experiments~\cite{Muller:2004tc}.

There are proposals for space based experiments (c.f.
~\cite{Bluhm:2003un,Lammerzahl:2004bm}) that will extend current
constraints from terrestrial experiments. Space based experiments
are ideal for testing Lorentz violation.  They can be better
isolated from contaminating effects like seismic noise.  In a
microgravity environment interferometers can run for much longer
periods of time as the cooled atoms in the system will not fall
out of the interferometer. As well, the rate of rotation can be
controlled. Sidereal variation experiments look for time dependent
effects due to rotations.  In space, the rate of rotation can be
better controlled, which allows the frequency of any possible
time-dependent signal to be tuned to achieve the best
signal-to-noise ratio.  Furthermore, space based experiments allow
for cavity and atomic clock comparison measurements to be combined
with time dilation experiments (as proposed in
OPTIS~\cite{Lammerzahl:2004bm}), thereby testing all the
fundamental assumptions of special relativity. The estimated level
of improvement from a space based mission such as OPTIS over the
corresponding terrestrial experiments is a few orders of
magnitude.

Another possibility for seeing a novel signal of Lorentz violation
is in GLAST~\cite{GLAST}.  GLAST is a gamma ray telescope that is
very sensitive to extremely high energy gamma ray bursts. As we
have mentioned, DSR evades almost all known high energy tests of
Lorentz invariance. If the theoretical issues are straightened out
and DSR does eventually predict a time of flight effect then GLAST
may be able to see it for some burst events. An unambiguous
frequency to time-of-arrival correlation linearly suppressed in
the Planck energy, coupled with the observed lack of birefringence
at the same order, will be a smoking gun for DSR, as other
constraints forbid such a construction in effective field
theory~\cite{Myers:2003fd}.

The question that must be asked at this juncture in regards to
Lorentz invariance is: when have we tested enough?  We currently
have bounds on Lorentz violation strong enough that there is no
easy way to put Lorentz violating operators of dimension $\leq 6$
coming solely from Planck scale physics into our field theories.
It therefore seems hard to believe that Lorentz invariance could
be violated in a simple way. If we are fortunate, the strong
constraints we currently have will force us to restrict the
classes of quantum gravity theories/spacetime models we should
consider. Without a positive signal of Lorentz violation, this is
all that can reasonably be hoped for.

\section{Acknowledgements}
\label{sec:acknowledgements} I would like to thank Steve Carlip,
Ted Jacobson, Stefano Liberati, Sayandeb Basu, and Damien Martin
for helpful comments on early drafts of this paper.  As well, I'd
like to thank Bob McElrath and Nemanja Kaloper for useful
discussions. This work was funded under DOE grant
DE-FG02-91ER40674.


\newpage

\bibliography{refs}

\end{document}